\documentclass[a4paper,11pt]{article}
\pdfoutput=1 

\usepackage{jheppub} 

\usepackage[T1]{fontenc} 
\usepackage{graphicx}
\usepackage{epstopdf}
\usepackage{bm}
\usepackage{epsfig}
\usepackage{graphics}
\usepackage{xspace}
\usepackage{hyperref}
\usepackage{slashed}
\usepackage{xcolor}
\usepackage{longtable}
\usepackage[small]{subfigure}
\usepackage[normalem]{ulem}
\usepackage{multirow}

\newcommand{\nl}{n_{\ell}}
\newcommand{\nf}{n_f}
\newcommand{\mbar}{\overline{m}}
\newcommand{\MSR}{\mathrm{MSR}}
\newcommand{\pole}{\mathrm{pole}}
\newcommand{\MSb}{\overline{\mathrm{MS}}}
\newcommand{\df}{{\rm d}}
\newcommand{\Ord}{\mathcal{O}}
\newcommand{\LQCD}{\Lambda_\mathrm{QCD}}

\usepackage{soul}

\title{\boldmath Bottom and Charm Mass determinations from global fits to $Q\overline{Q}$ bound states at N$^3$LO}

\preprint{\begin{flushright} IFT-UAM/CSIC-17-110\end{flushright}\vspace*{-1cm}}

\author[a,b]{Vicent Mateu}
\author[a]{Pablo~G.~Ortega}

\affiliation[a]{Departamento de F\'isica Fundamental and IUFFyM, Universidad de Salamanca\\Plaza de la Merced S/N, E-37008 Salamanca, Spain}
\affiliation[b]{Instituto de F\'isica Te\'orica UAM-CSIC\\C/ Nicol\'as Cabrera 13-15, Campus de Cantoblanco, E-28049 Madrid, Spain}

\emailAdd{pgortega@usal.es}
\emailAdd{vmateu@usal.es}

\abstract{
The bottomonium spectrum up to $n = 3$ is studied within Non-Relativistic Quantum Chromodynamics up
to N$^3$LO. We consider finite charm quark mass effects both in the QCD potential and the $\MSb$-pole
mass relation up to third order in the $\Upsilon$-scheme counting. The $u = 1/2$ renormalon of the static
potential is canceled by expressing the bottom quark pole mass in terms of the MSR mass. A
careful investigation of scale variation reveals that, while $n = 1, 2$ states are well behaved within
perturbation theory, $n = 3$ bound states are no longer reliable. We carry out our analysis in the
$\nl = 3$ and $\nl = 4$ schemes and conclude that, as long as finite $m_c$ effects are smoothly
incorporated in the MSR mass definition, the difference between the two schemes is rather small.
Performing a fit to $b\bar{b}$ bound states we find $\mbar_b(\mbar_b) = 4.216\,\pm\, 0.039$\,GeV. We extend
our analysis to the lowest lying charmonium states finding
$\mbar_c(\mbar_c)=1.273 \pm 0.054\,{\rm GeV}$. Finally, we perform simultaneous fits for $\mbar_b$ and
$\alpha_s$ finding $\alpha_s^{(\nf=5)}(m_Z)=0.1178\pm 0.0051$.
Additionally, using a modified version
of the MSR mass with lighter massive quarks we are able to predict the uncalculated
$\mathcal{O}(\alpha_s^4)$ virtual massive quark corrections to the relation between the $\MSb$ and
pole masses.
}

\begin{document} 
\maketitle
\flushbottom

\section{Introduction}\label{sec:intro}

The precise determination of hadron spectroscopy from fundamental principles  pursues to unveil QCD at its non-perturbative regime. The
non-perturbative nature of QCD at hadronic scales implied the development of phenomenological approaches such as quark 
models~\cite{Godfrey:2015dia,Eichten:1978tg,Godfrey:1985xj,Segovia:2016xqb,Segovia:2008zz} or,
more recently, computer-based calculations using Lattice QCD~\cite{Mohler:2017ibi,Prelovsek:2013cta,Liu:2012ze,Dowdall:2012ab}.
However, the unique properties of heavy quarkonium systems allow an entire
calculation in terms of non-relativistic QCD (NRQCD for short) in its weak coupling regime (that is, in pure perturbation theory). Effective Field
Theories (EFT for short) provide a clean separation of scales, enabling non-perturbative effects to be either factorized or treated in an
Operator Product Expansion.  Within perturbative NRQCD the quarkonium masses have been computed in recent years up to N$^3$LO precision,
that is up to $\mathcal{O}(m_Q\alpha_s^5)$ and $\mathcal{O}(m_Q\alpha_s^5\log\alpha_s)$~\cite{Fischler:1977yf,Billoire:1979ih,Schroder:1998vy,
Pineda:1997hz,Brambilla:1999qa,Kniehl:2002br,Penin:2002zv,
Smirnov:2008pn,Smirnov:2009fh, Anzai:2009tm}. The development of EFTs such as velocity NRQCD (vNRQCD)~\cite{Luke:1999kz} and potential 
NRQCD (pNRQCD)~\cite{Pineda:1997bj,Brambilla:1999xf}, which describe the interactions of a non-relativistic system with ultrasoft gluons,
organizing the perturbative expansions in $\alpha_s$ and the velocity of heavy quarks systematically, has been crucial to reach such accuracy.
Finally, a closed expression for arbitrary quantum numbers can be found in Ref.~\cite{Kiyo:2014uca}, which uses mathematical methods to
convert infinite sums into finite nested sums and transcendental numbers, and expresses the so called Bethe logarithm as a one-parameter
(numerical) integral.

The central object to compute the energy levels of $Q\overline{Q}$ states\,\footnote{In the rest of the paper $Q$ will stand for the heavy,
non-relativistic quark forming the bound states, while $q$ will refer to the next lighter quark.} is the QCD color-singlet static potential
$V_{\rm QCD}(r)$, which is known to very high accuracy\,: the leading correction, $\mathcal{O}(\alpha_s)$ with respect to the Coulomb term, was
obtained in \cite{Fischler:1977yf}, the $\mathcal{O}(\alpha_s^2)$ terms where computed in Refs.~\cite{Schroder:1998vy,Peter:1996ig}, while the
$\mathcal{O}(\alpha_s^3)$ corrections where calculated in Refs.~\cite{Brambilla:1999qa,Smirnov:2008pn,Smirnov:2009fh,Anzai:2009tm,Lee:2016cgz}.
It is well understood that the QCD static potential suffers from an $\Ord(\LQCD)$ renormalon ambiguity that exactly cancels that of the pole
mass~\cite{Pineda:1998id,Hoang:1998nz,Beneke:1998rk}, such that the static energy $E_{\rm stat}(r) = 2\,m_Q^{\rm pole} + V_{\rm QCD}(r)$ is
renormalon free.\footnote{For this cancellation to happen it is essential that the ambiguity of the pole mass and the potential are independent
of the mass and $r$, respectively.} This cancellation is essential to make the energy levels of quarkonia perturbative objects. For this
cancellation to take place, the pole mass has to be expressed in terms of so called short-distance masses. Previous studies of the bottomonium
and charmonium spectra used the renormalization-scale-dependent $\MSb$ mass $\mbar_Q(\mu)$ evaluated at
$\mu =\mbar_Q$~\cite{Brambilla:2001fw, Brambilla:2001qk,Kiyo:2013aea}. Although this scheme will make the renormalon
cancellation happen, it introduces a conceptual (and many times
also practical) problem. In the pole scheme, perturbative logs appearing in the theoretical expression of the quarkonium energy levels have
the following form: $\log[\,n\mu/(C_F\alpha_sm_Q)\,]$, with $n$ the principal quantum number. When expressing the pole mass in terms of
$\mbar_Q(\mbar_Q)$ a new class of logs appear, namely $\log(m_Q/\mu)$.\,\footnote{In practice these logs are present because even if the $\MSb$
mass is evaluated at the scale $\mu = \mbar_Q$, the relation of the pole mass to $\mbar_Q(\mbar_Q)$ has to be written as a series in powers
of $\alpha_s(\mu)$ for the renormalon cancellation to take place.} It is clear that no single choice of the scale $\mu$ will cause both logs to
vanish simultaneously. The origin of this mismatch can be traced to the fact that in NRQCD the mass of the heavy quark is a hard scale which has
been integrated out. This situation is resolved if instead a low-scale short-distance mass is used, such that the logs introduced when switching
scheme either have the form $\log(\mu/R)$, being $R$ an arbitrary scale the mass depends on (MSR mass~\cite{Hoang:2008yj,Hoang:2017suc},
Potential Subtracted~(PS) mass~\cite{Beneke:1998rk}, Renormalon Subtracted mass (RS)~\cite{Pineda:2001zq}, kinetic mass~\cite{Czarnecki:1997sz}
or jet mass~\cite{Jain:2008gb,Fleming:2007tv}), which can be chosen of the same order as (or equal to) $\mu$, or they are already of
the exact same form as the original ones (1S mass~\cite{Hoang:1998ng,Hoang:1998hm,Hoang:1999ye}). While the recent analysis of
Ref.~\cite{Ayala:2014yxa} employs the RS mass, we will make use of the MSR mass, motivated by the fact that it is straightforward to include
effects of lighter massive quarks \cite{Hoang:2017btd}.\footnote{Here we will use a modified version of the MSR mass in which the effects
of virtual massive quarks participate in the R-evolution.} At this point a comment is in order\,: while the problem just described is indeed critical
for the top quark, it is a lot less severe for the bottom and charm quarks, given that the hierarchy between their masses and the
respective non-relativistic scales is not very large. It is nevertheless theoretically more sound to use a low-scale mass.

The last piece needed to assemble the complete N$^3$LO theoretical expression in a short-distance scheme is the four-loop coefficient of the
relation between the $\MSb$ and pole masses, which has been computed recently~\cite{Marquard:2015qpa,Marquard:2016dcn}.
This term is necessary to compute the MSR mass to the same accuracy as the static potential in the so called $\Upsilon$-expansion 
scheme~\cite{Hoang:1998ng,Hoang:1998hm}. To make the list of theoretical ingredients complete one needs to include the effects of the finite
charm mass both in the static potential and the $\MSb$ to pole mass relation. Both are known up to $\mathcal{O}(\varepsilon^3)$\,: at
two~\cite{Gray:1990yh} and three~\cite{Bekavac:2007tk} loops for the latter; for any state at two loops~\cite{Eiras:2000rh} and for $n = 1$
states at three loops~\cite{Hoang:2000fm}, for the former. Therefore, if we are to include charm quark mass effects in the static potential,
we have to include them in the MSR mass as well. The MSR mass has so far been used only in phenomenological applications up
to $\mathcal{O}(\alpha_s)$, and always in the context of jet physics~\cite{Butenschoen:2016lpz, 
Hoang:2017kmk}. One (theoretical) motivation of our study is thus employing the MSR mass at $\mathcal{O}(\alpha_s^4)$ and in the context of a 
non-relativistic system. Furthermore, we want to test how efficiently the MSR mass deals with lighter massive flavors in a practical application.

Comparing the theoretical predictions for the bottomonium and charmonium masses with the corresponding experimental values one can determine
the bottom and charm quark masses with good accuracy in a clean environment. Precise values of these heavy quark masses are key to test the
validity of the Standard Model in the precision frontier, both in Flavor~\cite{Antonelli:2009ws} and Higgs Physics~\cite{Heinemeyer:2013tqa}.
Two recent analyses have determined the bottom quark mass from the $\Upsilon(1S)$ state~\cite{Ayala:2014yxa},\footnote{See~\cite{Ayala:2016sdn}
for an update of the analysis.} and bottom and charm quark masses from $n = 1$ states~\cite{Kiyo:2015ufa}. Both analyses use N$^3$LO
theoretical predictions and consider finite charm quark mass corrections. While~\cite{Ayala:2014yxa} uses the RS scheme for the heavy quark
mass and varies $\mu$ and $R$ independently (but one at a time) in 
a non-relativistic range ($1.5\,$GeV to $4\,$GeV), Ref.~\cite{Kiyo:2015ufa} uses the $\MSb$ scheme and the principle of minimal sensitivity,
setting $\mu$ at scales that are larger than the heavy quark mass itself ($5\,$GeV to $12\,$GeV for bottom, $2\,$GeV to $4.5\,$GeV for
charm), therefore relativistic. In our analysis we will use the MSR scheme varying both renormalization scales independently and simultaneously
around the value that minimizes perturbative logarithms (a non-relativistic scale). Additionally, based on the 
observation made in Ref.~\cite{Brambilla:2001qk}, that is, finite charm quark mass effects in bottomonium 
are closer to the decoupling ($m_c\to \infty$) than to the massless ($m_c\to 0$) limit, the authors of
Ref.~\cite{Ayala:2014yxa} argue that using $\nl=3$ active flavors makes this decoupling more explicit, and
therefore carry out their analysis in this scheme. On the other hand, Ref.~\cite{Kiyo:2015ufa} performs
the bulk of their analysis in the $\nl=4$ flavor scheme, and uses its difference to the $\nl=3$ result to 
estimate the uncertainties due to missing higher terms on the finite charm mass correction. We perform fits in both flavor schemes
but take the $\nl=3$ determination for our final result.

The main motivation to carry out our analysis is to include in the determination of the bottom quark mass states with principal quantum number
$n\ge 1$. It has been claimed that the gross spectrum of bottomonium up to $n = 4$ can be described purely within perturbation
theory~\cite{Brambilla:2001fw,Brambilla:2001qk,Kiyo:2013aea}. Those analyses are based on the $\MSb$ scheme for the bottom quark and on the
principle of minimal sensitivity to the renormalization scale, which in practice translates into settings the scale $\mu$ far from the (low energy)
values that would make perturbative logarithms vanish, and vary it in a very narrow range or towards larger scales. We shall show that the
scale of minimal sensitivity strongly depends on the order of perturbation theory considered, and also on the $\ell,s$ and $j$ quantum
numbers which do not appear in the perturbative logarithms. We preform a dedicated study of scale variation inspired by the usual lore,
that is, the argument of perturbative logarithms should vary roughly between $1/2$ and $2$, supplemented with the additional constraint that
the renormalization scale should not become smaller than $1\,$GeV, such that perturbation
theory is not jeopardized. We find that the central renormalization scale decreases as $n$ increases, becoming smaller than $1\,$GeV already at
$n = 3$. Even stopping scale variation at $1\,$GeV for $n = 3$, the predictions for the masses become unstable and the perturbative uncertainties
grow dramatically as compared to smaller values of $n$.

Our analysis would not be complete if non-perturbative effects are simply ignored.\footnote{We thank Thomas Rauh for bringing
our attention to this issue.} We will assume that the soft scale $mv$ is much larger than $\LQCD$, and that the ultra-soft scale
$mv^2$ is either much larger than $\LQCD$, or at the very worse, of the same order.
The analysis of Ref.~\cite{Pineda:1996uk} suggests that for states with $n=1$
we are in the former situation (therefore non-perturbative effects come in the form of local condensates),
while for $n=2$ likely falls in the latter (with non-perturbative effects manifesting themselves as
non-local condensates). In either situation the main contribution to the energy levels is pertubation
theory. Our analysis seems to indicate that for $n\ge3$ one has $mv^2 < \LQCD$, being
the static potential directly affected by non-local condensates. Our approach to non-perturbative effects
will be rather heuristic\,: given that the operator product expansion ties together perturbative and
non-perturbative physics, we will assume that as long as the perturbative expansion for the energy
levels is well behaved, non-perturbative effects do not heavily affect the perturbative result.
We will give extra validation to this ansatz by comparing the determination of the bottom quark mass
from fits to states with principal quantum numbers $n = 1$ and $n = 2$. From this analysis we can however not discard that for
$n = 2$ we are already in a situation in which perturbative and non-perturbative effects become entangled, and
until a fully fledged study of non-perturbative effects becomes available, it is not possible to draw further conclusions.
As emphasized in Ref.~\cite{Ayala:2014yxa}, one probably needs to have ultrasoft effects under control before dealing with
non-perturbative physics.

This article is organized as follows\,: In Sec.~\ref{sec:theory} we provide the theoretical expressions for quarkonia masses including
corrections from massive lighter quarks. In Sec.~\ref{sec:MSR} we review the MSR short-distance scheme for the mass of a heavy quark
and show how to include effects from massive lighter quarks in their R-evolution, explaining how to integrate those effects out at low scales.
In Sec.~\ref{sec:investigation} we perform an investigation of renormalization scale variation and the effects of finite charm mass in
bottomonium. The fitting procedure and the results for the determination of the bottom and charm quark masses, as well as simultaneous fits to $\mbar_b$ and $\alpha_s$, are presented in
Sec.~\ref{sec:fits}. We compare our results to previous determinations in Sec.~\ref{sec:comparison}. Our conclusions are contained 
in Sec.~\ref{sec:conclusions}. Some formulas are collected in App.~\ref{app:formulas}.

\section{Formula for $\mathbf{Q\overline{Q}}$ Bound States}\label{sec:theory}

\subsection{Analytic expression for massless quarks}\label{sec:massless}
The energy of a $Q\overline{Q}$ non-relativistic bound state characterized by the $(n,j,\ell,s)$ quantum numbers and with $\nl$ massless active flavors 
reads in the pole scheme~\cite{Penin:2002zv,Beneke:2005hg,Kiyo:2014uca}\,:
\begin{align}\label{eq:EXpole}
& E_X(\mu,\nl) = 2\,m_Q^{\rm pole}\!
\Bigg[1-\frac{C_F^2\,\alpha^{(\nl)}_s(\mu)^2}{8n^2}\sum_{i=0}^{\infty}\bigg(\frac{\alpha^{(\nl)}_s(\mu)}{4\pi}\bigg)^{\!\!i}\,
\varepsilon^{i+1}P_i(L_{\nl})\Bigg],\\
& L_{\nl}=\log\!\bigg(\frac{n\mu}{C_F\alpha_s^{(\nl)}(\mu)m_Q^{\rm pole}}\bigg)+ H_{n+\ell}\,,\qquad
 P_i(L) = \sum_{j=0}^i\,c_{i,j}\,L^j\,,\nonumber
\end{align}
with $H_{n}$ the harmonic number. In Eq~\eqref{eq:EXpole} $\varepsilon$ is a bookkeeping parameter that labels the various orders in the
$\Upsilon$-expansion.
The $c_{i,j}$ coefficients can be computed from the $c_{i,0}$ imposing $\mu$
independence of the energy states, what implies the following recursion relation\,:
\begin{align}
c_{k,j+1} = \frac{2}{j+1}\,\bigg\{\! (j+2)\,\beta_{k-1-j}\,c_{j,j} +\!\!\!
\sum_{i=j+1}^{k-1}\!\!\beta_{k-1-i}\big[(i+2)\,c_{i,j}-(j+1)\,c_{i,j+1}\big]\!\bigg\}\,,
\end{align}
where $\beta_i$ are the $(i+1)$-loop coefficients of the beta function defined in Eq.~\eqref{eq:alphaRGE}.
The $c_{i,0}$ coefficients have been calculated up to $i=3$ and their values can be found in Refs.~\cite{Brambilla:2001qk,Kiyo:2013aea}.
In general they depend on the quantum numbers $(n,j,\ell,s)$ and $c_{3,0}$ also depends on $\log(\alpha_s)$.
Eq.~\eqref{eq:EXpole} inherits the $u=1/2$ renormalon of the static potential, what makes it unfit for
a precision determination of heavy quark masses. This can be resolved by expressing the pole mass in terms
of a short-distance mass. Moreover, if large logs of the ratio of the non-relativistic scale and the quark
mass are to be avoided, a low-scale short-distance mass should be used. In our analysis we employ the two
versions of the MSR mass discussed in Ref.~\cite{Hoang:2017suc}. Let us consider a generic short-distance mass whose relation to the pole
scheme is expressed as
\begin{align}
m_Q^{\rm pole} = m_Q^{\rm SD}\Bigg[1 + \sum_{n=1}\varepsilon^n\,\delta_n^{\rm SD} \bigg(\frac{\alpha^{(\nl)}_s(\mu)}{4\pi}\bigg)^{\!\!n}\Bigg]\,,
\end{align}
where $m_Q^{\rm SD}$ depends on some energy scale denoted generically by $R$, and $\delta_n^{\rm SD}$ contain powers of $\log(\mu/R)$.
One then has that
\begin{align}
L_{\nl} &= L_{\rm SD} - \sum_{i=1}\varepsilon^i\delta_i^{L}\bigg(\frac{\alpha^{(\nl)}_s(\mu)}{4\pi}\bigg)^{\!\!i}\,,\quad
L_{\rm SD} = \log\!\bigg(\frac{n\mu}{C_F\alpha_s^{(\nl)}(\mu)m_Q^{\rm SD}}\bigg)+ H_{n+\ell}\,,\nonumber\\
\delta_{i+1}^{L} & = \delta_{i+1}^{\rm SD}-\frac{1}{i+1}\sum_{j=1}^i \,j\,\delta_j^L\,\delta_{i+1-j}^{\rm SD}\,.
\end{align}
This relation can be used to convert the sum that appears in Eq.~\eqref{eq:EXpole} to a short-distance scheme\,:
\begin{align}\label{eq:logSD}
&\sum_{i=0}^{\infty}\bigg(\frac{\alpha^{(\nl)}_s(\mu)}{4\pi}\bigg)^{\!\!i} \varepsilon^{i+1}P_i(L_{\nl}) =
 \sum_{i=0}^{\infty}\bigg(\frac{\alpha^{(\nl)}_s(\mu)}{4\pi}\bigg)^{\!\!i}\varepsilon^{i+1} P_i^{\rm SD}(L_{\rm SD})\,,\\
&  P_j^{\rm SD} = \sum_{i=1}^{j+1}\,\,\sum_{k=0}^{{\rm min}(i-1,j+1-i)} \!\! \!\!c_{i,k}^{\rm SD}\,\delta_{k,j+1-i}^L\,, \nonumber\\
& c_{i,k}^{\rm SD} = (-1)^k\sum_{j=k}^{i-1}\binom{j}{k}\,c_{i,j}\,L_{\rm SD}^{j-k}\,,\nonumber\\
& \delta_{1,i}^L = \delta_i^L\,,\quad \delta_{n+1,j}^L = \sum_{i=1}^{j-n}\delta_i^L\,\delta_{n,j-i}^L\,.\nonumber
\end{align}
Here we have defined the $\delta_{i,j}^L$ coefficients from $\big(\sum_{i=1} x^i \delta_i^L \big)^j = \sum_{i=j}x^i \delta_{j,i}^L$, which
can be calculated from the recursion relation in the last line of Eq.~\eqref{eq:logSD}. The only thing left to express the energy levels of quarkonium
in a short-distance scheme is converting the global factor $m_Q^{\rm pole}$\,:
\begin{align}\label{eq:EXSD}
E_X(\mu,\nl) & = 2\,m_Q^{\rm SD}\,\Bigg\{1 + \sum_{i=1}\varepsilon^i\bigg(\frac{\alpha^{(\nl)}_s(\mu)}{4\pi}\bigg)^{\!\!i}
\bigg[\delta_i^{\rm SD} - F\,P_{i-1}^{\rm SD} - (1-\delta_{i,1}) \,F \sum_{j=1}^{i-1}\delta^{\rm SD}_j P_{i-j-1}^{\rm SD} \bigg]\Bigg\}\,,\nonumber\\
F & = \frac{\pi\, C_F^2\alpha^{(\nl)}_s(\mu)}{2n^2}\,,
\end{align}
where $\delta_{1,i}$ is the Kronecker symbol taking the value $1$ for $i=1$ and zero for higher values of $i$.

Having in mind that ultimately we want to determine the mass of the heavy quark $Q$, one can devise perturbative expansions based
on Eq.~\eqref{eq:EXSD} in which $m_Q^{\rm SD}$ (or even $\mbar_Q$) is singled out. We use two such expansions, which are in complete
correspondence to the ``linearized'' and ``linearized iterative'' methods defined in Ref.~\cite{Dehnadi:2011gc}, and denote them
as \textit{expanded out} and \textit{iterative}. The former corresponds to\,:
\begin{align}\label{eq:expanded}
m_Q^{\rm SD} = \frac{E_X^{\rm exp}}{2}\Bigg[1+ \sum_{i=1}\varepsilon^i\bigg(\frac{\alpha^{(\nl)}_s(\mu)}{4\pi}\bigg)^{\!\!i}
A^{\rm exp}_i(m_Q^{\rm SD},R,\mu,)\Bigg]\,,
\end{align}
directly obtained from Eq.~\eqref{eq:EXSD} isolating the global factor $m_Q^{\rm SD}$, with $E_X^{\rm exp}$ the experimental
value for the bound state and $A_i^{\rm exp}$ some easy-to-calculate coefficients. One can remove the $m_Q^{\rm SD}$ dependence
that still persists on the $A_i^{\rm exp}$ in an iterative way, by introducing lower-order expressions for $m_Q^{\rm SD}$ in
higher-order ones and expanding consistently in $\varepsilon$. This way one obtains the iterative expansion\,:
\begin{align}\label{eq:iterative}
m_Q^{\rm SD} = \frac{E_X^{\rm exp}}{2}\Bigg[1+ \sum_{i=1}\varepsilon^i\bigg(\frac{\alpha^{(\nl)}_s(\mu)}{4\pi}\bigg)^{\!\!i}
A^{\rm iter}_i(E_X^{\rm exp},R,\mu,)\Bigg]\,.
\end{align}
There is still some residual $m_Q$ dependence on the right hand side due to the threshold matching conditions of $\alpha_s$.
This can be eliminated numerically by iterating Eq.~\eqref{eq:iterative} a few times.

\subsection{Finite Charm Mass Effects on Bottomonium}\label{sec:charm}
Equation~\eqref{eq:EXpole} considers that the $\nl$ active flavors that contribute to the energy of a $Q\overline Q$ bound state are all massless.
However, massive charm quark loops contribute both to the binding energy and the relation of the pole and short-distance masses, which 
should be properly accounted for. In this section we consider both $m_Q$ $m_q$ in the pole scheme, since switching any of them
to a short-distance scheme is trivial. Whenever we use a short-distance scheme for $m_Q$, we will take $m_q$  in the $\MSb$ scheme,
and more specifically we will only use $\mbar_q\equiv \mbar_q^{(\nl)}(\mbar_q^{(\nl)})$. Massive lighter quarks effects on the relation
between the pole and other short-distance schemes will be discussed in Secs.~\ref{sec:MSbarcharm}, 
\ref{sec:MSRcharm} and \ref{sec:MSRnl}.

Modifications to the Coulomb potential due to non-zero lighter quark masses have been calculated up to $\mathcal{O}(\varepsilon^3)$.
The direct consequence of such corrections in the Coulomb potential is an energy-shift effect in the heavy quarkonium mass.
%
In the pole scheme they read\,:
\begin{align}\label{eq:EXpoleCorr}
E_X(\mu,\nl,m_Q^{\rm pole},m_q^{\rm pole}) = E_X(\mu,\nl,m_Q^{\rm pole})+\varepsilon^2\delta E_X^{(1)}+\varepsilon^3\delta E_X^{(2)}\,.
\end{align}
If Eq.~\eqref{eq:EXpoleCorr} is expressed in the $\nl$ scheme, where the massless limit is realized, these corrections obviously vanish
if $m_q\to 0$. Likewise, if Eq.~\eqref{eq:EXpoleCorr} is written in the $(\nl-1)$ scheme, the decoupling limit is realized and the corrections
should vanish if $m_q\to\infty$. However, in the $\nl\,(\nl-1)$ scheme the decoupling (massless) limit is not manifest. We can use this
information to constrain the form of $\delta E_X^{(i)}$ in the $\nl$ scheme for large values of $m_q$ requiring that, once $\alpha_s^{(\nl)}$ 
is expressed in terms $\alpha_s^{(\nl-1)}$ through the threshold relation in Eq.~\eqref{eq:threshold} the decoupling limit is reached.
This strategy was used in Ref.~\cite{Brambilla:2001qk}.

The $\varepsilon^2$ correction has been calculated in Ref.~\cite{Eiras:2000rh}, for arbitrary quantum numbers, and in the $\nl$ scheme it
reads\,:
\begin{align}\label{eq:deltEx1}
 &\delta E_X^{(1)}(\nl)=\alpha^{(\nl)}_s(\mu)^3\,\delta E_1(\rho),\\
 &\delta E_1(\rho)=\frac{C_F^2}{3\,(4 \pi )\,n^2}  \Bigg[\!-\frac{1}{(2 n-1)!}\sum _{k=0}^{n-\ell-1} \binom{n-\ell-1}{k}
 \binom{n+\ell}{k+2 \ell+1} \rho ^{2 (n-k-\ell-1)} f_D(n,\ell,k,\rho )\nonumber\\
 &-\pi  n \rho ^3 \left((n-\ell-1) (n+\ell)+\frac{1}{3} (n+1) (2 n+1)\right)+\rho^2 \bigg((n-\ell-1) (n+\ell)+(2 n+1) n\bigg)\nonumber\\
 &-\frac{3\pi}{2} n \rho +2 \ln \left(\frac{2}{\rho }\right)-2\, H_{n+\ell}\Bigg],\nonumber
\end{align}
where $\rho = (2 \,n\, m_c)/[m_b C_F \alpha^{(\nl)}_s(\mu)]$ and $f_D(n,\ell,k,\rho)$ is defined in Eq.~\eqref{eq:dE1fD}. In the $(\nl-1)$ scheme
one simply has
\begin{align}
\delta E_X^{(1)}(\nl - 1)=\alpha^{(\nl-1)}_s(\mu)^3 \bigg[\delta E_1(\rho) - \lim_{\rho\to \infty} \delta E_1(\rho)\bigg]\,,
\end{align}
such that the decoupling limit is manifest.

\begin{figure}[h]
\center
\includegraphics[width=0.5\textwidth]{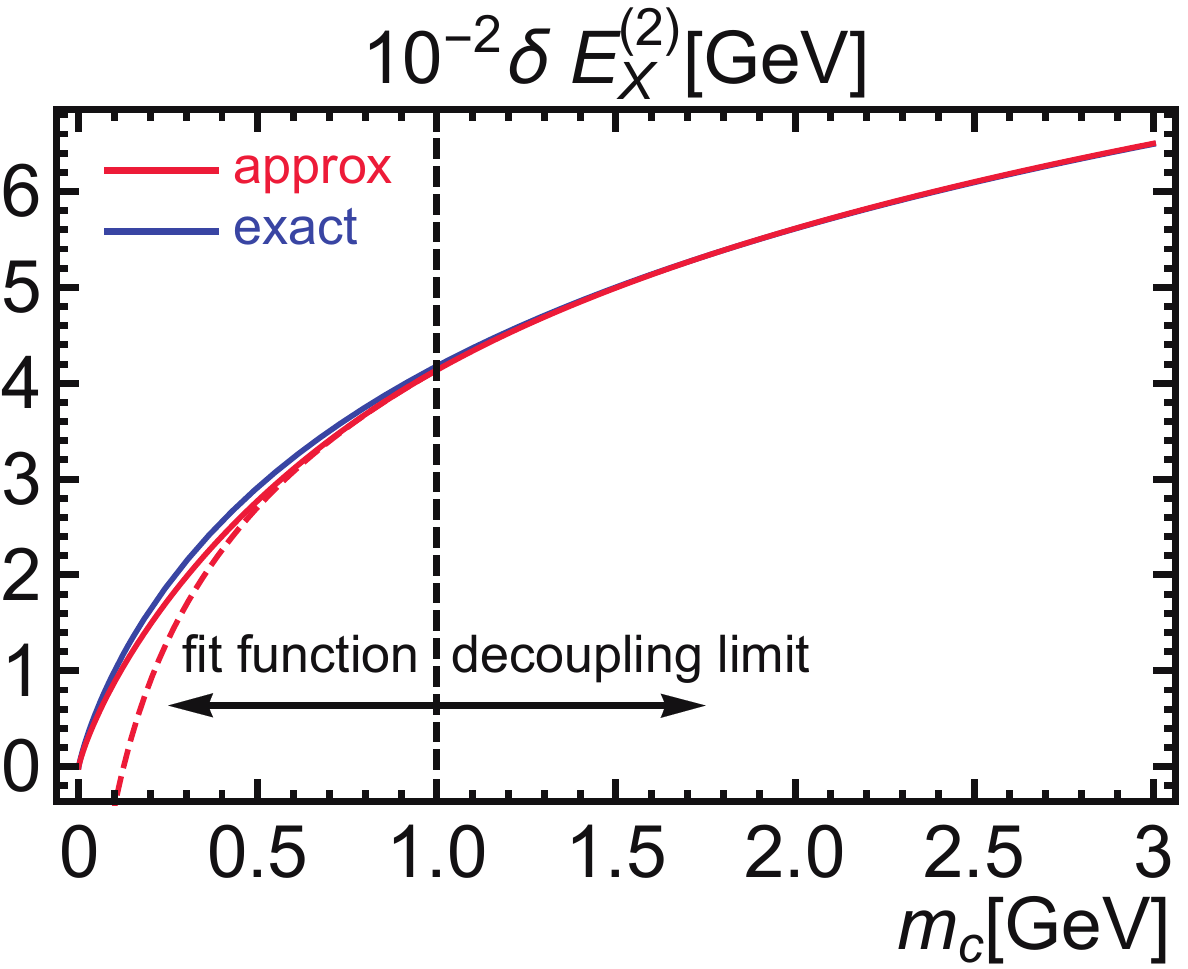}
\caption{\label{fig:CharmCorr2} Finite charm quark mass corrections to the $\Upsilon(1S)$ mass at $\mathcal{O}(\varepsilon^3)$
in the pole scheme. We take the values $m_b=4.2\,$GeV, $\mu=2\,$GeV and $\alpha_s^{(\nf = 5)}(m_Z)=0.118$. The blue line corresponds to the
exact result, while the red solid line is the approximation described in the text. The red dashed line shows the decoupling limit
approximation.}
\end{figure}

The exact $\mathcal{O}(\varepsilon^3)$ finite charm mass corrections to the binding energy have been computed
for the $n=1$ bottomonium energy levels in Ref.~\cite{Hoang:2000fm}, and for states with arbitrary $n$ and $\ell=0$ in
Ref.~\cite{Beneke:2014pta}, remaining unknown for other sets of quantum numbers. Since these corrections for $n>1$ are
prohibitively slow, in order to include them for non-$1S$ states we will use the results of the analysis carried out in
Ref.~\cite{Brambilla:2001qk}, where the authors concluded that for $n=1$ the $m_c\to \infty$ limit is an excellent
approximation, which becomes even better for large values of the principal quantum number. The formula for such limit
in the pole mass scheme is detailed in Eq.~\eqref{eq:dEX2inf}. We will supplement this approximation with the requirement that
in the $\nl$ scheme the correction must vanish if $m_c \to 0$. Our approach uses the decoupling approximation for values of $m_c$
larger than $m_c^*(n)$, where $m_c^*(n)=(1\,{\rm GeV})/\sqrt{n}$, and the fit function \mbox{$f(m_c) = m_c\,[\,a + b\,\log(m_c)\,]$} for $m_c<m_c^*$,
where the coefficients $a$ and $b$ are adjusted such that the transition is smooth at the junction point. This functional form
enforces the massless limit, and the dependence of $m_c^*$ in $n$ takes the decoupling approximation as exact for lower values of
$m_c$. We have tested our approach for $n=1$ and found that it reproduces the exact known result within a few percent, as can be
appreciated in Fig.~\ref{fig:CharmCorr2}. In this figure we also show the decoupling approximation below the junction point
as a red dashed line.

\section{The MSR scheme}\label{sec:MSR}
As mentioned in the Introduction, the use of the pole mass in perturbative calculations is ineffective,
not only for its bad perturbative behavior 
but also because of the existence of confinement, which hides the quark propagator pole in the non-perturbative regime.
%

Alternatively, the $\MSb$ mass is an adequate scheme for physical situations that involve energies much larger than the heavy quark.
Ideally, for the heavy quarkonium, one would like to extend such scheme for energies below $m_Q$ without losing
the good infrared properties of the $\MSb$ scheme. At such scales the evolution of the logarithms in the
regular $\MSb$ mass is unphysical and behaves badly. A more adequate short-distance scheme is the MSR mass.
The $\MSR$ scheme, first introduced in Ref.~\cite{Hoang:2008yj}, and extensively discussed in Ref.~\cite{Hoang:2017suc},
is a natural extension of the $\MSb$ mass for renormalization scales below the heavy quark mass.
The definition of the $\MSR$ mass, expressed as the $\MSR$-pole mass difference, is derived directly from the $\MSb$ pole mass relation,
exploiting the fact that the renormalon ambiguity is independent of the value of the heavy quark mass. Let us start form the perturbative
relation of the $\MSb$ and pole masses\,:
\begin{align}\label{eq:msbarpoleseries}
& \delta \mbar_Q(\mbar_Q) \equiv m_Q^{\pole} - \mbar_Q = \mbar_Q\,\sum_{n=1}^\infty\,a_n^{\overline{\rm MS}} (\nl,n_h)\,\bigg(\frac{\alpha_s^{(n_\ell+n_h)}(\mbar_Q)}{4\pi}\bigg)^{\!\!n} \,,
\end{align}
where $n_f = \nl + n_h$ and we have defined $\mbar_Q\equiv \mbar_Q^{\,(\nf)}(\mbar_Q^{\,(\nf)})$. The strong coupling constant
is expressed in the $\MSb$ scheme with the usual renormalization group equation\,:
\begin{align}\label{eq:alphaRGE}
\frac{{\df} \alpha_s^{(n_f)}(\mu)}{\df \ln \mu}=-\,2\,\alpha_s^{(n_f)}(\mu)\sum_{n=0}^\infty
\beta_n(n_f)\left(\frac{\alpha_s^{(n_f)}(\mu)}{4\pi}\right)^{\!\!n+1}.
\end{align}
For later use we split the two- and three-loop terms into their various flavor components
\begin{align}\label{eq:aFlav}
&a_2^{\overline{\rm MS}} = a_2^{({\rm g})} + \nl\,a_2^{(\nl)} + n_h\,a_2^{(n_h)}\,,\\
&a_3^{\overline{\rm MS}} = a_3^{({\rm g})} + \nl\,a_2^{(\nl)} + n_h\,a_2^{(n_h)} + \nl^2\,a_2^{(\nl^2)} + n_h^2\,a_2^{(n_h^2)}
+ \nl\, n_h\,a_2^{(\nl\,n_h)}\,.\nonumber
\end{align}
In contrast to the $\MSb$ mass, which only depends logarithmically in the scale $\mu$, the $\MSR$ mass has a logarithmic and
linear dependence on R\,:
\begin{align}\label{eq:MSRdef}
  \delta m_Q^\MSR \equiv
m_Q^{\rm pole}-m_Q^\MSR(R) & = R\sum_{n=1}^\infty a_{n,0}(\nl)\left(\frac{\alpha^{(\nl)}_s(R)}{4\pi}\right)^{\!\!n}\\
& =R\sum_{n=1}^\infty \sum_{k=0}^n a_{n,k}(\nl)\left(\frac{\alpha^{(\nl)}_s(\mu)}{4\pi}\right)^{\!\!n}\ln^k\!\left(\frac{\mu}{R}\right),\nonumber
\end{align}
where $a_{n,k}$ are derived from Eq.~\eqref{eq:msbarpoleseries}, and are different for the Practical and Natural versions of the
MSR mass (see Sec.~\ref{sec:MSRmassless}). The advantages of this mass scheme is that it can be safely employed for scales
$R<m_Q$ when a clean treatment of
virtual massive-quark effects is relevant, as the virtual effects of the massive flavor is integrated out from the $\MSR$ mass expression.

While the first line of Eq.~\eqref{eq:MSRdef} defines the MSR mass and is used to derive its anomalous
dimension, the second line is more useful when implementing the MSR scheme in a series expressed in powers of
$\alpha_s(\mu)$. Since the MSR mass is $\mu$-independent, the $a_{n,k}$ coefficients can be computed from
the $a_{k,0}$ with the following recursion relation\,:
\begin{equation}
a_{n,k} = \frac{2}{k}\,\sum_{i=k}^{n-1}\,i\,a_{i,k-1}\,\beta_{n-1-i}\,.
\end{equation}
The R dependence of the $\MSR$ mass is described by the following renormalization group equation\,:
\begin{equation}\label{eq:Revol}
-\,\frac{\df}{\df R}m_Q^{\MSR}(R)=\gamma^R[\alpha_s^{(\nl)}(R)]=\sum_{n=0}^\infty \gamma_n^R \left(\frac{\alpha_s^{(\nl)}(R)}{4\pi}\right)^{\!\!n+1},
\end{equation}
where $\gamma_n^R$ are the R-anomalous dimension coefficients~\cite{Hoang:2017suc},
which can be calculated as follows\,:
\begin{align}
\gamma_n^R = a_{n+1,0}-2\sum_{j=0}^{n-1} (n-j)\,\beta_j\, a_{n-j,0}\,.
\end{align}
Since the ambiguity of Eq.~\eqref{eq:MSRdef} is R independent, the R-anomalous dimension is automatically renormalon-free. This RGE provides a 
systematic reordering of the terms in the asymptotic series related to the $\mathcal{O}(\Lambda_{\rm QCD})$ renormalon ambiguity\,:
\begin{align}\label{eq:MSRsum}
m_Q^{\rm MSR}(R_2) - m_Q^{\rm MSR}(R_1) = \!\int_{R_1}^{R_2}\!\df R\,\gamma_n^R[\alpha_s^{(\nl)}(R)]\,,
\end{align}
which results in the summation of powers of $\log(R_1/R_0)$ to all orders in perturbation theory.\,\footnote{This
class of logarithms was first summed up for the RS mass in \cite{Bali:2003jq} through the inverse Borel transform
of the asymptotic series.}

\subsection{Natural and Practical MSR schemes}\label{sec:MSRmassless}
As we discussed in Sec.~\eqref{sec:MSR}, in contrast to the $\MSb$ mass,
the $\MSR$ mass is a suitable scheme for applications requiring scales lower than the
heavy quark mass. For that reason, the UV effects of the quark Q can be integrated out,
changing from a scheme with $n_\ell+1$ dynamical flavors to one with only $n_\ell$.
There are, hence, two alternative but equivalent ways to achieve this transformation.
On the one hand, the threshold matching relation for the strong coupling can be used
to rewrite $\alpha_s^{(n_\ell+1)}$ in terms of $\alpha_s^{(n_\ell)}$
\begin{align}\label{eq:threshold}
\alpha_s^{(\nl + 1)}(\mbar_Q) = \alpha_s^{(\nl)}(\mbar_Q)\!\Bigg[1 + \sum_{n=2}\xi_n(\nl)\left(\frac{\alpha^{(\nl)}_s(\mu)}{4\pi}\right)^{\!\!n}\Bigg],
\end{align}
which once implemented into Eq.~\eqref{eq:msbarpoleseries} is used to define the \emph{Practical $\MSR$ mass}\,:
\begin{align}
m_Q^{\rm pole}-m_Q^{\rm MSRp}(R)= R\sum_{n=1}^\infty a_n^{\rm MSRp}(n_\ell)\left(\frac{\alpha_s^{(n_\ell)}(R)}{4\pi}\right)^{\!\!n}.
\end{align}
The $a_n^{\rm MSRp}$  coefficients can be computed as follows\,:
\begin{align}
a^{\rm MSRp}_{n,0}(\nl,n_h) = \sum_{i = 1}^n a_i^{\overline{\rm MS}} (\nl,n_h)\,\xi_{i-1,n}(\nl)\,,
\end{align}
where we have included the $n_h$ argument in the MSRp scheme coefficients for later convenience. The coefficients $\xi_{i,j}$ are defined as 
$(1+\sum_{j=1}\xi_j \,x^j)^i = 1 + \sum_{j=1} \xi_{i,j}\,x^j$, and therefore fulfill $\xi_{1,i} = \xi_i$, where we define $\xi_0=\xi_{i,0}=1$. 
For $i>1$  they can be computed with the following recursion relation\,:
\begin{align}
 \xi_{n+1,j}(\nl) = \xi_j(\nl) + \xi_{n,j}(\nl) + \sum_{i=1}^{j-1}\xi_i(\nl)\,\xi_{n,j-i}(\nl)\,.
\end{align}
On the other hand, the virtual loop corrections of the heavy quark Q can be directly integrated out of the $\MSb$ to pole relation taking $n_h=0$
in Eq.~\eqref{eq:msbarpoleseries}, which leads to the definition of the \emph{Natural $\MSR$ mass}:
\begin{align}
m_Q^{\rm pole}-m_Q^{\rm MSRn}(R)= R\sum_{n=1}^\infty a_n^{\MSb}(n_\ell,0)\left(\frac{\alpha_s^{(n_\ell)}(R)}{4\pi}\right)^{\!\!n},
\end{align}
that is, $a^{\rm MSRn}_{n,0}(\nl) \equiv a_n^{\overline{\rm MS}} (\nl,0) $.

Each materialization of the $\{n_\ell+1\to n_\ell\}$ scheme change has its advantages, depending on the specific applications.
The Natural MSR mass only exhibits corrections arising from gluons and massless quarks, being its relation with the $\MSb$ mass
mediated by a non-trivial expression\,:
\begin{align}
& m_Q^{\rm MSRn}(\mbar_Q)-\mbar_Q = {\rm HQS}(\mbar_Q,\nl, 1)\,,\\
& {\rm HQS}(\mbar_Q,\nl, n_h)\equiv\mbar_Q\sum_{n=2}\Big[a_k^{\rm MSRp}(\nl,n_h)-a_k^{\MSb}(n_\ell,n_h-1)\Big]\!
\left(\frac{\alpha_s^{(n_\ell)}(\overline{m}_Q)}{4\pi}\right)^{\!\!n}\,.\nonumber
\end{align}
Alternatively, the MSRp\,-$\MSb$ relation is more direct, as the Practical $\MSR$ mass is equal
to the $\MSb$ mass at the scale of the mass to all orders in perturbation theory:
\begin{align}
  m_Q^{\rm MSRp}(\mbar_Q) = \mbar_Q(\mbar_Q)\,.
\end{align}
Finally, we note that the $\MSb$ scheme used in Refs.~\cite{Brambilla:2001fw, Brambilla:2001qk,Kiyo:2013aea} exactly
corresponds to the MSRp mass with $R=\mbar$. Therefore their theoretical expressions are contained in ours.

\subsection[$\MSb$ mass with light fermion masses]
{$\mathbf\MSb$ mass with light fermion masses}\label{sec:MSbarcharm}
%
%
Corrections from massive lighter quarks modify the relation of the pole mass to other short-distance schemes. In this section we shall focus the discussion to the $\MSb$ mass, though for our numerical analysis the MSR scheme will be used.

The corrections from massive lighter quarks to the $\MSb$-pole mass relation start at $\mathcal{O}(\alpha_s^2)$ and vanish in the
$m_q\to 0$ limit. Hence the corrected relation can be expressed as\,:
\begin{align}\label{eq:MSRbcharm}
& m_Q^{\rm pole}-\mbar_Q \equiv \delta m_Q(\mbar_Q) \,+\, \mbar_Q\,\Delta^{\MSb}_{\mbar_c}(\mbar_Q,\xi)\,,\\
& \Delta^{\MSb}_{\mbar_c}(\mbar_Q,\xi) = \sum_{n=2} \bigg(\frac{\alpha_s^{(\nl+n_h)}(\mbar_Q)}{4\pi}\bigg)^{\!\!n}\nonumber
\Delta^{\MSb}_n(\nl,n_h,\xi)\,,
\end{align}
where $\delta m_Q$ is given in Eq.~\eqref{eq:msbarpoleseries} and 
$\xi=\mbar_q/\mbar_Q$. 
The functions $\Delta_i$ obey two constraints\,: $\Delta^{\MSb}_n(\nl,n_h,0) = 0$ and
$\Delta^{\MSb}_n(\nl,n_h,1) = a_n^{\MSb}(\nl-1,n_h+1) - a_n^{\MSb}(\nl,n_h)$.\footnote{These relations will be refereed to as
``the $\xi=0$ constraint'' and ``the $\xi=1$ constraint'', respectively, in the rest of the paper.}
The $\varepsilon^2$ term for the non-zero charm mass corrections has been calculated exactly in 
Ref.~\cite{Gray:1990yh}\,:\,\footnote{Eq.~\eqref{eq:Delta2} is written in a way in which each term is manifestly real
for any value of $\xi$ between $0$ and $1$.}
%
%
%
\begin{align}\label{eq:Delta2}
&\Delta_2^{\MSb}(\xi)=
\frac{16}{3} \bigg[\ln^2(\xi)+\frac{\pi^2}{6}-\xi^2 \bigg(\frac{3}{2}+\ln(\xi)\bigg)+(1+\xi) (1+\xi^3) \bigg({\rm Li}_2(-\xi)-\frac{1}{2}\ln^2(\xi)+\nonumber\\
&+\ln(\xi)\ln(1+\xi)+\frac{\pi^2}{6}\bigg)-(1-\xi) (1-\xi^3) \bigg(\frac{\pi^2}{6}+\frac{1}{2}\ln^2(\xi)+{\rm Li}_2(1-\xi)\bigg)\bigg].
\end{align}
This correction, together with its linear approximation [\,see Eq.~\eqref{eq:appform}\,], are shown in Fig.~\ref{fig:Delta2}.
The $\xi = 1$ constraint in these corrections takes the form $\Delta_2(1) = a_2^{(n_h)} - a_2^{(\nl)}$. The $\varepsilon^3$ correction
can be split in three contributions\,:
\begin{equation}\label{eq:Delt3split}
\Delta_3^{\MSb}(\xi)(\nl,n_h,\xi) = \Delta_3^{(g)}(\xi) +\nl\, \Delta_3^{(\nl)}(\xi)+ n_h\, \Delta_3^{(n_h)}(\xi)\,.
\end{equation}
%
The $\xi = 1$ constraint can be decomposed in the following relations\,:
\begin{align}
\Delta_3^{(g)}(1)\,& = a_3^{(\nl^2)}+a_3^{(n_h^2)}+a_3^{(n_h)}-a_3^{(\nl)}-a_3^{(\nl\,n_h)}\,,\\
\Delta_3^{(\nl)}(1) & = a_3^{(\nl\,n_h)}-2\,a_3^{(\nl^2)}\,,\nonumber\\
\Delta_3^{(n_h)}(1) & = 2\,a_3^{(n_h^2)}-a_3^{(\nl\,n_h)}\,.\nonumber
\end{align}
The exact expression for $\Delta_3^{(g,\nl,n_h)}$ has been  analytically calculated in Ref.~\cite{Bekavac:2007tk}, but it is terribly
lengthly, therefore impractical for a numerical implementation. Ref.~\cite{Bekavac:2007tk} provides $8$ terms of the expansion for small
$\xi$, which can be combined with the $\xi=1$ constraint in a Pad\`e parametrization to provide an approximation which is accurate to 8
significant digits across the range $0\le \xi \le 1$, which is enough for our purposes.
Our parametrizations are explicitly shown in Eq.~\eqref{eq:PadeEQ}. In Fig.~\ref{fig:Delta3} $\Delta_3^{\MSb}(\xi)(\nl=4,n_h=0,\xi) $ is shown, together with the approximation given in Eq.~\eqref{eq:approx}.

\subsection{MSR mass with light active fermion masses}\label{sec:MSRcharm}
To include effects from lighter massive quarks we follow the same approach that was
used in Ref.~\cite{Pietrulewicz:2014qza} to implement R-evolution in the gap parameter of
the soft function when secondary heavy quarks are produced through gluon splitting.
This implementation in the MSR mass happens to satisfy Heavy Quark Symmetry (HQS for short)
exactly. HQS states that in the limit of an infinitely heavy quark, low-energy QCD effects are
flavor independent. This limit has to be taken only in the virtual effects, as only quantum
fluctuations are integrated out. If one looks into $m_Q^{\rm pole} - m_Q^{\MSb}$ in this limit,
$n_h$ becomes $0$, what effectively converts this series to the MSRn scheme,
and $m_q/m_Q$ tends to $0$, what makes the lighter $q$ quark massless. For $m_q^{\rm pole} - m_q^{\MSb}$ 
one has that $m_q\to 0$ in the virtual effects implies $n_h\to0$ and $\nl - 1\to\nl$.
Therefore in this limit $m_Q^{\rm pole} - m_Q^{\rm MSRn}(\mbar_q) = m_q^{\rm pole} - m_q^{\MSb}$.
With our definition this equality will be satisfied even for finite values of the heavy quark mass.
Although there is no theoretical advantage in having (for the MSRn mass) heavy quark symmetry
(accidentally) exact, and the
impact of this feature in the final results is small, it has a practical advantage, as it ensures a smooth transition
to an MSR mass in which the lighter quark $q$ is also integrated out.

\begin{figure*}[t]
	\center
	\subfigure[]
	{\label{fig:Delta3}\includegraphics[width=0.312\textwidth]{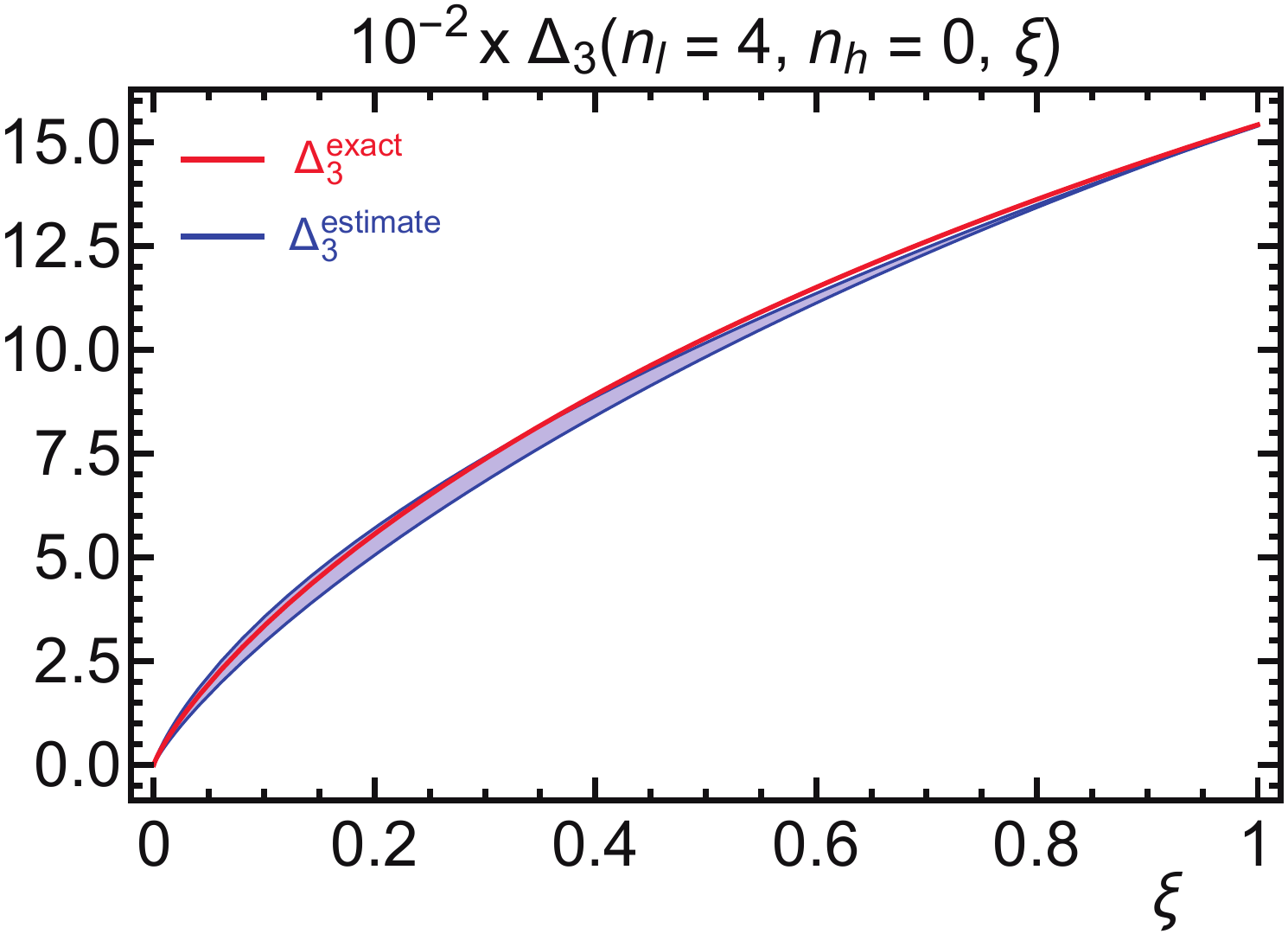}~~}
	\subfigure[]
	{\label{fig:Delta4}\includegraphics[width=0.3\textwidth]{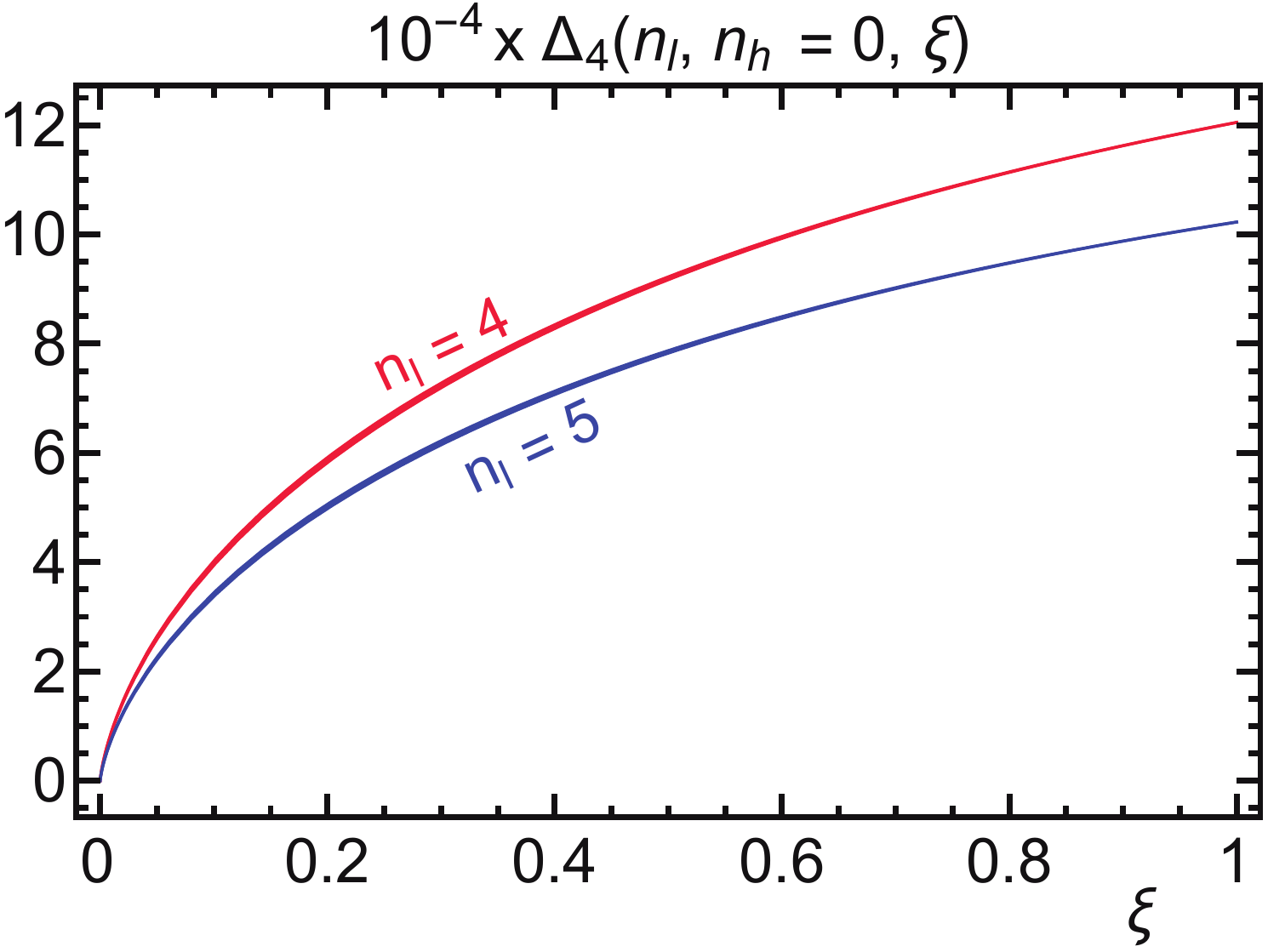}~~}
	\subfigure[]
	{\label{fig:Delta2}\includegraphics[width=0.3\textwidth]{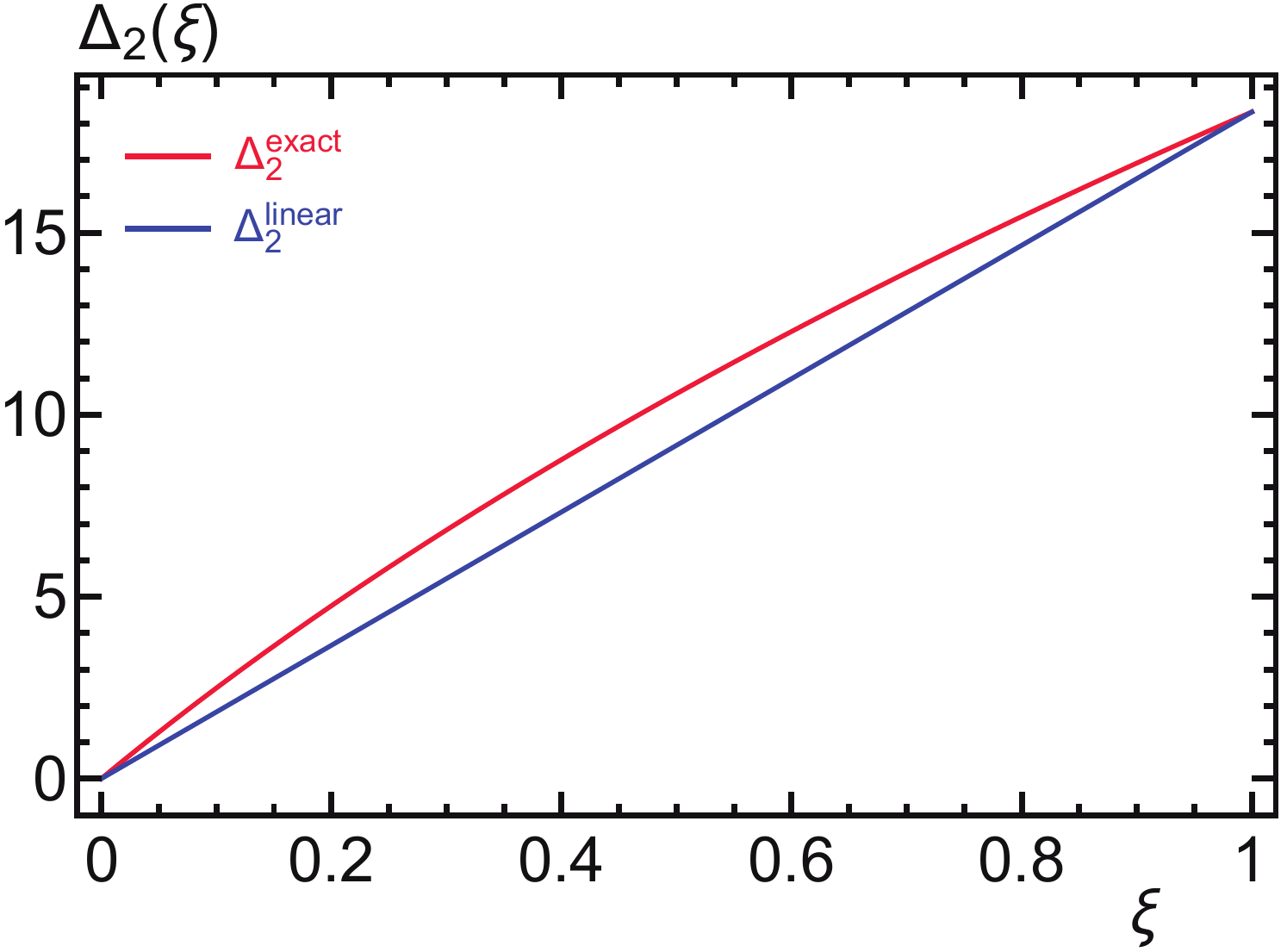} }
 \caption{\label{fig:DeltaEstimate} (a) Estimate of the three-loop massive lighter quark effects on the relation between the $\MSb$ and
pole masses for $\nl=4$ (blue band) versus the exact result (red line). (b) Estimate of the four-loop term $\Delta_4$ for  $\nl=4$ (upper
red band) and $\nl=5$ (lower blue band). The uncertainty bands are estimated from the difference of two consecutive orders of
the approximation. (c) Lighter massive quark correction to the 
$\mathcal{O}(\alpha_s^2)$ term of the $\MSb$-pole relation. In red the exact form of Eq.~\eqref{eq:Delta2}  is shown, while blue shows
the linear approximation of Eq.~\eqref{eq:appform}.}
\end{figure*}

Our definition of the MSR mass including non-zero lighter quark masses reads\,:\,\footnote{In Ref.~\cite{Hoang:2017btd}
the MSR mass with light massive quarks was implemented differently\,:
$\delta m_Q^\MSR(R,\mbar_q) = \delta m_Q^\MSR(R) + \mbar_Q\,\Delta_{\mbar_q}(\mbar_Q,\mbar_q/\mbar_Q)$, which is well suited to study
the large-order behavior of the series. With this definition the
R-evolution equation is the same as for massless lighter quarks, but one might have potentially large logs of the form $\log(R/\mbar_Q)$
when implementing the MSR scheme in a perturbative series. 
With this definition one needs to include Heavy Quark Symmetry breaking corrections to the MSRn mass
when integrating out the massive lighter quark, whereas with our implementation these corrections
are absent.}
\begin{align}\label{eq:MSRdefMC}
&\delta m_Q^\MSR(R,\overline{m}_q) = \delta m_Q^\MSR(R) + R\,\Delta_{\mbar_q}(R,\xi_R)\,,\\
&\Delta_{\mbar_q}(R,\xi_R) = \sum_{k=2}\Delta_{\mbar_q}^{(k)}(\xi_R)
\left(\frac{\alpha_s^{(n_\ell)}(R)}{4\pi}\right)^{\!\!k}\,,\nonumber
\end{align}
where $\xi_R=\mbar_q/R$. In the MSRn scheme one has $\Delta_{\mbar_q}^{(n)} = \Delta^{\MSb}_n|_{n_h=0}$. For the MSRp scheme, at two loops one 
has $\Delta_{\mbar_q}^{(2)} = \Delta^{\MSb}_2$ and at three loops $\Delta_{\mbar_q}^{(3)} = \Delta^{\MSb}_3|_{n_h=1}$ (nontrivial modifications
in the MSRp scheme start at $4$ loops). 
The nice property of Eq.~\eqref{eq:MSRdefMC} is that in the case of the MSRn scheme one has exact Heavy Quark Symmetry\,:
\begin{align}
& m_Q^{\rm pole} - m_Q^{\rm MSRn}(\mbar_q) = \mbar_q\sum_{n=1}\Big[a_n^{\MSb}(\nl,0) + 
\Delta^{\MSb}_n(\nl,0,1)\Big]\left(\frac{\alpha_s^{(n_\ell)}(\mbar_q)}{4\pi}\right)^{\!\!n}\\
& = \mbar_q\sum_{n=1} a_n^{\MSb}(\nl - 1,1)\left(\frac{\alpha_s^{(n_\ell)}(\mbar_q)}{4\pi}\right)^{\!\!n} =
m_q^{\rm pole} - \mbar_q\,,\nonumber
\end{align}
where we have used the $\xi=1$ constraint. 
For the MSRp scheme HQS is not exact, but the correction is calculable in perturbation theory and thanks to the $\xi=1$ constraint
happens to be identical to the MSRn to $\MSb$ matching condition\,:
\begin{align}
&m_Q^{\rm pole} - m_Q^{\rm MSRp}(\mbar_q) - \big[m_q^{\rm pole} - \mbar_q\big]= {\rm HQS}(\mbar_q,\nl-1, 2)\,.
\end{align}
Therefore HQS corrections can be calculated even if the lighter quark finite mass effects have not been computed yet,
as long as the massless corrections are known.
\begin{figure*}[t]
	\center
	\subfigure[]
	{\label{fig:DelGam2}\includegraphics[width=0.465\textwidth]{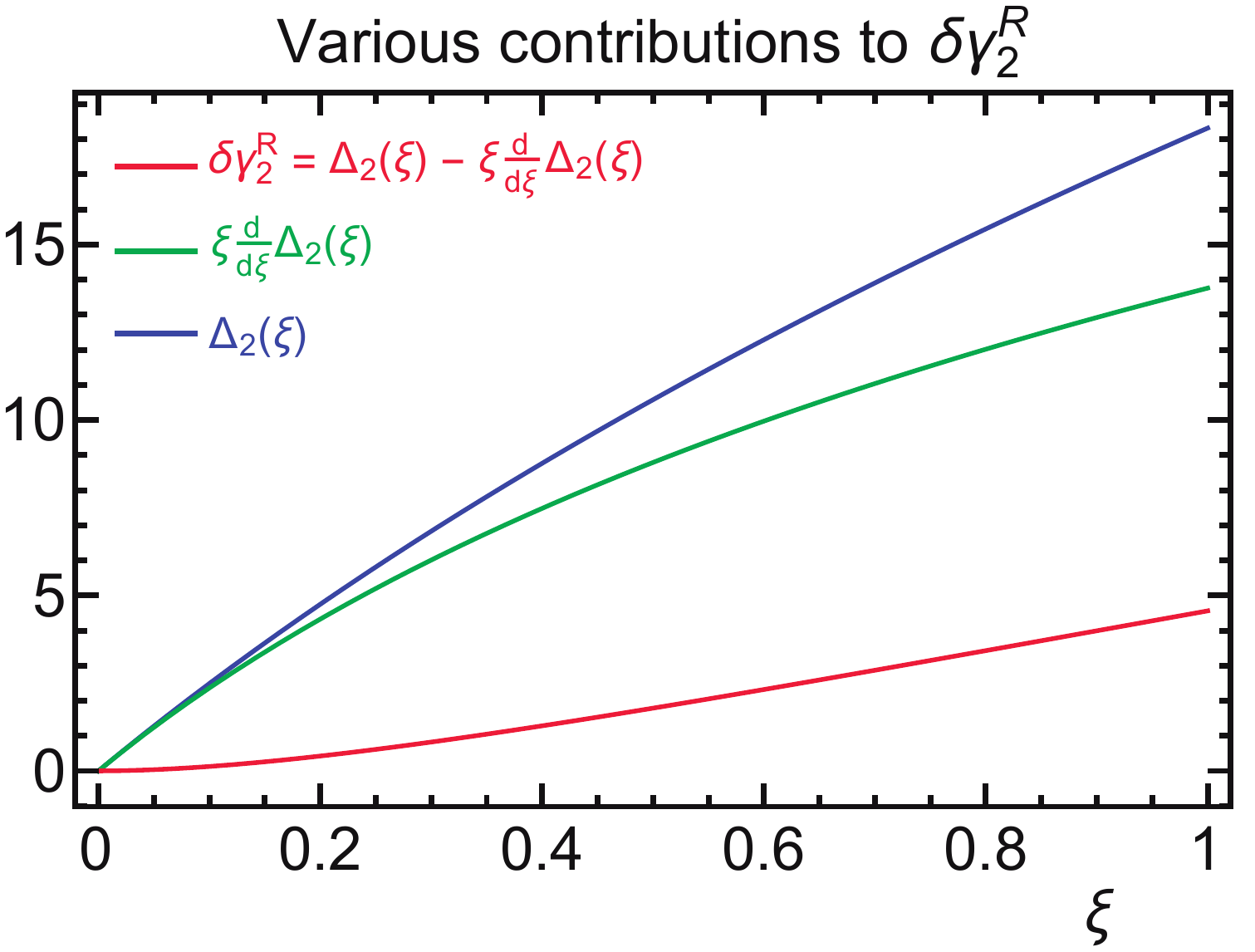}~~}
	\subfigure[]
	{\label{fig:DelGam3}\includegraphics[width=0.487\textwidth]{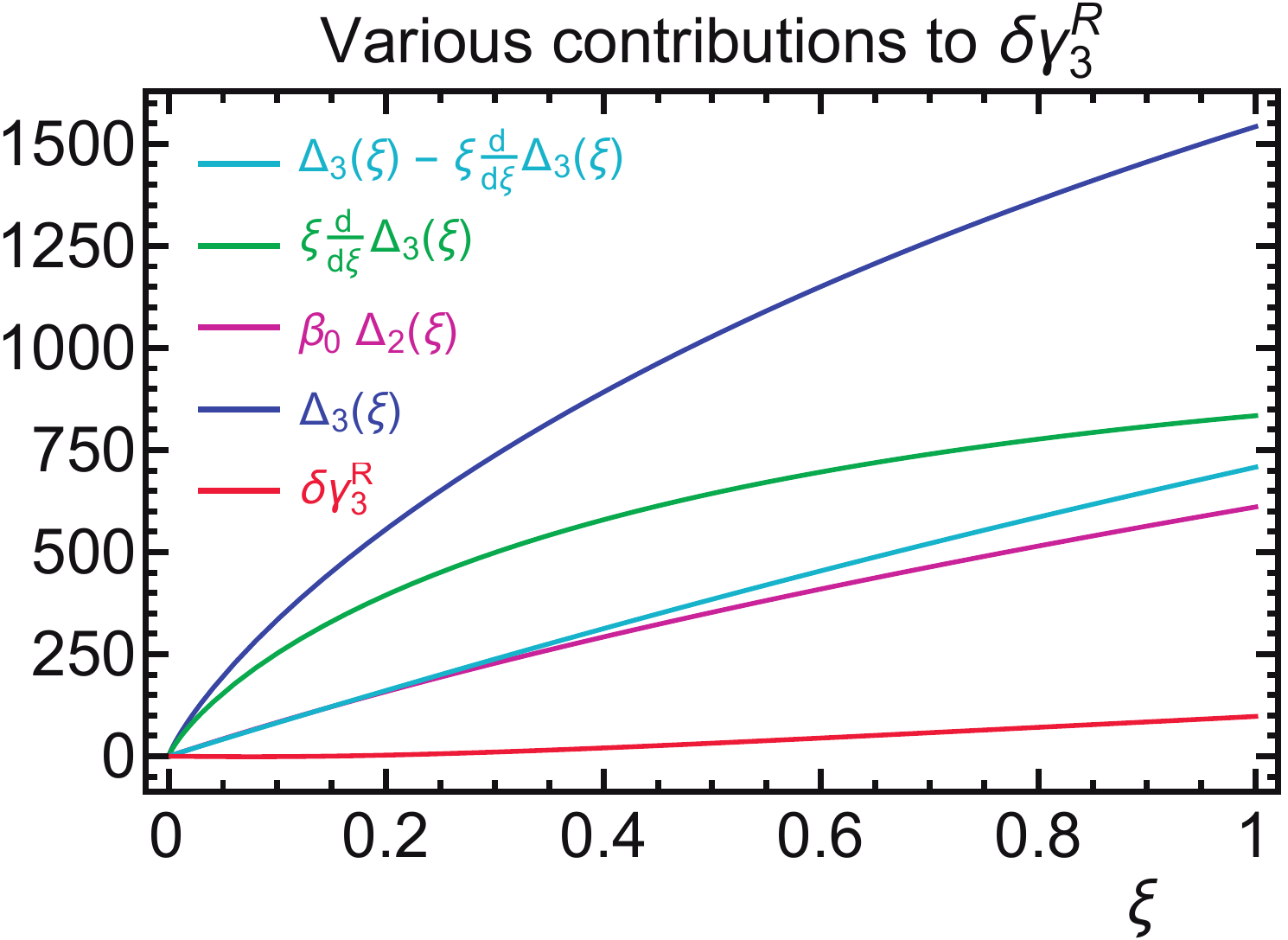} }
 \caption{\label{fig:DeltaGammaPieces} Finite lighter quark mass corrections to the R-anomalous dimension at two and three loops.
The various pieces in which $\delta\gamma_2^R$ (left panel) and $\delta\gamma_3^R$ (right panel) are decomposed are shown separately.
These pieces correspond to the terms shown in Eq.~\eqref{eq:RevolMC}.}
\end{figure*}
The R-anomalous dimension is modified as a consequence of the finite lighter quark mass\,:
\begin{align}\label{eq:RevolMC}
-\,\frac{\rm d}{{\rm d}R}m_Q^{\MSR}(R) &\,=\, \gamma^R[\alpha_s^{(\nl)}(R)] + \delta\gamma^R[\xi_R,\alpha_s^{(\nl)}(R)] \,,\\
 \delta\gamma^R[\xi_R,\alpha_s^{(\nl)}(R)]& \,=\, \sum_{n=1}\delta\gamma_n^R(\xi_R) \left(\frac{\alpha_s^{(\nl)}(R)}{4\pi}\right)^{\!\!n+1}\,,\nonumber\\
 \delta\gamma_n^R(\xi_R)  &\,= \, \Delta_n(\xi_R)-\xi_R\,\frac{\df \Delta_n(\xi_R)}{\df \xi_R}
-2\sum_{j=0}^{n-2} (n-j)\,\beta_j\, \Delta_{n-j}(\xi_R) .\nonumber
\end{align}
This equation can be integrated either numerically as it stands, or recast into a form that factors out $\LQCD$, as explained in
Ref.~\cite{Pietrulewicz:2014qza}. Given that the $\mathcal{O}(\alpha_s^4)$ finite lighter quark mass corrections are unknown,
we will cut the sum of the second term strictly at $n=3$, even if there are lower order $\Delta$ terms that contribute at $n=4$ in the
sum over $j$. Such decision is adopted because the inclusion of such terms could overestimate the real size of that correction if
large cancellations happen. These cancellations are indeed expected since $\Delta_n$ are afflicted by a renormalon but $\delta\gamma_R^n$
are not. In Fig.~\ref{fig:DeltaGammaPieces} this is graphically depicted at two and three loops, where it is seen that the effect is
larger at smaller $\xi$. Since the cancellations are expected to be larger for high values of $n$ (where the renormalon dominates),
we can estimate $\Delta_4(r)$ by requiring that $\delta\gamma_3^R(\xi) = 0$. In general one would have\,:
\begin{align}\label{eq:appform}
\Delta_n(\xi)&\approx \xi\, \Delta_n(1)+ 2\,\xi\,\sum_{j=0}^{n-2} (n-j)\,\beta_j\int_\xi^1\df x\,\frac{\Delta_{n-j}(x)}{x^2}\,,
\end{align}
where $\Delta_n(1)$ are known from the $\xi=1$ constraint. Eq.~\eqref{eq:appform} automatically satisfies the $\xi=1$ and $\xi=0$ constraints.
If this approximation is used from $n=1$ one gets the following predictions\,:
\begin{align}\label{eq:approx}
\Delta_2(\xi)^{\rm app}&= \xi\, \Delta_2(1)\,,\\
\Delta_3(\xi)^{\rm app}&= \xi\, \Delta_3(1) - 4\,\beta_0\,\Delta_2(1)\,\xi\,\log(\xi)\,,\nonumber\\
\Delta_4(\xi)^{\rm app}&= \xi\, \Delta_4(1) - [\,4\,\beta_1\,\Delta_2(1)+6\,\beta_0\,\Delta_r(1)\,]\,\xi\,\log(\xi) +
12\,\beta_0^2\,\Delta_2(1)\,\xi\,\log^2(\xi)\,.\nonumber
\end{align}
This approximation corresponds to the estimate made in Eqs.~(3.12) and (3.13) of Ref.~\cite{Hoang:2017btd} if one sets $\mu=\mbar_Q$
[\,see Eq.~(3.14) of that reference\,]. Since $\Delta_2$ and $\Delta_3$ are exactly known, we can use them to make a better prediction for
$\Delta_4$, which we call $\Delta_4^{\rm best}$. To test this method, one can also use the known expression of $\Delta_2$ to predict
$\Delta_3$, calling it again $\Delta_3^{\rm best}$, and consider the difference of $\Delta_3^{\rm best}$ and $\Delta_3^{\rm app}$ as
an estimate of the uncertainty of this procedure. We observe that $\Delta_3^{\rm best}$ is a better approximation to $\Delta_3$ than
$\Delta_3^{\rm app}$, and that $\Delta_3$ is contained in the uncertainty band, although quite close to its upper bound. This is depicted
in Fig.~\ref{fig:Delta3}.
Finally one can compute the estimate of $\Delta_4$ which uses the exact form of $\Delta_2$ and $\Delta_3^{\rm best}$, which we call
$\Delta_4^{\rm good}$. We observe that $\Delta_4^{\rm good}$ is closer to $\Delta_4^{\rm best}$ than $\Delta_4^{\rm app}$.
The difference between $\Delta_4^{\rm good}$ and $\Delta_4^{\rm best}$ will be used to estimate the uncertainty on
$\Delta_4^{\rm best}$, which happens to be $2\%$ at worst across the whole spectrum for either $4$ or $5$ light flavors. Our estimate is
shown graphically in Fig.~\ref{fig:Delta4}. Our prediction for $\Delta_4$ and its uncertainty band can be parametrized with the following fit
functions\,:
\begin{align}
\Delta_4(\nl=4) & = \xi  \,[\,120472 + 35 \,\xi +(2201\, \xi +20909) \log ^2(\xi )+ (-3768 \,\xi -76707) \log (\xi )\,]\nonumber\\
&\pm \,\xi\,[\,5\, \xi+(-2667\, \xi -375)   \log ^2(\xi )+(5459 \,\xi -5430)\,  \log (\xi )\,]\,,\nonumber\\[2mm]
\Delta_4(\nl=5) & = \xi  \,[\,102238 + 33\, \xi +(1966\, \xi +17620) \log ^2(\xi )-(3011 \xi +66608) \log (\xi )\,]\nonumber\\
&\pm \,\xi\,[\,7\, \xi - (2208\, \xi + 388)  \log^2(\xi ) + (4873\, \xi - 4811)  \log (\xi )\,]\,.
\end{align}
The method of requiring a vanishing R-anomalous dimension is not as powerful in predicting the massless coefficients, and for those
one needs to know at least the one-loop term. For instance, for $\nl=4$, and assuming that all orders lower than the one that is
predicted are known, one gets\,:
\begin{align}
& a_2^{\rm \MSb}(\nl=4,0)=89[130]\,, & a_3^{\rm \MSb}(\nl=4,0) &= 5400\pm 1900\,,\quad \\
& a_4^{\rm \MSb}(\nl=4,0)=330000\pm 36000\,, &a_5^{\rm \MSb}(\nl=4,0) &=(232\pm 7)\times 10^5\,,\nonumber
\end{align}
there the quoted uncertainty corresponds to the difference of the central value and the prediction that 
assumes that the previous order is also estimated with this method. The exact results for
$a_n^{\rm \MSb}(\nl=4,0)$ for $n = 2,\, 3,\, 4$ are $130.128,\, 4582.54,\, 214100$.

Finally, we note that the uncertainty associated to the absence of the $\Delta_4$, $\delta\gamma_3^R$ and $\delta E_X^{(3)}$ terms
can be estimated by comparing the schemes with $\nl$ and $(\nl - 1)$ dynamical flavors.

\subsection[MSR mass in the $n_\ell$ flavor scheme]
{MSR mass in the ${\mathbf n_\ell}$ flavor scheme}\label{sec:MSRnl}
The MSR mass described in Sec.~\ref{sec:MSRcharm} accounts for the charm mass corrections explicitly, as it considers the charm quark
a dynamical flavor. However there are physical situations in which, due to the energy scales involved, it is convenient to integrate
out the charm quark as well. For bottomonium, the charm mass is sufficiently large compared to the non-relativistic scales to assume
one is close to the decoupling limit, i.e., $\mbar_c\to\infty$. Indeed, the closeness of the exact finite-charm mass corrections of
Sec.~\ref{sec:charm} to the decoupling limit for scales around $1$ to $4$\,GeV points to such conclusion. In that case, the charm
quark can be integrated out and one should switch to an $(\nl\!-\!1)$-flavor MSR scheme with the charm quark decoupled to all orders in 
$\Upsilon$ expansion, reducing to a theory with $(n_\ell - 1)$ dynamical flavors.

\begin{figure}
\center
\includegraphics[width=0.53\textwidth]{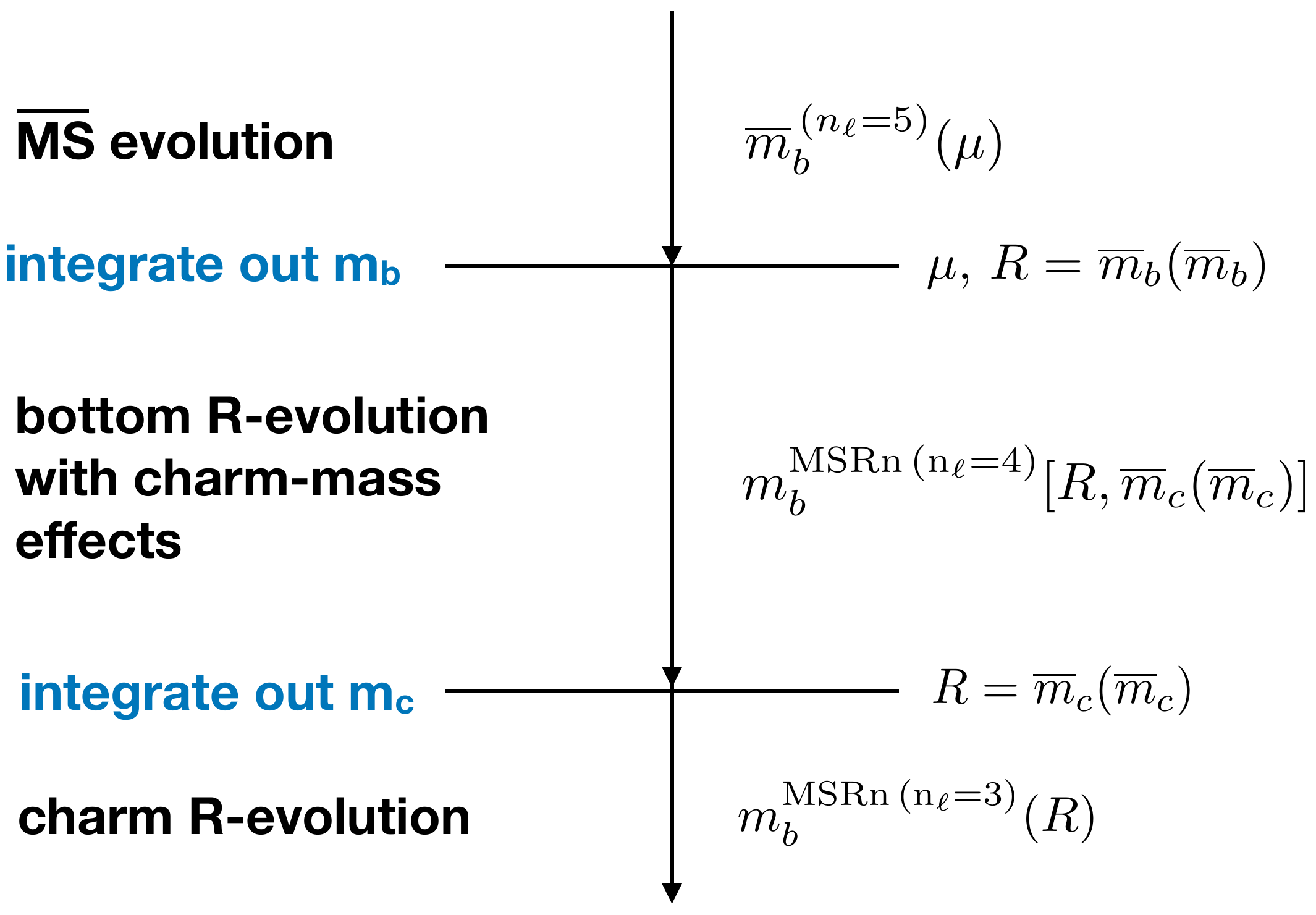}
 \caption{\label{fig:MSRRunMatch} Graphical representation of the sequence of R-running and matching that one needs to carry out to
evaluate the MSR$^{(\nl )}$ and MSR$^{(\nl - 1)}$ masses.}
\end{figure}

We define the MSR mass with $\nl$ active flavors, either Natural or Practical, as follows\,:
\begin{align}\label{eq:MSRnl-1}
m_Q^{\rm pole} - m_Q^{{\rm MSR}\,(\nl-1)}(R) = m_q^{\rm pole} - m_q^{\rm MSR}(R)\,,
\end{align}
where it is understood that $m_q^{\rm MSR}$ has $(\nl-1)$ dynamical flavors. Therefore, the R-evolution in this scheme does not include
charm mass effects, and its relation to the pole mass has only $(\nl - 1)$ flavors. One needs to connect the MSR$^{(\nl - 1)}$ mass to $\mbar_Q$,
and if large logs of $\mbar_Q/\mbar_q$ are to be avoided, one should not directly subtract Eqs.~\eqref{eq:MSRnl-1} and \eqref{eq:MSRbcharm}.
\mbox{R-evolution} can sum up these logs if the relation is organized as follows\,:
\begin{align}\label{eq:MSRnlrunmatch}
& m_Q^{{\rm MSR}\,(\nl-1)}(R) \,=\, \Big[m_Q^{{\rm MSR}\,(\nl-1)}(R) - m_Q^{{\rm MSR}\,(\nl-1)}(\mbar_q)\Big] \,+ \,
 \Big[ m_Q^{{\rm MSR}\,(\nl-1)}(\mbar_q) - m_Q^{\rm pole}\Big] \nonumber \\
& + \,\Big[m_Q^{\rm pole} - m_Q^{\rm MSR'}(\mbar_q)\Big] \nonumber
 + \,\Big[ m_Q^{\rm MSR'}(\mbar_q) - m_Q^{\rm MSR'}(\mbar_Q)\Big] \,+\, \Big[ m_q^{\rm MSR'}(\mbar_Q) - \mbar_Q\Big] + \mbar_Q\\[4mm]
& =\, \Big[m_q^{\rm MSR}(R) - m_q^{\rm MSR}(\mbar_q)\Big] \,+ \,\Big[ m_q^{\rm MSR}(\mbar_q) - \mbar_q\Big] +
\Big[\mbar_q - m_q^{\rm pole}\Big]\nonumber
 + \,\Big[m_Q^{\rm pole} - m_Q^{\rm MSR'}(\mbar_q)\Big]\\
& + \,\Big[ m_Q^{\rm MSR'}(\mbar_q) - m_Q^{\rm MSR'}(\mbar_Q)\Big] \,+\, \Big[ m_Q^{\rm MSR'}(\mbar_Q) - \mbar_Q\Big] + \mbar_Q\,,
\end{align}
where one can use different versions of the MSR mass in the $\nl$ and $(\nl - 1)$ schemes, hence we distinguish them with a prime.
We have used Eq.~\eqref{eq:MSRnl-1} in the first
and second terms in square brackets of the first line to get to the second line.
One uses R-evolution to sum up logs of $R/\mbar_q$ and $\mbar_q/\mbar_Q$ in the first and fifth terms in square brackets, respectively.
The second and one-to-last terms in square brackets are the $\MSb$-MSR matching conditions, which are zero in the Practical scheme. The
sum of the third and fourth terms in square brackets correspond to the HQS breaking expression, which is zero in the Natural scheme.
In practice one can write Eq.~\eqref{eq:MSRnlrunmatch} as
\begin{align}
m_Q^{{\rm MSR}\,(\nl-1)}(R) \,=\, m_q^{\rm MSR}(R) - \mbar_q + {\rm HQS'}(\mbar_q,\nl - 1,2) + m_Q^{\rm MSR'}(\mbar_q)\,,
\end{align}
where $m_q^{\rm MSR}$ and $m_Q^{\rm MSR'}$ include matching to the $\MSb$ mass and R-evolution, and \mbox{HQS = 0} if MSR' is the Natural scheme.
The sequence of R-running and matching can be better visualized in Fig.~\ref{fig:MSRRunMatch}.

In Ref.~\cite{Ayala:2014yxa} the finite charm quark mass corrections to a three-flavor bottom quark $\MSb$ mass are computed in 
fixed-order, without R-evolution. In our notation that computation corresponds to $\mbar_b\,-\,m_b^{{\rm MSRp}\,(\nl-1)}(\mbar_b)$.
One could decompose this difference into various pieces that sum up logarithms
\begin{align}
&\mbar_b\,-\,m_b^{{\rm MSRp}\,(\nl-1)}(\mbar_b) = \Big[m_b^{\rm MSRp}(\mbar_b) - m_b^{\rm MSRp}(\mbar_c)\Big]\, + \,
\Big[m_b^{\rm MSRp}(\mbar_c) - m_b^{\rm pole}\Big]\\
& +\,\Big[m_b^{\rm pole} - m_b^{\rm MSRp\,(\nl - 1)}(\mbar_c)\Big] +
\Big[m_b^{\rm MSRp\,(\nl - 1)}(\mbar_c) - m_b^{\rm MSRp\,(\nl - 1)}(\mbar_b)\Big]\nonumber\\[4mm]
& = \Big[m_b^{\rm MSRp}(\mbar_b) - m_b^{\rm MSRp}(\mbar_c)\Big] + {\rm HQS}(\mbar_c) +
\Big[m_c^{\rm MSRp}(\mbar_c) - m_c^{\rm MSRp}(\mbar_b)\Big]\,,\nonumber
\end{align}
where we have used that $\mbar_b = m_b^{\rm MSRp}(\mbar_b)$. The first and last terms in brackets sum up logarithms of $\mbar_b/\mbar_c$. Since
these are not very large, we find that our result is not in contradiction with Ref.~\cite{Ayala:2014yxa}. Taking $\mbar_b = 4.2\,$GeV,
$\mbar_c = 1.3\,$GeV, $\alpha_s^{(\nf=5)}(m_Z)=0.1181$ we find $\mbar_b\,-\,m_b^{{\rm MSRp}\,(\nl-1)}(\mbar_b) = 0.05\pm 0.05\,$GeV, where the
error comes from scale variation in the R-evolution.\footnote{Ref.~\cite{Hoang:2017kmk} computes the similar quantity
$\mbar_b\,-\,m_b^{{\rm MSRn}\,(\nl-1)}(\mbar_b)$, finding an effect which is also compatible with zero. We find $0.04\pm 0.04\,$GeV, using
our slightly different MSR definition. Furthermore, in our uncertainty estimate of HQS we probe scales a factor of two smaller (larger) than
their lowest (highest) scale.}

%


\section{Scale Variation and Charm Quark Mass Effects}\label{sec:investigation}
In this section we perform a numerical investigation of the scale variation that one needs to perform to properly estimate uncertainties
coming from missing uncalculated higher-order terms.\,\footnote{For the numerics in this article we run $\alpha_s$ with the
five-loop beta function, and we match at the various quark mass thresholds taking the five-loop expression with $\mu_q=\mbar_q$.
For the MSR mass we perform 4-loop massless R-evolution (and matching to $\mbar_q$) and include finite charm mass corrections at three loops.
The matching between the MSR$^{(\nl=4)}$ and MSR$^{(\nl=3)}$ masses is performed at four loops.}
These unknown terms appear in the perturbative expansion for the quarkonium energy levels as well as on the relation between the
pole and $\MSb$ masses. The way one estimates the former is varying the scale $\mu$, while varying $R$ provides an estimate
for the latter. The range in which these two parameters should be varied depends on the principal quantum number $n$. For $\mu$ this
is easy to see, since $n$ appears in the argument of the perturbative logarithms. Since there are logarithms of $\mu/R$ in the MSR mass
subtractions, $R$ has to be of the same order of $\mu$, and that is how the $R$ variation inherits its $n$ dependence. In any case we observe
that changing the value of $R$ in a given range causes variations on the predicted masses much smaller than varying $\mu$ in the same range.
Therefore we interpret that $\mu$ variation estimates the error coming from missing higher order terms in the theoretical expression
for the bound-state masses, whereas $R$ variation accounts for the arbitrariness that one has in connecting the relativistic
$R\sim \mbar_Q$ and non-relativistic regimes $R\sim\mbar_Q\alpha_s C_F$, in the same way that one uses the MSR mass to smoothly match the
1S mass into the $\MSb$ one. For simplicity we use the same range of variation for both $\mu$ and $R$.

\begin{figure*}[t]
	\center
	\subfigure[]
    {\label{fig:MassScale1}\includegraphics[width=0.483\textwidth]{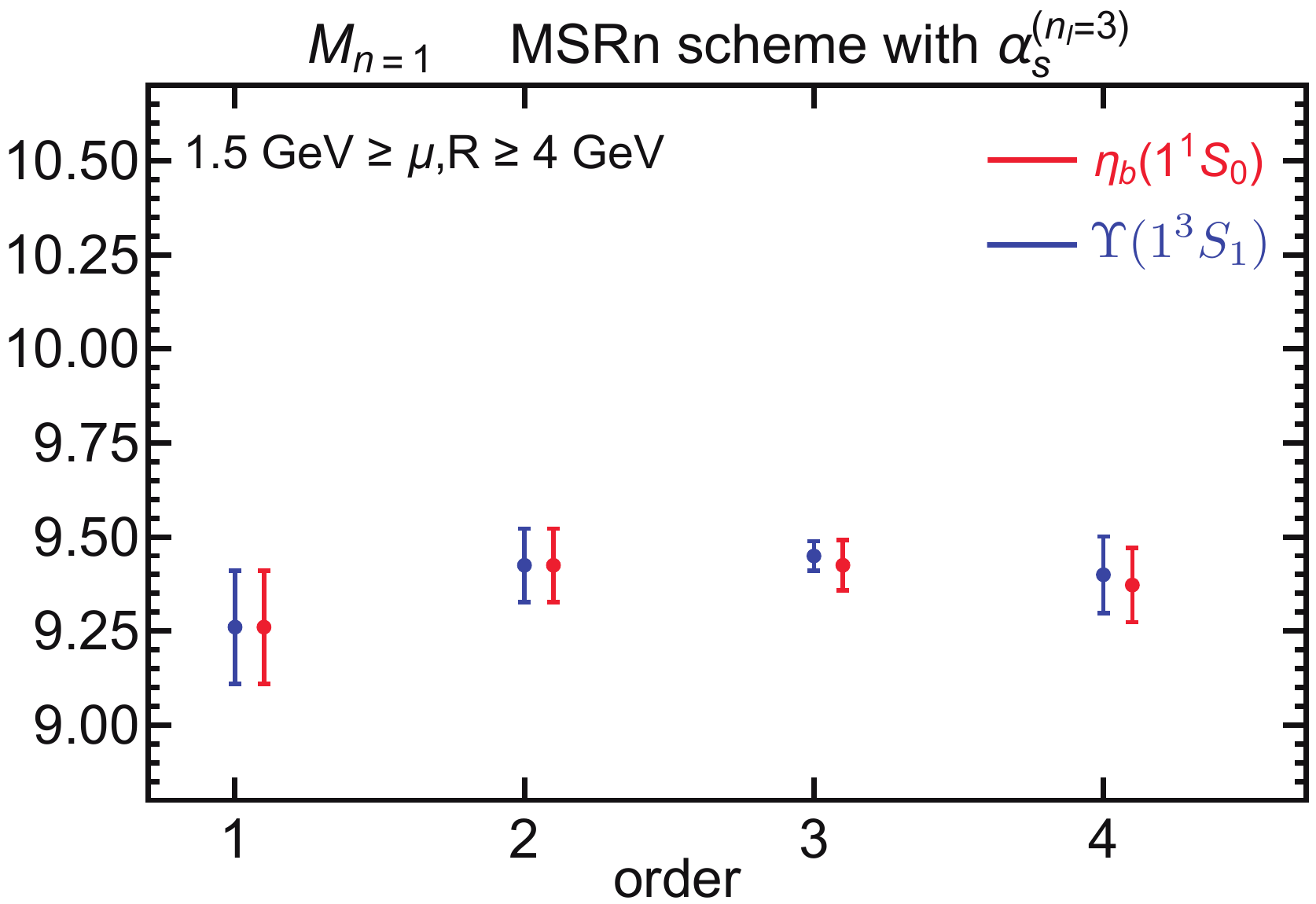}~~}
	\subfigure[]
	{\label{fig:MassScale2}\includegraphics[width=0.483\textwidth]{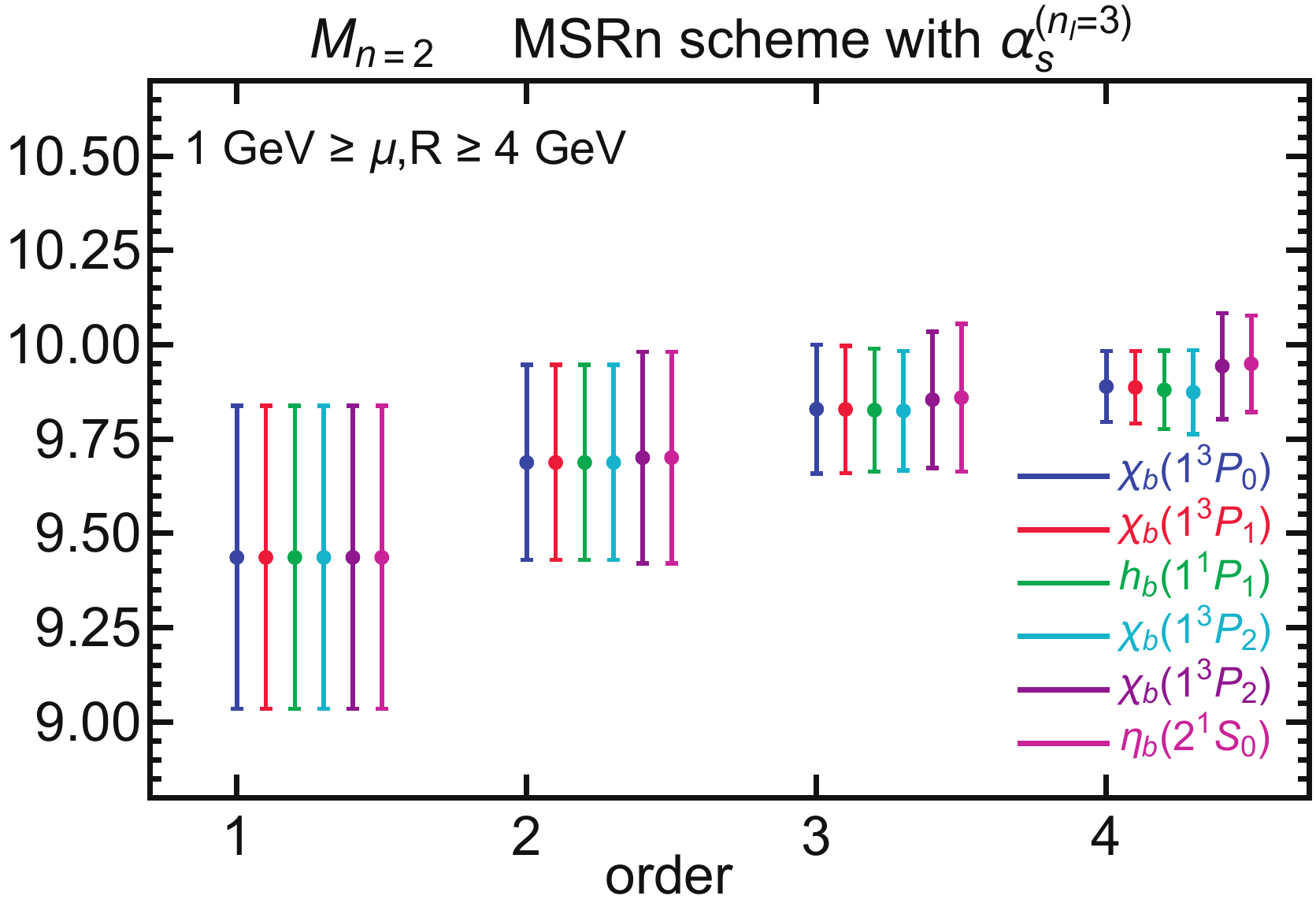}}
	\subfigure[]
	{\label{fig:MassScale4}\includegraphics[width=0.483\textwidth]{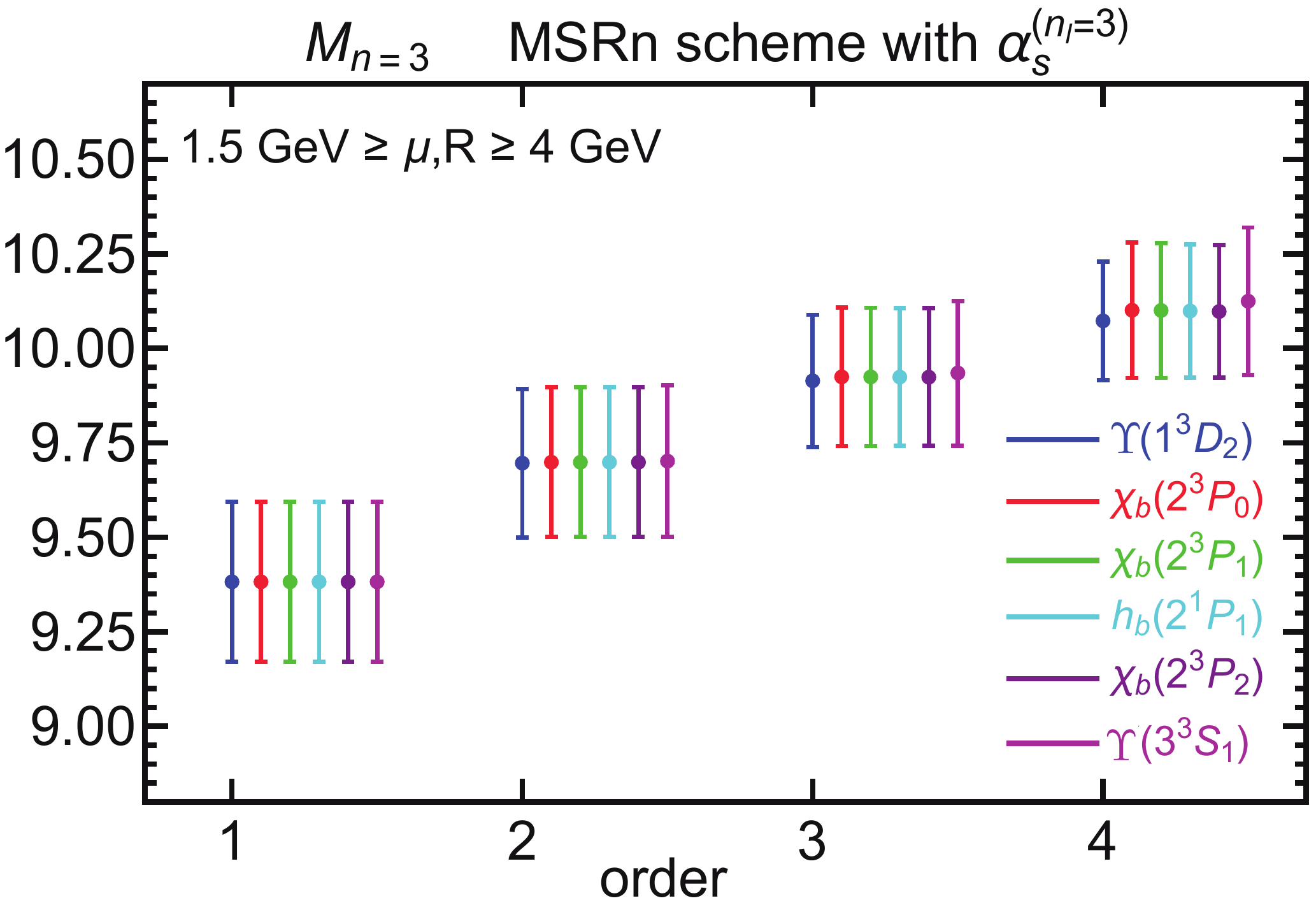}~~}
	\subfigure[]
	{\label{fig:MassScale3}\includegraphics[width=0.483\textwidth]{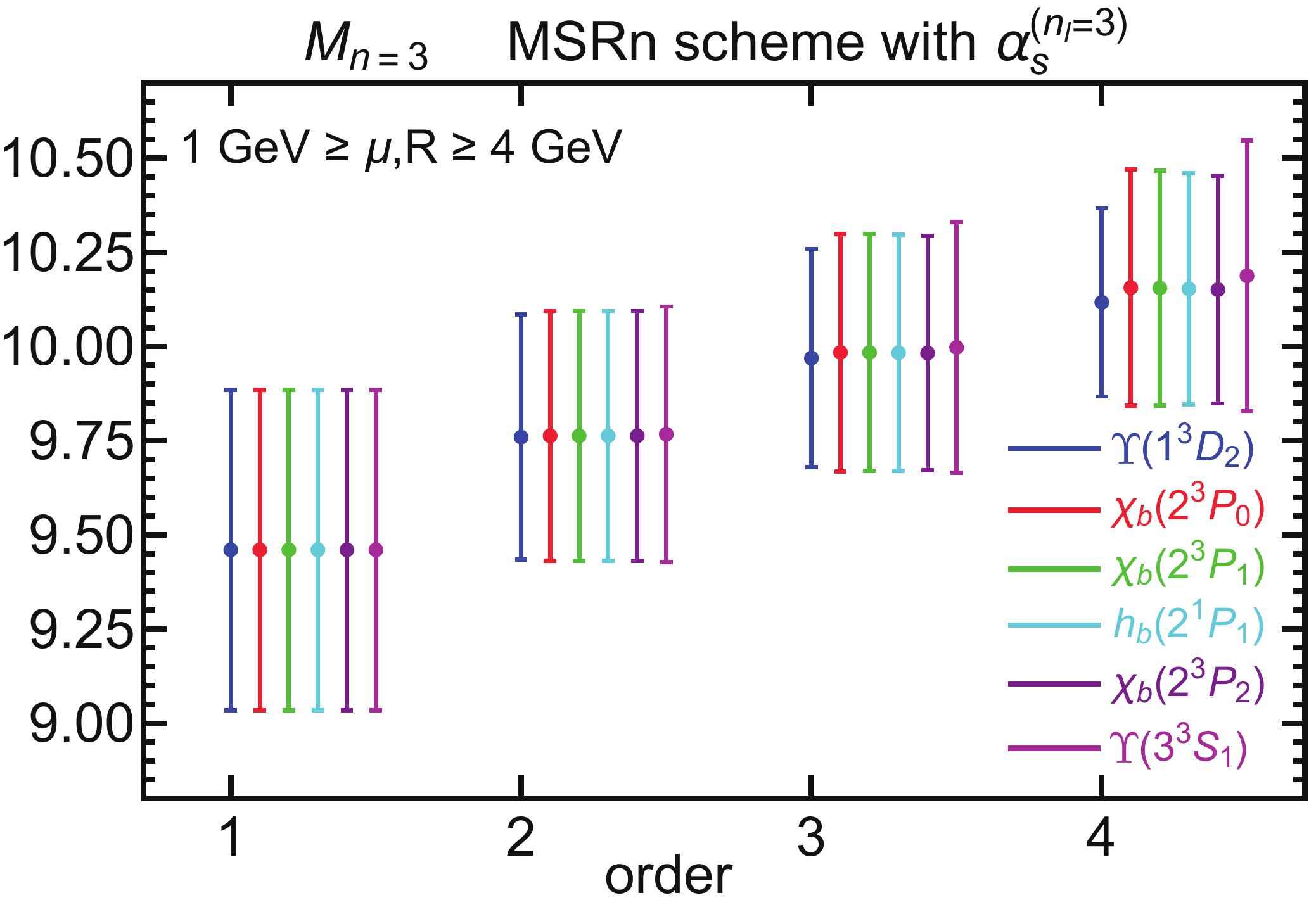} }
 \caption{\label{fig:MassScales} Prediction for the bottomonium masses up to $n = 3$ for various perturbative orders. The top figures
show the two $n=1$ states with a scale variation of $\mu$ and $R$ between $1.5\,$GeV and $4\,$GeV (left plot) and $n = 2$ with variation
between $1\,$GeV and $4\,$GeV (right plot). The two lower plots have the same scale variation as the corresponding upper plots,
but both show results for $n = 3$.}
\end{figure*}

We will start with the usual procedure to figure out the ``natural'' value of the renormalization scale\,: $\mu_{\rm nat}$ has to
be such that the argument of the perturbative logarithms equals unity. To estimate the range of variation we can simply require that the argument
of the logarithms becomes $2^{\pm \phi}$\,: $(\mu_{\rm nat} \pm \Delta\mu)\sim 2^{\pm \phi}C_F\alpha_s^{(\nl)}(\mu)m_Q/n$, where the specific
value of $m_Q$ depends on the scheme one is using, and one can adjust the value of $\phi$ or even take different values for upwards or
downwards variations. For this numerical exercise we take $m_Q=m_Q^{\rm MSR}(R=\mu)$ and half of the average of the masses of states with 
principal quantum number $n$ (experimental values). Both choices yield similar results\,: for bottomonium and taking $\phi = 0.5$ we find
for $n=1,2$ the following ranges\,: $\mu_{n=1}\sim1.9^{+1.6}_{-0.4}$\,GeV, $\mu_{n=2}\sim1.25\pm 0.25$\,GeV. Most of the variation from
this ranges comes from the lower edge, therefore we simplify the scale variations to $1.5\,{\rm GeV}\ge\mu_{n=1}\ge 4\,{\rm GeV}$ and
$1\,{\rm GeV}\ge\mu_{n=2}\ge 4\,{\rm GeV}$. We observe that using these ranges the estimated uncertainty bars make the results compatible
order by order in perturbation theory, as can be seen in Figs.~\ref{fig:MassScale1} and \ref{fig:MassScale2}. We observe that for $n=1$
a turnover takes place between N$^2$LO and N$^3$LO, as the error bar increases when increasing the order of perturbation theory, contrary
to the expectations. We observe that this behavior persists even if the lower scale in the variation range is reduced to $1\,$GeV.
For $n=2$ this does not happen and the minimal uncertainty is achieved at the highest perturbative order, as expected. It is hard to pin
down the origin of this behavior, but one could imagine that some sort of accidental numerical cancellation happens at N$^2$LO and $n=1$,
which makes the renormalization scale variation method underestimate the actual uncertainty due to missing higher order terms. This behavior
is obviously transmitted to the extraction of the heavy quark masses.

The same exercise for
$n=3$ yields a lower bound below $1\,$GeV, presumably too low for perturbation theory to work. Therefore we try the same variations as
for $n=1,2$, and as shown in Figs.~\ref{fig:MassScale2} and \ref{fig:MassScale4} none of these choices is satisfactory\,: the $n=1$
variation makes the N$^3$LO prediction incompatible with the LO and NLO results; on the other hand the $n=2$ variation makes the error
bars equally large for all orders. We therefore conclude that perturbation theory is not applicable for values of $n$ equal or larger
than $3$.

\begin{figure*}[t]
	\center
	\subfigure[]
	{\label{fig:charmScale}\includegraphics[width=0.491\textwidth]{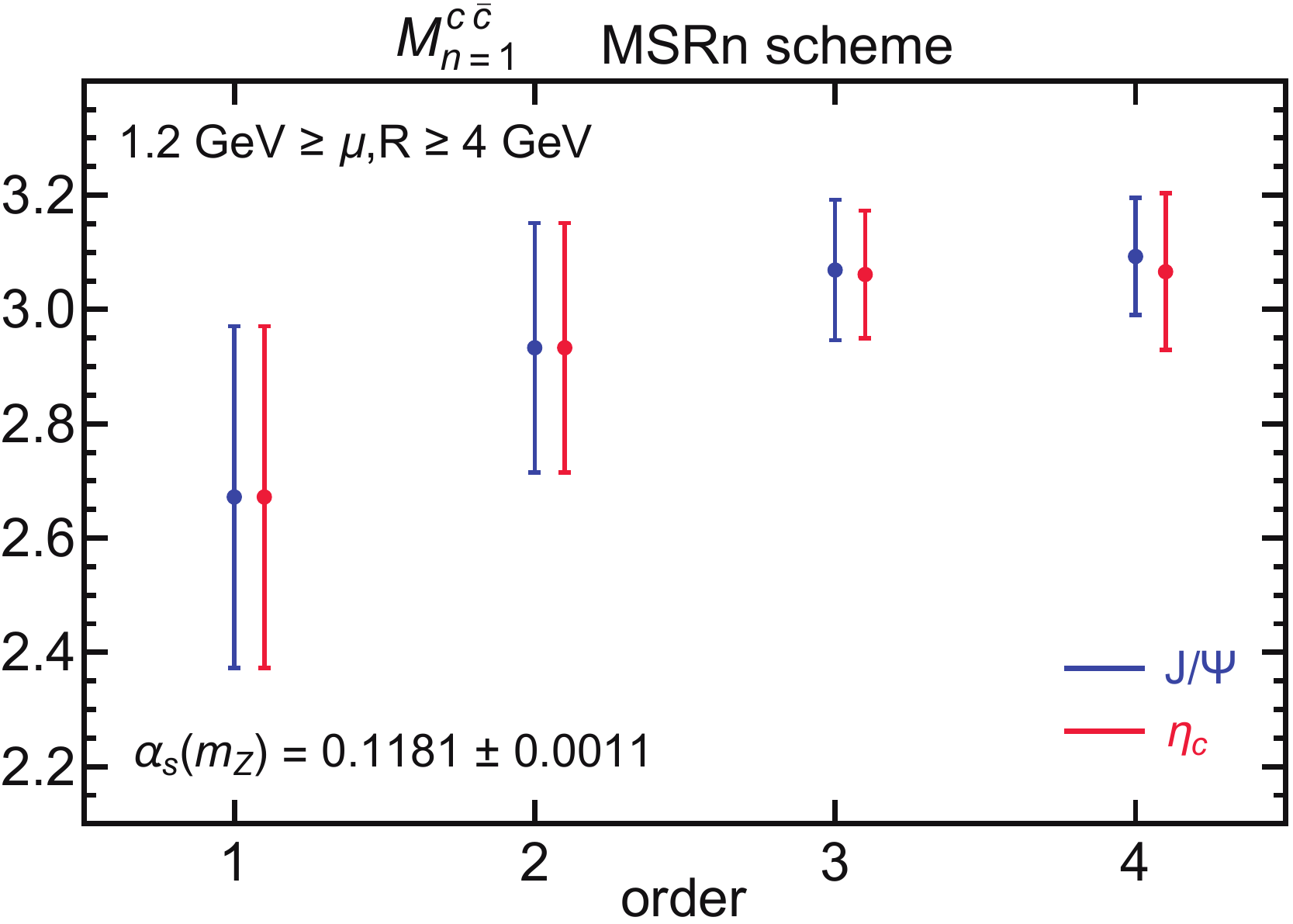}~~}
	\subfigure[]
	{\label{fig:mcSchemes}\includegraphics[width=0.475\textwidth]{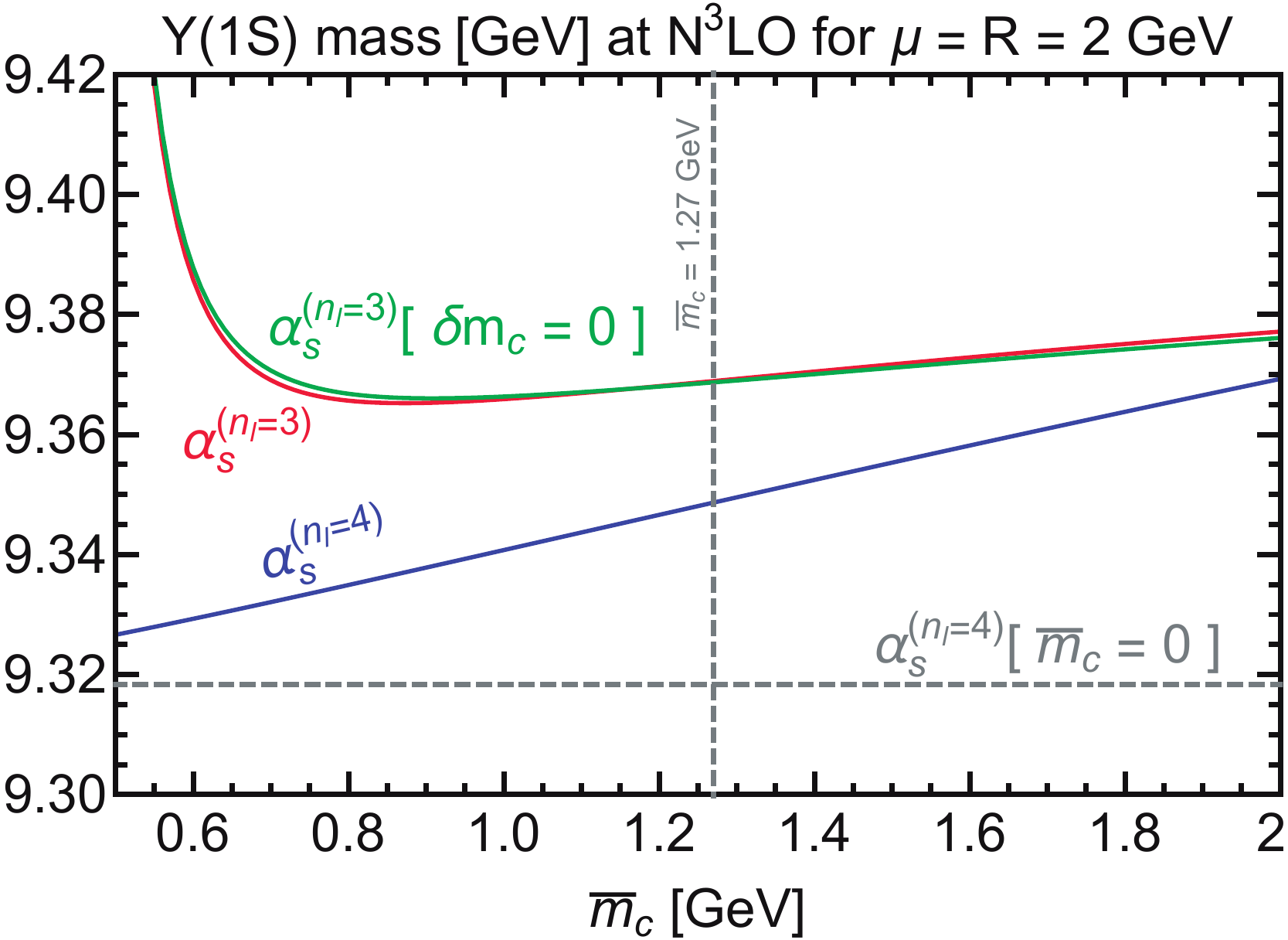} }
 \caption{\label{fig:charmSchemeScale} Left panel\,: Mass of the $n=1$ charmonium states at various perturbative orders.
Blue and red colors correspond to the masses of the $J/\Psi$ and $\eta_c$ particles, respectively. The error bars correspond
to a scale variation between $1.2\,$GeV and $4\,$GeV. Right panel\,: Mass of the $\Upsilon(1^3S_1)$ bottomonium bound state in
the MSRn scheme as a function of the charm mass in the $\MSb$ scheme. The blue and red lines show the $\alpha_s^{(\nl=4)}$ and
$\alpha_s^{(\nl=3)}$ schemes, respectively. The green line corresponds to the $\alpha_s^{(\nl=3)}$ scheme with no further finite
charm mass corrections implemented. The horizontal gray dashed line shows the $\mbar_c=0$ approximation in the $\alpha_s^{(\nl=4)}$
scheme, while the vertical dashed gray line signals the world average value for the charm quark mass.}
\end{figure*}

The same analysis for charmonium again yields values for the natural scale $\mu$ below $1\,$GeV. In this case we take a completely
heuristic approach and choose the scale variation such that there is order-by-order convergence and agreement. We find that varying
the scales in the range $1.2\,{\rm GeV}\ge\mu_{\rm charm}\ge 4\,{\rm GeV}$ gives a pattern very similar to the $n=1$ bottomonium states,
as can be seen by comparing Figs.~\ref{fig:MassScale1} and~\ref{fig:charmScale}.

\begin{figure*}[t]
	\center
	\subfigure[]
	{\label{fig:MinimalBottom1}\includegraphics[width=0.31\textwidth]{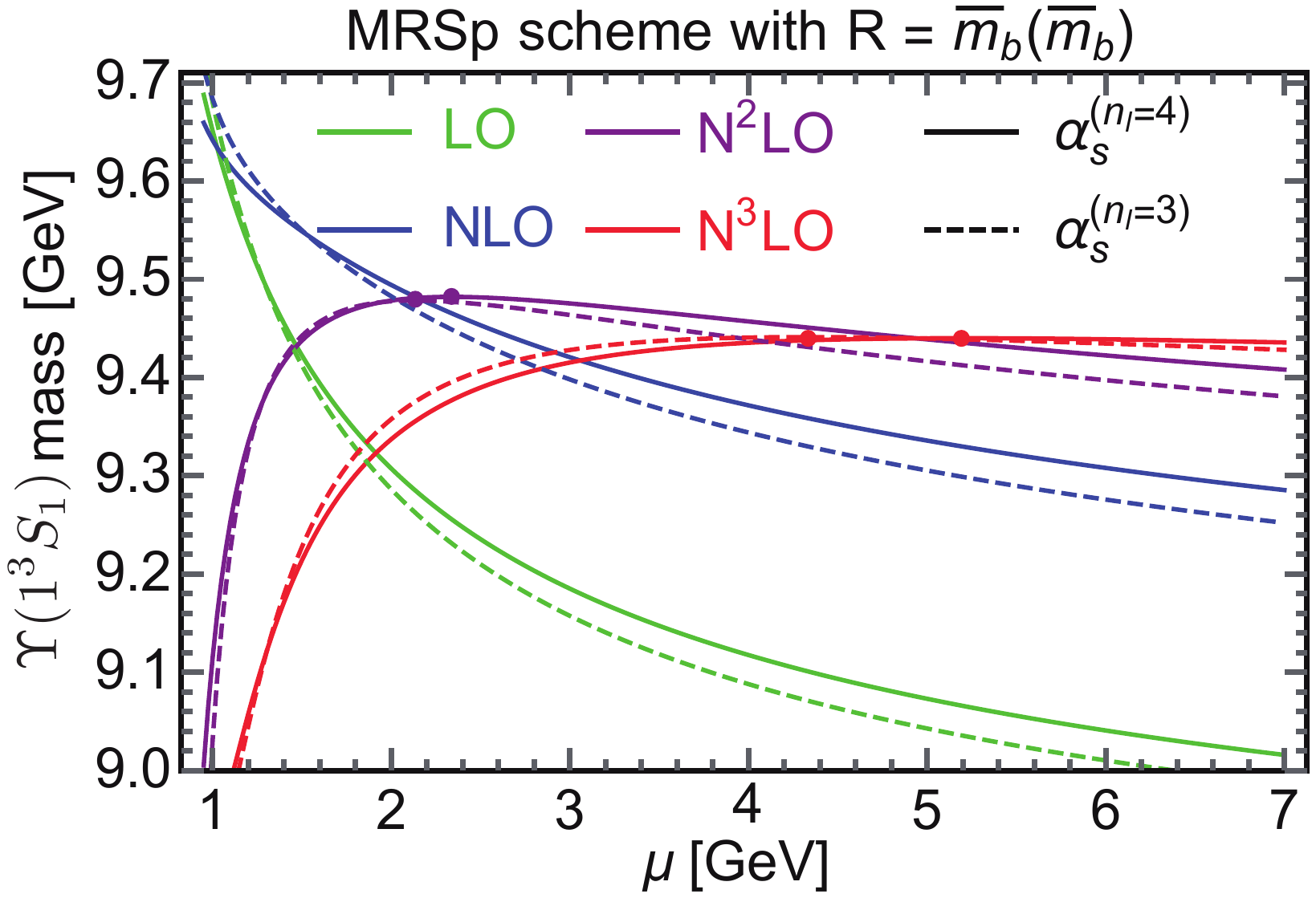}~~}
	\subfigure[]
	{\label{fig:MinimalBottom2}\includegraphics[width=0.32\textwidth]{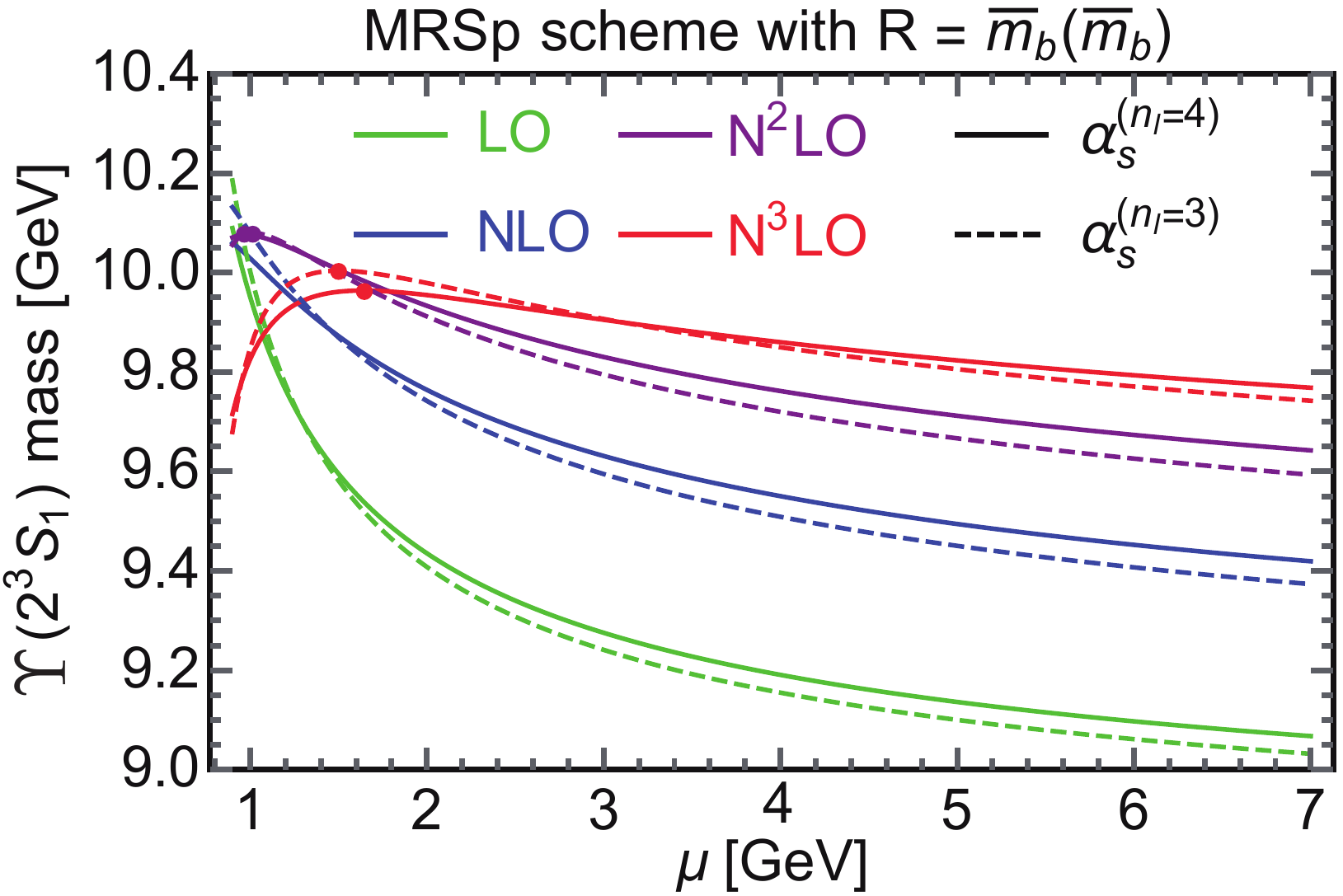}~~}
	\subfigure[]
	{\label{fig:MinimalBottom2}\includegraphics[width=0.31\textwidth]{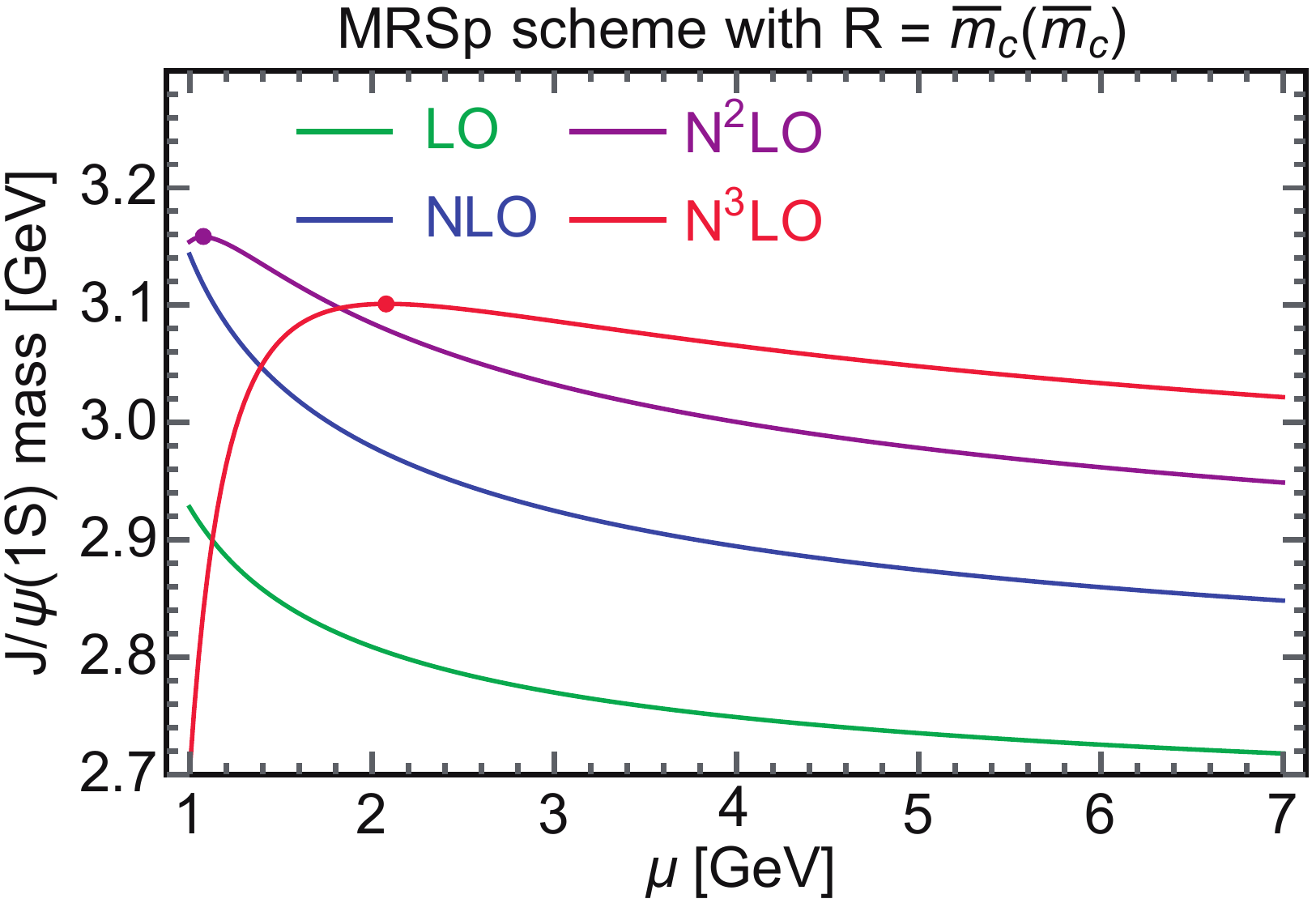} }
 \caption{\label{fig:Minimal} Canonical scales according to the principle of minimal sensitivity. The left~(a), center~(b) and
right~(c) panels show the bottomonium $\Upsilon(1^3S_1)$ and $\Upsilon(2^3S_1)$ states, and the charmonium $J/\psi(1S)$ state,
respectively. In (a) and (b) the solid line shows the $\alpha_s^{(\nl=4)}$ scheme, whereas dashed lines correspond to the
$\alpha_s^{(\nl=3)}$ scheme. Green, blue, purple and red colors represent LO, NLO, N$^2$LO and N$^3$LO predictions, respectively.
Whenever a maximum is attained, its position is marked with a fat colored dot.}
\end{figure*}

Finally let us compare our considerations for natural renormalization scales with the so called Principle of Minimal Sensitivity
(PMS for short). For this exercise we consider the MSRp scheme with $R = \mbar_Q$. This mimics the $\MSb$ scheme for charmonium and for
bottomonium in the $\alpha_s^{(\nl=4)}$ scheme. For bottomonium in the $\alpha_s^{(\nl=3)}$ scheme it is similar to the $\MSb$ scheme,
but with R-evolution summing up logarithms of $\mbar_c/\mbar_b$. We make this choice because it is the most similar to what is used in
Refs.~\cite{Brambilla:2001qk,Kiyo:2015ufa}, while the outcome of this analysis in the MSRn scheme is very similar. Our results are
shown in Fig.~\ref{fig:Minimal}. We find that the PMS largely depends on the perturbative order one is considering\,: only in the two
highest orders a maximum is attained, and between these two differences as large as $3\,$GeV for $n=1$ bottomonium states occur, and
around $1\,$GeV in the other cases we studied. There is a moderate dependence on the number of flavors one is using ($1\,$GeV at worst).
In the case of bottom, the highest order $\MSb$ scheme with $\nl = 3$ shows both a minimum and a maximum for the ground state.
For $n=1$ the maximum position is biased towards large (relativistic) scales.

%
%

The last thing we want to explore in this section is the $m_c$ dependence of the bottomonium binding energies and the various ways in which
the charm quark mass corrections can be implemented. As discussed already in Sec.~\ref{sec:charm}, one can assume a massless charm
quark to which one adds non-zero charm quark mass corrections [\,what we call the $\alpha_s^{(\nl=4)}$ scheme\,], or one can assume a theory 
with a decoupled (that is, integrated out) charm quark to which one adds corrections to the decoupling limit [\,what we call the
$\alpha_s^{(\nl=3)}$ scheme\,]. In Refs.~\cite{Brambilla:2001qk,Ayala:2014yxa}
it has been argued that the latter is the most accurate description, because the decoupling limit is much closer to the physical situation
than the massless limit. Here we confirm this claim, and actually show that it holds for almost any value of the charm quark mass, as
can be seen in Fig.~\ref{fig:mcSchemes}\,: the decoupling limit and the $\alpha_s^{(\nl=3)}$ scheme are much close to one another than the
massless limit and the $\alpha_s^{(\nl=4)}$ scheme for most values of the charm mass, except for very small values of $m_c$ where
(as expected) the two former blow up. The main reason for the proximity of the decoupling limit and the $\alpha_s^{(\nl=3)}$ scheme is that
for both of them the dependence on the charm quark mass comes mainly from $\alpha_s^{(\nl=3)}$ and $m_b^{{\rm MSR}\,(\nl=3)}$,
whereas the massless limit has no dependence on $\mbar_c$ at all, while the $\alpha_s^{(\nl=4)}$ scheme
depends on the charm quark mass through corrections to the static energy and the $\MSb$ pole dependence.

%
%
\section{Fits to Experimental Data}\label{sec:fits}
We have created two independent computer codes, one (slow) written in Mathematica~\cite{mathematica} and 
another one (fast) written in Fortran 2008~\cite{gfortran}. Both codes agree within $8$ significant digits.
The duplication of the code allows a cross-check of all the formulas and procedures described above,
giving confidence in our results. For our numerical analysis we mainly use the Fortran code.

The aim of our code is to compute the perturbative mass of a heavy quarkonium state
with arbitrary quantum numbers [\,Eq.~\eqref{eq:EXpole}\,] in the MSR mass scheme\,\footnote{Our code can also be run in the
pole and $\MSb$ schemes.} as a function of the $\MSb$ mass of the heavy quark and two renormalization scales $\mu$ and $R$.
For a given dataset, the value of $\mbar_Q$ is extracted from the minimum of the following $\chi^2$ function, which depends on
a set of pairs of renormalization scales $\{\mu_n,R_n\}$, one for each principal quantum number that appears in the dataset\,:
\begin{align}
\chi^2(\{\mu_n,R_n\})=\sum_i\left(\frac{M_i^{\rm exp}-M_i^{\rm pert}(\mu_i,R_i,\mbar_Q)}{\sigma_i^{\rm exp}}\right)^2,
\end{align}
where the sum extends to the individual states in the given dataset and $M_i^{\rm exp}$ and $\sigma_i^{\rm exp}$ are the experimental
masses and errors extracted from the PDG~\cite{Patrignani:2016xqp}.
Such fit would give us a best-fit $\MSb$ mass value as a function of $\mu_n$ and $R_n$.

One could think the optimal way of including theoretical uncertainties in our fits is by adding these as part of the $\chi^2$ functions.
The problem is that theoretical errors are highly correlated among various states, as stems from the following facts\,:
\begin{enumerate}
\item All masses are determined from the same static potential. Therefore whatever the unknown N$^4$LO static potential is, it is the
same for all states. Varying the renormalization scale $\mu$ uncertainties coming from this  and higher-order unknown terms are estimated.
Therefore the value of $\mu$ has to be varied in a correlated way for all states included in the dataset.
\item All masses in a short-distance scheme use the same $\MSb$-pole mass relation to cancel the factorially divergent behavior, and no
matter how the $\mathcal{O}(\alpha_s^5)$ correction looks like, it is going to the same for all states. Uncertainties coming from this and
higher order uncalculated terms are estimated varying the scale $R$. Therefore the value of R has to be varied in a correlated way for
all states in the dataset.
\item All masses depend on the same value of $\alpha_s^{(\nf=5)}(m_Z)$ and $\mbar_c$, therefore uncertainties associated to these two quantities
are obviously $100 \%$ correlated.
\end{enumerate}
We performed a numerical study to estimate the theoretical error matrix and found the correlation is very close to $100\%$.
Given that perturbative errors greatly dominate over experimental uncertainties, a fit including such a highly correlated error
matrix is severely affected by the so-called d'Agostini bias~\cite{DAgostini:1993arp}, resulting in a global best-fit value which
is considerably lower than individual determinations from each state in the dataset. Two possible ways out of this problem are\,:
a)~perform global fits with $\chi^2$ functions which depend on $\mu$ and $R$;
b)~determine the heavy quark masses from fits to individual states and a posteriori perform a weighted average of the best-fit results. For
the latter one has to compute the weighted average using only the experimental errors, since the highly correlated theoretical errors cannot
judge the quality of individual results (and would again cause the bias towards small values). The theoretical uncertainty
is then estimated as the average of the individual theoretical uncertainties. In a more formal way,
this is achieved by using for the weighted average a theoretical correlation matrix in
which each entry is substituted by the average of all the entries in the original matrix. This procedure is quite standard, and it is
motivated by the idea that the parameter one is fitting for is unique, and therefore its theoretical error is unique as well, being the
average of its various estimates the least biased approximation. A theoretical error matrix exactly proportional to the unit matrix avoids
the d'Agostini bias and yields as the result of the fit precisely the weighted average with experimental uncertainties only.
For our analysis we take approach a) as our default and use b) as a validation of our results. We find both methods are in quite good
agreement, although in same cases b) yields slightly larger perturbative uncertainties.

In order to estimate the dependence of $\mbar_Q$ with the renormalization scales we vary them, within reasonable ranges,
in an independent way. The specific energy windows for each scale depend on the convergence behavior for each heavy quarkonium
state. As already argued in Sec.~\ref{sec:investigation}, for $n=1$ states one should vary the renormalization scales between 
$(\mu_1,R_1)\in[\,1.5,4\,]$\,GeV, whereas for states with $n=2$ one should rather take the ranges $(\mu_2,R_2)\in[\,1,4\,]$\,GeV.
For $n=3$ states
perturbation theory is already badly behaved such that scales around $1$\,GeV are too low and the perturbative series starts to
fail, which significantly increases the error associated to the scales. For that reason, the ranges used for $n=3$ will be
$(\mu_3,R_3)\in[\,1.5,4\,]$\,GeV, which still yields uncertainty bars much larger than for $n=1,2$ states.
In any case $n=3$ states are only used as an illustration and do not enter our final numbers. Since the $\mu_n $ and $R_n$ scales have to
be varied together to account for theory correlations we use the following parametrization\,: $\mu_2(\mu) = \mu$, $R_2(R) = R$,
$\mu_{1,3}(\mu) = 1.5\,{\rm GeV} + 2.5\,(\mu-1\,{\rm GeV})/3$, $R_{1,3}(\mu) = 1.5\,{\rm GeV} + 2.5\,(R-1\,{\rm GeV})/3$,
and vary $\mu$ and $R$ between $1\,$GeV
and $4\,$GeV. This linear rescaling for $n=1,3$ implements the correct correlation to the $n=2$ state avoiding unnaturally low scales,
and makes the $\chi^2$ function dependent on the global $\mu$ and $R$ only. The central value for $\mbar_Q$
will be calculated as the average of all the best-fits on a ($\mu,R$) grid.


From the $\chi^2$ minimization we extract the uncertainty coming from the
experimental error of the heavy quarkonium masses, denoted as $\Delta^{\rm exp}$.
Around the best-fit value of a given ($\mu,R$) pair, the $\chi^2$ function can be approximated by
\begin{align}
\chi^2 \approx \chi^2_{\rm min} + \left(\frac{\mbar_Q-\mbar_Q^{\rm BF}}{\Delta^{\rm exp}}\right)^2.
\end{align}
The central value for the experimental error will be taken as the average of all the $\Delta^{\rm exp}$ in the ($\mu,R$) grid
(which happen to be all very similar). Due to the high accuracy of the PDG heavy quarkonium masses up to $n=3$, such uncertainty
is very small.

The largest source of uncertainty will be that associated to the variation of the scales, dubbed $\Delta^{\rm pert}$.
Its value will be taken as the difference of the maximum and minimum best-fit values in the grid\,:
\begin{align}
  \Delta^{\rm pert}=\frac{1}{2}\left[\,\max(\mbar_{Q}^{\rm BF}(\mu,R))-\min(\mbar_{Q}^{\rm BF}(\mu,R))\,\right].
\end{align}

Finally, the errors associated to the uncertainty on the strong coupling $\Delta^{\alpha_s}$ (using the 2016 PDG world average
$\alpha_s^{(\nf=5)}(m_Z)=0.1181\pm0.0011$~\cite{Patrignani:2016xqp}) and, for the bottom mass fits, the charm mass $\Delta^{\mbar_c}$
($\mbar_c=1.28\pm0.03$\,GeV~\cite{Patrignani:2016xqp}) will be computed by repeating the fits using the 1-$\sigma$ upper and
lower values for these quantities, and taking the half of the differences of the respective resulting best-fit values.\,\footnote{For
charmonium fits we proceed in the same way with the bottom quark mass PDG value to determine the associate error $\Delta^{\mbar_b}$.}

For our fits we have followed two complementary methods, which we have carried out independently as an extra check.
In our first strategy we created grids in $\mbar_Q$ for the theoretical prediction of each individual state.
We created different grids for each pair $\{\mu,R\}$ in the grid and theoretical setup. The grids are then combined into
$\chi^2$ functions for the various datasets, which are minimized to fit for the bottom mass. The second strategy computes the
bottom mass from individual states numerically solving the equation $M_{\rm theo} = M_{\rm exp}$. We find that this equation can
be solved iteratively to arbitrary precision, and that with $10$ iterations one obtains $10$ significant digits. In order
to perform global fits to multiple states, in this second strategy we directly minimize the $\chi^2$ function numerically using the
COMPASS\_SEARCH~\cite{compass} routine. Both methods agree in more than three significant digits.

\renewcommand{\arraystretch}{1.2}\setlength{\LTcapwidth}{\textwidth}
\setlength{\tabcolsep}{5.5pt}
\begin{table}[htb!]\centering
	\begin{tabular}{|l|ccccc|ccccc|}\hline
		&\multicolumn{5}{c|}{$\alpha_s^{(\nl=4)}$} &\multicolumn{5}{c|}{$\alpha_s^{(\nl=3)}$}\\\hline
		States &  $\mbar_b$ & $\Delta^{\rm exp}$ & $\Delta^{\rm pert}$ & $\Delta^{\alpha_s}$ & $\Delta^{\mbar_c}$
		&  $\mbar_b$ & $\Delta^{\rm exp}$ & $\Delta^{\rm pert}$ & $\Delta^{\alpha_s}$ & $\Delta^{\mbar_c}$\\
		\hline
		$\eta_b(1S)$ & $4.219$ & $0.0011$ & $0.045$ & $0.005$ & $0.0004$ & $4.210$ & $0.0011$ & $0.046$ & $0.006$ & $0.0001$ \\ 
		$\Upsilon(1S)$ & $4.234$ & $0.0001$ & $0.048$ & $0.005$ & $0.0004$ & $4.226$ & $0.0001$ & $0.048$ & $0.007$ & $0.0002$ \\ 
		$\chi_{b0}(1P)$ & $4.195$ & $0.0002$ & $0.039$ & $0.020$ & $0.0006$ & $4.187$ & $0.0002$ & $0.050$ & $0.020$ & $0.0010$ \\ 
		$\chi_{b1}(1P)$ & $4.207$ & $0.0002$ & $0.036$ & $0.020$ & $0.0006$ & $4.200$ & $0.0002$ & $0.046$ & $0.021$ & $0.0010$ \\ 
		$h_b(1P)$ & $4.209$ & $0.0004$ & $0.036$ & $0.021$ & $0.0006$ & $4.201$ & $0.0004$ & $0.045$ & $0.021$ & $0.0010$ \\ 
		$\chi_{b2}(1P)$ & $4.214$ & $0.0002$ & $0.035$ & $0.021$ & $0.0006$ & $4.206$ & $0.0002$ & $0.044$ & $0.021$ & $0.0010$ \\ 
		$\eta_b(2S)$ & $4.233$ & $0.0018$ & $0.044$ & $0.023$ & $0.0006$ & $4.222$ & $0.0018$ & $0.065$ & $0.023$ & $0.0012$ \\ 
		$\Upsilon(2S)$ & $4.241$ & $0.0001$ & $0.040$ & $0.023$ & $0.0006$ & $4.230$ & $0.0001$ & $0.059$ & $0.024$ & $0.0012$ \\ 
		$\Upsilon(1D)$ & $4.244$ & $0.0006$ & $0.057$ & $0.027$ & $0.0006$ & $4.238$ & $0.0006$ & $0.072$ & $0.028$ & $0.0014$ \\ 
		$\chi_{b0}(2P)$ & $4.265$ & $0.0003$ & $0.063$ & $0.029$ & $0.0006$ & $4.257$ & $0.0003$ & $0.080$ & $0.029$ & $0.0015$ \\ 
		$\chi_{b1}(2P)$ & $4.274$ & $0.0002$ & $0.064$ & $0.029$ & $0.0006$ & $4.267$ & $0.0002$ & $0.081$ & $0.029$ & $0.0015$ \\ 
		$h_b(2P)$ & $4.276$ & $0.0005$ & $0.064$ & $0.029$ & $0.0006$ & $4.269$ & $0.0005$ & $0.081$ & $0.029$ & $0.0015$ \\ 
		$\chi_{b2}(2P)$ & $4.280$ & $0.0002$ & $0.064$ & $0.029$ & $0.0006$ & $4.272$ & $0.0002$ & $0.081$ & $0.029$ & $0.0015$ \\ 
		$\Upsilon(3S)$ & $4.309$ & $0.0002$ & $0.070$ & $0.030$ & $0.0006$ & $4.301$ & $0.0002$ & $0.088$ & $0.030$ & $0.0016$ \\ 
		\hline
		$n = 1$ & $4.234$ & $0.0017$ & $0.048$ & $0.005$ & $0.0004$ & $4.225$ & $0.0018$ & $0.048$ & $0.007$ & $0.0002$ \\ 
		$n = 2$ & $4.219$ & $0.0081$ & $0.033$ & $0.021$ & $0.0006$ & $4.210$ & $0.0078$ & $0.042$ & $0.022$ & $0.0011$ \\ 
		$n = 3$ & $4.282$ & $0.0080$ & $0.065$ & $0.029$ & $0.0006$ & $4.275$ & $0.0079$ & $0.083$ & $0.029$ & $0.0015$ \\ 
		$L=P$ & $4.207$ & $0.0039$ & $0.036$ & $0.020$ & $0.0006$ & $4.199$ & $0.0039$ & $0.046$ & $0.021$ & $0.0010$ \\ 
		$n \le 2$ & $4.225$ & $0.0084$ & $0.030$ & $0.016$ & $0.0005$ & $4.216$ & $0.0089$ & $0.034$ & $0.017$ & $0.0008$ \\ 
		$n \le 3$ & $4.240$ & $0.0100$ & $0.021$ & $0.019$ & $0.0005$ & $4.231$ & $0.0106$ & $0.032$ & $0.020$ & $0.0010$ \\ 
		\hline
	\end{tabular}
	\caption{\label{tab:mbmassMSRn} Results for $\mbar_b$ fits for different set of states in the MSRn scheme at N$^3$LO. The first
		$14$ rows correspond to fits to individual states, whereas the last $6$ rows show global fit results.
		Columns 2 to 6 and 7 to 11 show the $\alpha_s^{(\nl=4)}$ and $\alpha_s^{(\nl=3)}$
		schemes, respectively. Within each flavor-scheme block, the uncertainty is split in columns second to fifth into experimental,
		perturbative, due to uncertainty in $\alpha_s^{(\nf=5)}(m_Z)$ and due to uncertainty in $\mbar_c(\mbar_c)$, respectively.
		For the various global fits the error associated to the experimental
		uncertainty (that is, the error coming from the fit) is rescaled by the factor $\sqrt{\chi^2_{\rm min}/{\rm d.o.f.}}$
		All masses and errors are expressed in GeV. }
\end{table}

\subsection[Bottomonium Fits and Determination of $\mbar_b$]
{Bottomonium Fits and Determination of $\mathbf{\mbar_b}$}\label{sec:fitsmb}
\begin{figure*}[t]
	\center
	\subfigure[]
    {\label{fig:Bottom-mass-states}\includegraphics[width=0.5\textwidth]{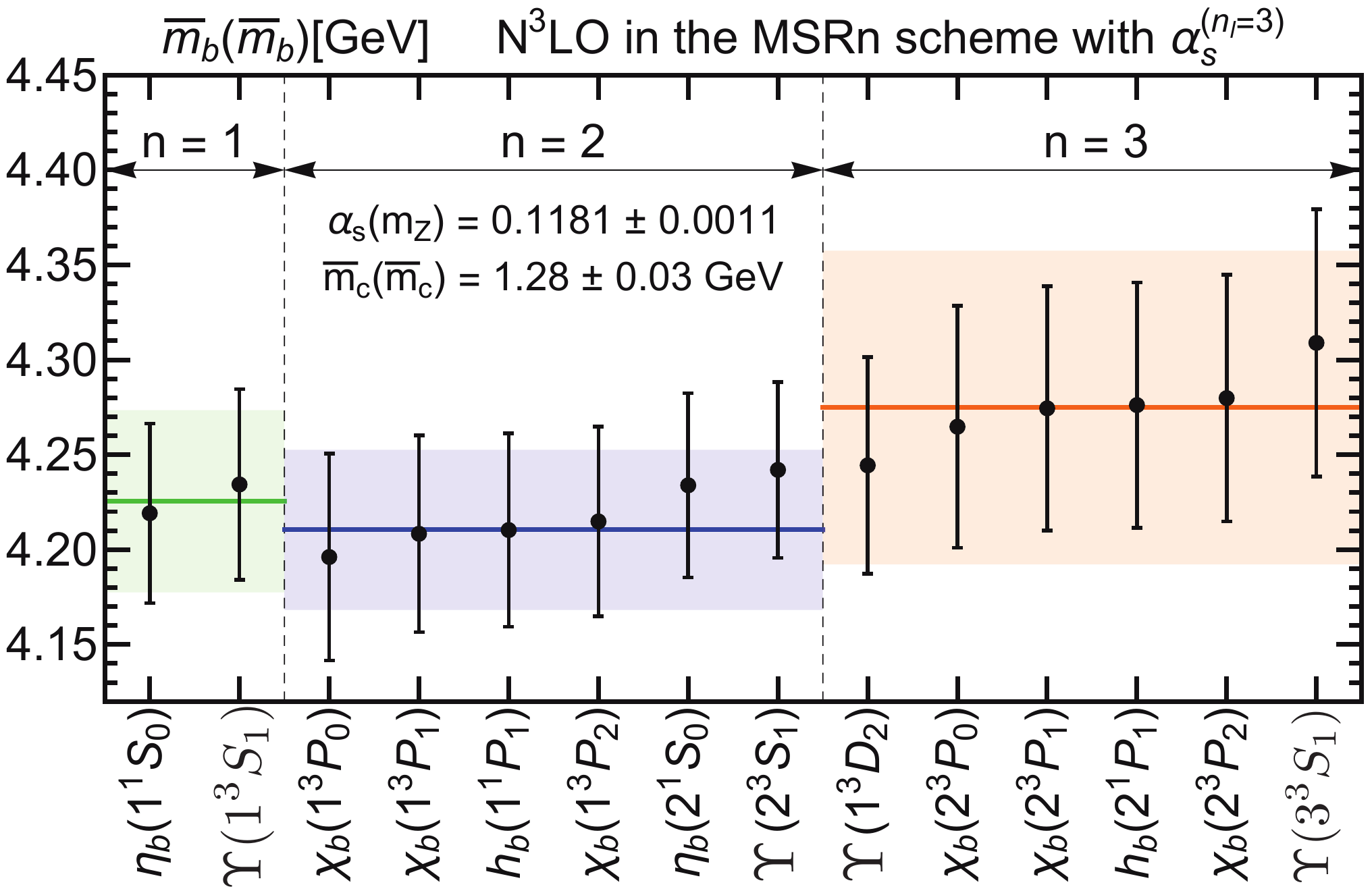}~~~}
	\subfigure[]
{\label{fig:Bottom-mass-orders}\includegraphics[width=0.45\textwidth]{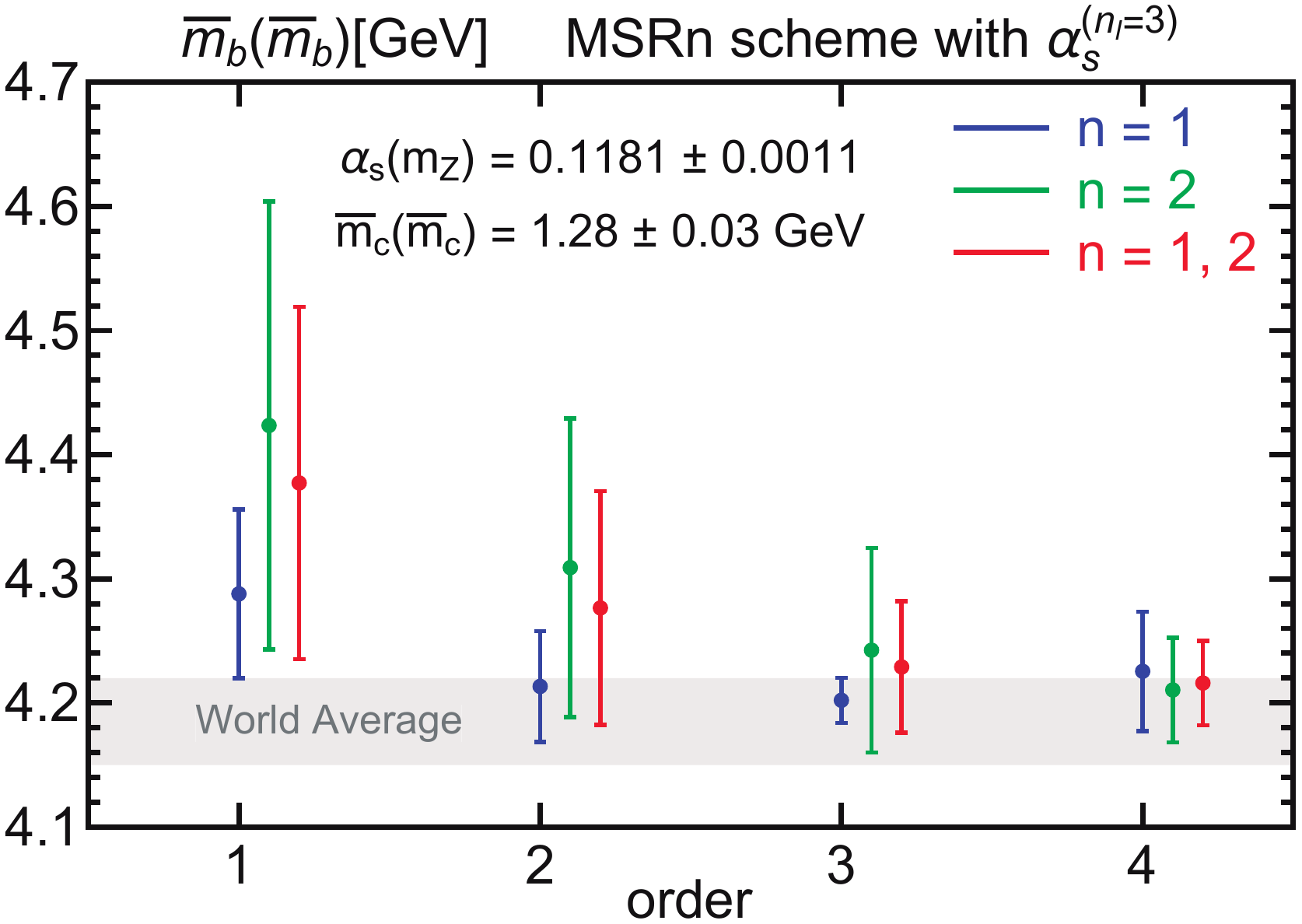}}
 \caption{\label{fig:bottomResults} Left panel\,: Bottom quark mass determinations from individual fits to bottomonium states
with principal quantum number $n\le3$ (black dots with error bars) and global fits to $n = 1$ (green band), $n = 2$
(blue band), and $n = 3$ states (orange band). All computations at N$^3$LO in the MSRn scheme with $\nl = 3$ active flavors.
Right panel\,: Global fits to $n = 1$ (blue), $n=2$ (green) and $n=1,\,2$ (red) as a function of the perturbative order N$^n$LO.
The PDG world average value is shown with a gray band. Both panels\,: only perturbative uncertainties shown. }
\end{figure*}

\begin{figure*}[t]
	\center
	\subfigure[]
    {\label{fig:Bottom-mass-flavors}\includegraphics[width=0.48\textwidth]{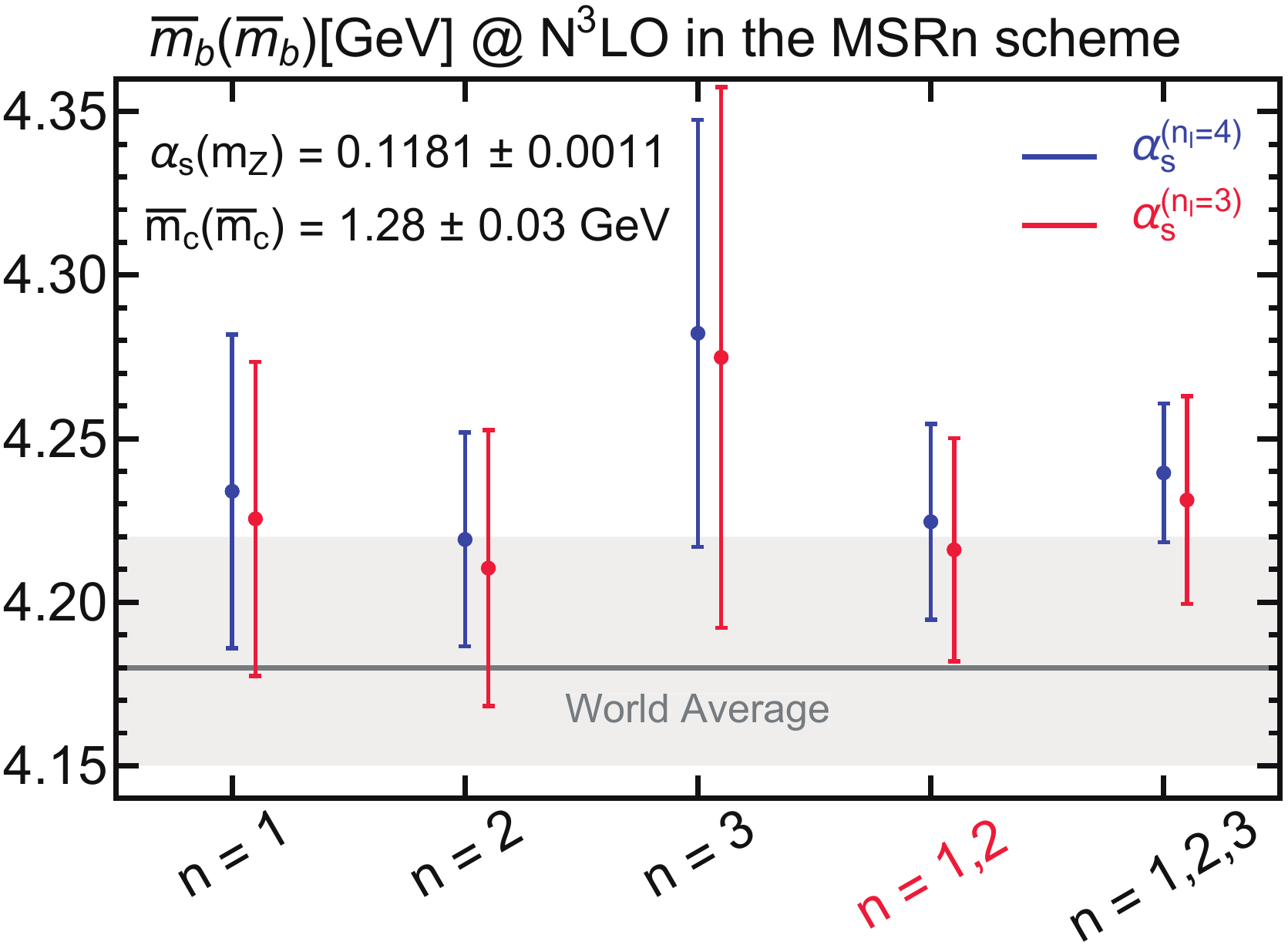}~~~}
	\subfigure[]
{\label{fig:BottomoniumPrediction}\includegraphics[width=0.46\textwidth]{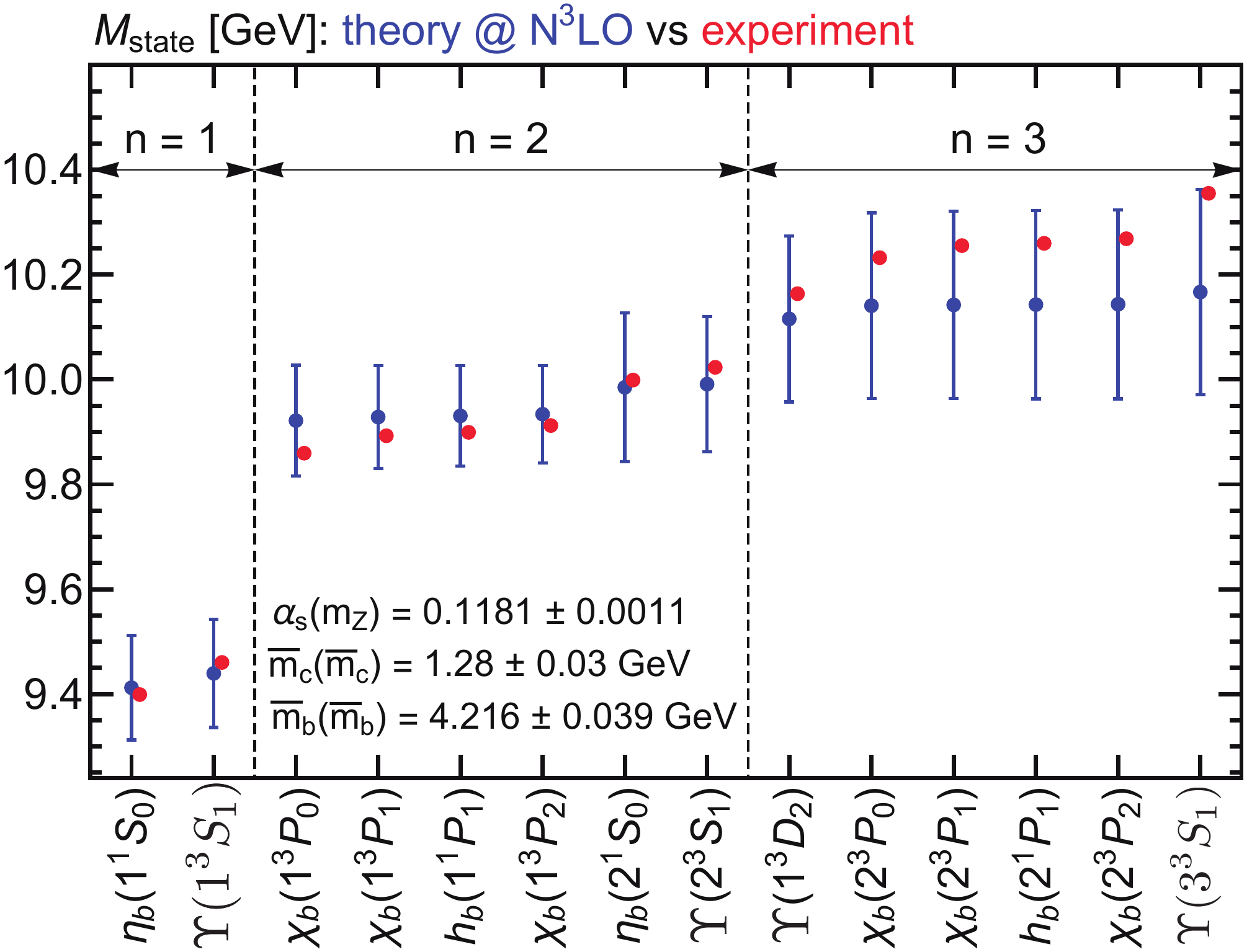}}
 \caption{\label{fig:bottomResults2} Left panel\,: Bottom quark mass determined at N$^3$LO in the MSRn scheme with $\nl=4$ (blue) and
$\nl=3$ (red) flavors from global fits to $n=1$, $n=2$, $n=3$, $n=1,2$ and $n=1,2,3$ states. Right panel\,: Prediction of the
bottomonium energy levels at N$^3$LO (in blue) vs the corresponding experimental values (in red).
Experimental error bars are too small to be visible.}
\end{figure*}

In this section we present the core result of our analysis, the determination of the bottom quark mass. We provide results for fits
to $14$ individual bottomonium states (all states with $n\le3$), as well as for global fits, for which we create a few datasets\,:
\begin{enumerate}
\item Set$_{n=1}$ = \{ $\eta_b(1S)$, $\Upsilon(1S)$ \}.
\item Set$_{n=2}$ = \{ $\chi_{b0}(1P)$, $\chi_{b1}(1P)$, $h_b(1P)$, $\chi_{b2}(1P)$, $\eta_b(2S)$, $\Upsilon(2S)$ \}.
\item Set$_{n=3}$ = \{ $\Upsilon(1D)$, $\chi_{b0} (2 P)$, $\chi_{b1}(2P)$, $h_b(2P)$, $\chi_ {b2}(2P)$,
$\Upsilon(3S)$ \}.
\item Set$_{L=P}$ = \{ $\chi_{b0}(1P)$, $\chi_{b1}(1P)$, $h_b(1P)$, $\chi_{b2}(1P)$ \}.
\item Set$_{n\le2}$ = Set$_{n=1}\;\cup$ Set$_{n=2}$.
\item Set$_{n\le3}$ = Set$_{n=1}\;\cup$ Set$_{n=2}\;\cup$ Set$_{n=3}$.
\end{enumerate}
Our results at N$^3$LO are collected in Tables~\ref{tab:mbmassMSRn}~and~\ref{tab:mbmassMSRp} for the MSRn and MSRp schemes,
respectively, and in Fig.~\ref{fig:Bottom-mass-states} for the MSRn scheme with $\nl=3$ active flavors. The table presents the results
using perturbative expansions in terms of $\alpha_s^{(\nl=4)}$ and $\alpha_s^{(\nl=3)}$, shown in the left
and right blocks, respectively. In columns second to fifth we split the error in its various contributions\,: experimental,
perturbative, and due to the uncertainties in $\alpha_s^{(\nf=5)}(m_Z)$ and $\mbar_c(\mbar_c)$. We observe that the perturbative
uncertainty clearly dominates over the rest, followed by the $\alpha_s^{(\nf=5)}(m_Z)$ error. The error due to the uncertainty on
the charm mass is negligible. The
experimental uncertainty deserves a more detailed explanation\,: for individual fits it is negligibly small in all cases, but
for global fits we modify it to account for the large values of the $\chi^2_{\rm min}$.
For instance, in the Set$_{n \le 2}$ we get $\chi^2_{\rm min}/{\rm d.o.f.} = 21300$, indicating that the theoretical accuracy is
much less than the experimental precision. In order to correct for this deficiency of our fits,
we inflate our experimental fit uncertainties by a factor such that $\chi^2_{\rm min}/{\rm d.o.f.}$ becomes unity. This procedure
cannot be applied to individual fits as for those both $\chi^2_{\rm min}$ and d.o.f are identically zero. Lower orders for
Set$_{n = 1}$, Set$_{n = 2}$ and Set$_{n \le 2}$ in the MSRn scheme with $\nl=3$ active flavors are shown in
Fig.~\ref{fig:Bottom-mass-orders}. Finally in Fig.~\ref{fig:Bottom-mass-flavors} we show the results of various global fits at
N$^3$LO in the MSRn scheme both with $\nl=4$ and $\nl=3$ active flavors.
\begin{table}[htb!]\centering
	\begin{tabular}{|l|ccccc|ccccc|}\hline
		&\multicolumn{5}{c|}{$\alpha_s^{(\nl=4)}$} &\multicolumn{5}{c|}{$\alpha_s^{(\nl=3)}$}\\\hline
		States &  $\mbar_b$ & $\Delta^{\rm exp}$ & $\Delta^{\rm pert}$ & $\Delta^{\alpha_s}$ & $\Delta^{\mbar_c}$
		&  $\mbar_b$ & $\Delta^{\rm exp}$ & $\Delta^{\rm pert}$ & $\Delta^{\alpha_s}$ & $\Delta^{\mbar_c}$\\
		\hline
		$\eta_b(1S)$ & $4.219$ & $0.0011$ & $0.047$ & $0.005$ & $0.0004$ & $4.210$ & $0.0011$ & $0.048$ & $0.006$ & $0.0001$ \\ 
		$\Upsilon(1S)$ & $4.234$ & $0.0001$ & $0.050$ & $0.005$ & $0.0004$ & $4.226$ & $0.0001$ & $0.050$ & $0.007$ & $0.0002$ \\ 
		$\chi_{b0}(1P)$ & $4.196$ & $0.0002$ & $0.054$ & $0.020$ & $0.0006$ & $4.189$ & $0.0002$ & $0.069$ & $0.020$ & $0.0010$ \\ 
		$\chi_{b1}(1P)$ & $4.208$ & $0.0002$ & $0.052$ & $0.020$ & $0.0006$ & $4.201$ & $0.0002$ & $0.066$ & $0.021$ & $0.0010$ \\ 
		$h_b(1P)$ & $4.210$ & $0.0004$ & $0.051$ & $0.020$ & $0.0006$ & $4.203$ & $0.0004$ & $0.065$ & $0.021$ & $0.0010$ \\ 
		$\chi_{b2}(1P)$ & $4.215$ & $0.0002$ & $0.050$ & $0.020$ & $0.0006$ & $4.207$ & $0.0002$ & $0.063$ & $0.021$ & $0.0010$ \\ 
		$\eta_b(2S)$ & $4.234$ & $0.0018$ & $0.049$ & $0.022$ & $0.0006$ & $4.223$ & $0.0018$ & $0.066$ & $0.023$ & $0.0012$ \\ 
		$\Upsilon(2S)$ & $4.242$ & $0.0001$ & $0.046$ & $0.023$ & $0.0006$ & $4.232$ & $0.0001$ & $0.060$ & $0.023$ & $0.0012$ \\ 
		$\Upsilon(1D)$ & $4.244$ & $0.0006$ & $0.057$ & $0.027$ & $0.0006$ & $4.238$ & $0.0006$ & $0.072$ & $0.028$ & $0.0014$ \\ 
		$\chi_{b0}(2P)$ & $4.265$ & $0.0003$ & $0.064$ & $0.028$ & $0.0006$ & $4.257$ & $0.0003$ & $0.081$ & $0.029$ & $0.0015$ \\ 
		$\chi_{b1}(2P)$ & $4.274$ & $0.0002$ & $0.064$ & $0.029$ & $0.0006$ & $4.267$ & $0.0002$ & $0.081$ & $0.029$ & $0.0015$ \\ 
		$h_b(2P)$ & $4.276$ & $0.0005$ & $0.065$ & $0.029$ & $0.0006$ & $4.269$ & $0.0005$ & $0.082$ & $0.029$ & $0.0015$ \\ 
		$\chi_{b2}(2P)$ & $4.280$ & $0.0002$ & $0.065$ & $0.029$ & $0.0006$ & $4.273$ & $0.0002$ & $0.082$ & $0.029$ & $0.0015$ \\ 
		$\Upsilon(3S)$ & $4.309$ & $0.0002$ & $0.070$ & $0.030$ & $0.0006$ & $4.301$ & $0.0002$ & $0.089$ & $0.030$ & $0.0016$ \\ 
		\hline
		$n = 1$ & $4.234$ & $0.0017$ & $0.050$ & $0.005$ & $0.0004$ & $4.226$ & $0.0018$ & $0.050$ & $0.007$ & $0.0002$ \\ 
		$n = 2$ & $4.220$ & $0.0081$ & $0.048$ & $0.021$ & $0.0006$ & $4.212$ & $0.0078$ & $0.060$ & $0.022$ & $0.0011$ \\ 
		$n = 3$ & $4.282$ & $0.0080$ & $0.066$ & $0.029$ & $0.0006$ & $4.275$ & $0.0079$ & $0.083$ & $0.029$ & $0.0015$ \\ 
		$L=P$ & $4.208$ & $0.0039$ & $0.052$ & $0.020$ & $0.0006$ & $4.201$ & $0.0039$ & $0.065$ & $0.021$ & $0.0010$ \\ 
		$n \le 2$ & $4.225$ & $0.0083$ & $0.041$ & $0.016$ & $0.0005$ & $4.217$ & $0.0088$ & $0.048$ & $0.016$ & $0.0008$ \\ 
		$n \le 3$ & $4.240$ & $0.0100$ & $0.029$ & $0.019$ & $0.0005$ & $4.232$ & $0.0105$ & $0.034$ & $0.020$ & $0.0010$ \\ 
		\hline
	\end{tabular}
	\caption{\label{tab:mbmassMSRp} Same as Table~\ref{tab:mbmassMSRn} for the MSRp scheme. }
\end{table}
\setlength{\tabcolsep}{6pt}

Following the observations made in Ref.~\cite{Ayala:2014yxa}, which we confirmed in Sec.~\ref{sec:investigation}, we favor the expansion
with $\nl=3$ flavors over $\nl = 4$, and take it for our final determination. In any case, the perturbative uncertainties are slightly
larger for $\nl=3$ [\,see Fig.~\ref{fig:Bottom-mass-flavors}\,], so our choice is conservative. Furthermore, we consider Set$_{n\le 2}$
as our default,
as all states contained in it can be accurately described within perturbation theory, and is less biased than taking only ground states.
Finally, we observe than the difference between the MSRn and MSRp schemes is at the sub-MeV level for the central values,
($0.02\%$) but the MSRn scheme yields perturbative uncertainties $13\,$MeV ($28\%$) smaller than the MSRp. Accordingly we consider
the MSRn scheme (which is also theoretically cleaner) as our default, finding our final result for the bottom mass\,:
\begin{align}\label{eq:bottomFinal}
\mbar_b(\mbar_b) & = 4.216 \pm 0.009_{\rm exp} \pm 0.034_{\rm pert}\pm 0.017_{\alpha_s}\pm 0.0008_{\,\mbar_c}\,{\rm GeV}\\
& = 4.216 \pm 0.039\,{\rm GeV} \nonumber\,.
\end{align}
If one uses a correlated scale variation setting $\mu = R$ the central value decreases by $1\,$MeV and the perturbative
error shrinks to $18\,$MeV. If we set $R=4\,$GeV the result reads \mbox{$\mbar_b = 4.233\pm 0.010_{\rm pert}\pm \ldots$}
with a central value $17\,$MeV larger and perturbative error roughly a factor of $3$ smaller.
We have performed fits with $\mbar_c = 0$ and in the decoupling limit. The former is $30\,$MeV above the result in
Eq.~\eqref{eq:bottomFinal} while the latter is $4.7\,$MeV below. The result in the $\nl=4$ scheme is $8.6\,$MeV higher than
our final result with $\nl=3$. If we add this difference to our uncertainties to account for effects of missing
higher-order charm mass corrections we would get a total error of $40\,$MeV. If we add the difference between Set$_{n=1}$ and
Set$_{n=2}$ as an estimate of non-perturbative effects our total uncertainty would raise to $43\,$MeV ($44\,$MeV if both
effects are considered simultaneously). At the sight of this, we believe our uncertainty of $39\,$MeV is conservative enough.

\begin{figure*}[t]
	\center
	\subfigure[]
    {\label{fig:Bottom-mass-alpha}\includegraphics[width=0.47\textwidth]{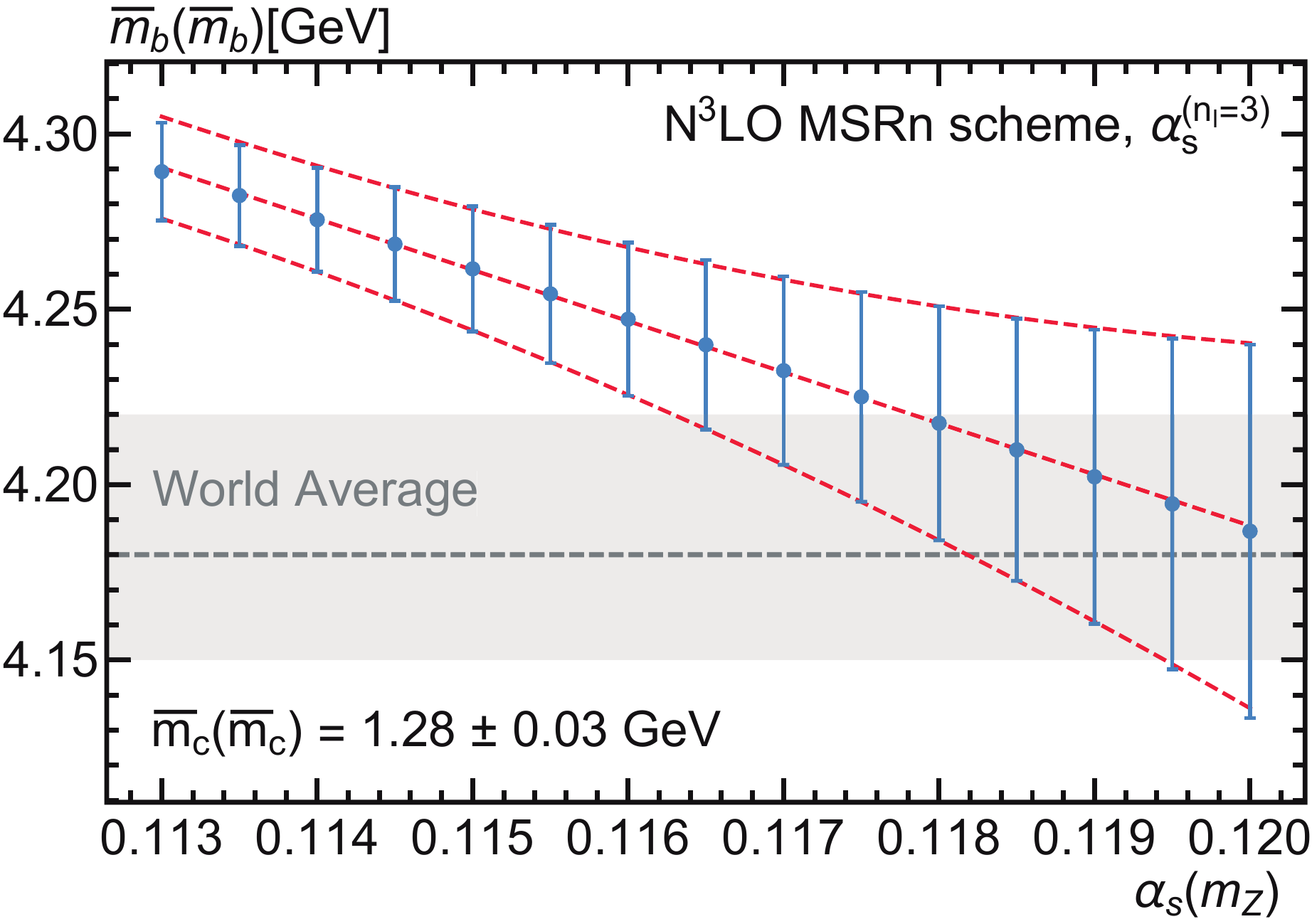}~~~}
	\subfigure[]
{\label{fig:Charm-mass-alpha}\includegraphics[width=0.47\textwidth]{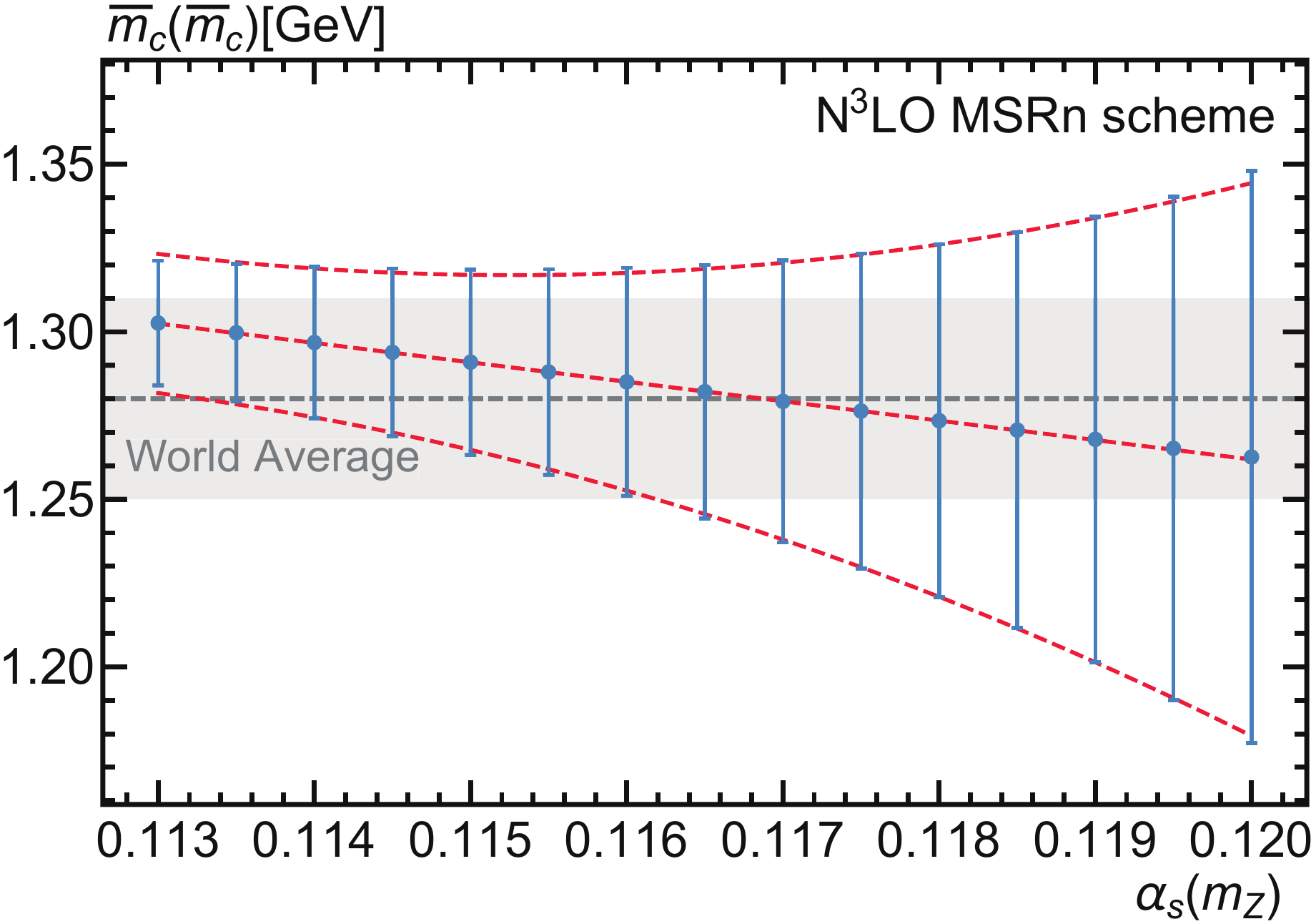}}
 \caption{\label{fig:alpha} Dependence of the bottom (left panel) and charm (right panel) quark masses on the value of $\alpha_s^{(\nf=5)}(m_Z)$.
The determinations correspond to global fits for the $n=1,2$ (bottom) and $n=1$ (charm) states. The MSRn scheme is used at
N$^3$LO order, in both cases in terms of $\alpha_s^{(\nl=3)}$.}
\end{figure*}

In Fig.~\ref{fig:BottomoniumPrediction} and Table~\ref{tab:Resbottom} the predictions for $n\le3$ bottomonium masses are shown
at N$^3$LO using our best-fit value for the bottom quark mass. From the results it is evident that the experimental precision is far superior
than the theoretical accuracy. Finally we explore the dependence of our final result on the value of $\alpha_s^{(\nf=5)}(m_Z)$, finding
\begin{align}\label{eq:bottomFinalAlpha}
\mbar_b(\mbar_b) & = 4.216 - 14.587\, (\alpha_s - 0.1181)\pm 0.0005_{\rm exp} \pm 0.0008_{\,\mbar_c}\\
& \pm [\,0.0341332 + 7.91281\, (\alpha_s - 0.1181) +  798.251\, (\alpha_s - 0.1181)^2\, ]_{\rm pert}\,{\rm GeV}  .\nonumber
\end{align}
The central value rapidly decreases as $\alpha_s^{(\nf=5)}(m_Z)$ grows, but the perturbative error mildly increases [\,see Fig.~\ref{fig:Bottom-mass-alpha}\,].

\begin{table}
  \begin{center}
  \begin{tabular}{|l|cclr|}
    \hline 
    State & $n$ & $^{2s+1}\ell_j$ & $M_{\rm exp}$ [GeV] & $M_{\rm pert}^{(\nl=3)}$ [GeV] \\ 
    \hline
  $\eta_b(1S)$    & $1$ & $^1S_0$ & $9.3990(23)$ & $9.41 \pm 0.10$  \\
  $\Upsilon(1S)$  & $1$ & $^3S_1$ & $9.46030(26)$ & $9.44 \pm 0.10$\\
  $\chi_{b0}(1P)$ & $2$ & $^3P_0$ & $9.85944(52)$ & $9.92 \pm 0.11$ \\
  $\chi_{b1}(1P)$ & $2$ & $^3P_1$ & $9.89278(40)$ & $9.93 \pm 0.10$\\
  $h_b(1P)$       & $2$ & $^1P_1$ & $9.8993(8)$   & $9.93 \pm 0.10$\\
  $\chi_{b2}(1P)$ & $2$ & $^3P_2$ & $9.91221(40)$ & $9.93 \pm 0.09$ \\
  $\eta_b(2S)$    & $2$ & $^1S_0$ & $9.999(4)$    & $9.99 \pm 0.14$ \\
  $\Upsilon(2S)$  & $2$ & $^3S_1$ & $10.02326(31)$& $9.99 \pm 0.13$ \\
  $\Upsilon(1D)$  & $3$ & $^3D_2$ & $10.1637(14)$  & $10.12\pm 0.16$ \\
  $\chi_{b0}(2P)$ & $3$ & $^3P_0$ & $10.2325(6)$   & $10.14\pm 0.18$ \\ 
  $\chi_{b1}(2P)$ & $3$ & $^3P_1$ & $10.25546(55)$ & $10.14\pm 0.18$ \\
  $h_b(2P)$       & $3$ & $^1P_1$ & $10.2598(12)$  & $10.14\pm 0.18$ \\
  $\chi_{b2}(2P)$ & $3$ & $^3P_2$ & $10.26865(55)$ & $10.14\pm 0.18$ \\
  $\Upsilon(3S)$  & $3$ & $^3S_1$ & $10.3552(5)$   & $10.17\pm 0.20$ \\
    \hline 
  \end{tabular}
  \end{center}
  \caption{\label{tab:Resbottom} Masses of the bottomonium states, up to $n=3$, calculated with the perturbative
    formula of Eq.~\eqref{eq:EXpole} and with $\nl=3$ dynamical flavors.
    The results are computed with the final bottom quark fit value of $\mbar_b = 4.216$\,GeV.}
\end{table}

\subsection[Charmonium Fits and Determination of $\mbar_c$]
{Charmonium Fits and Determination of $\mathbf{\mbar_c}$}\label{sec:fitsmc}
\begin{figure*}[t]
	\center
	\subfigure[]
    {\label{fig:Charm-mass-orders}\includegraphics[width=0.45\textwidth]{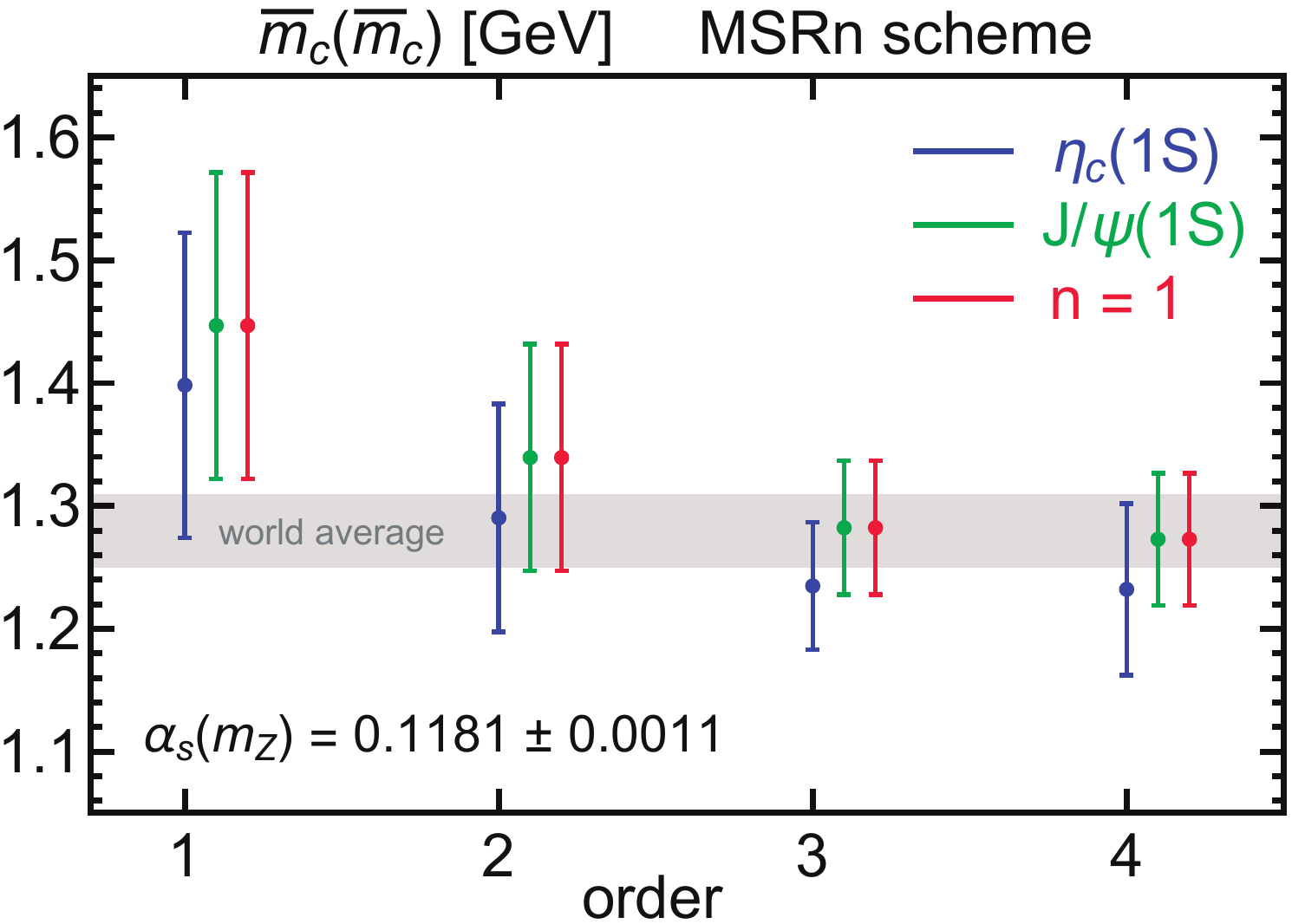}~~~}
	\subfigure[]
{\label{fig:CharmoniumPrediction}\includegraphics[width=0.465\textwidth]{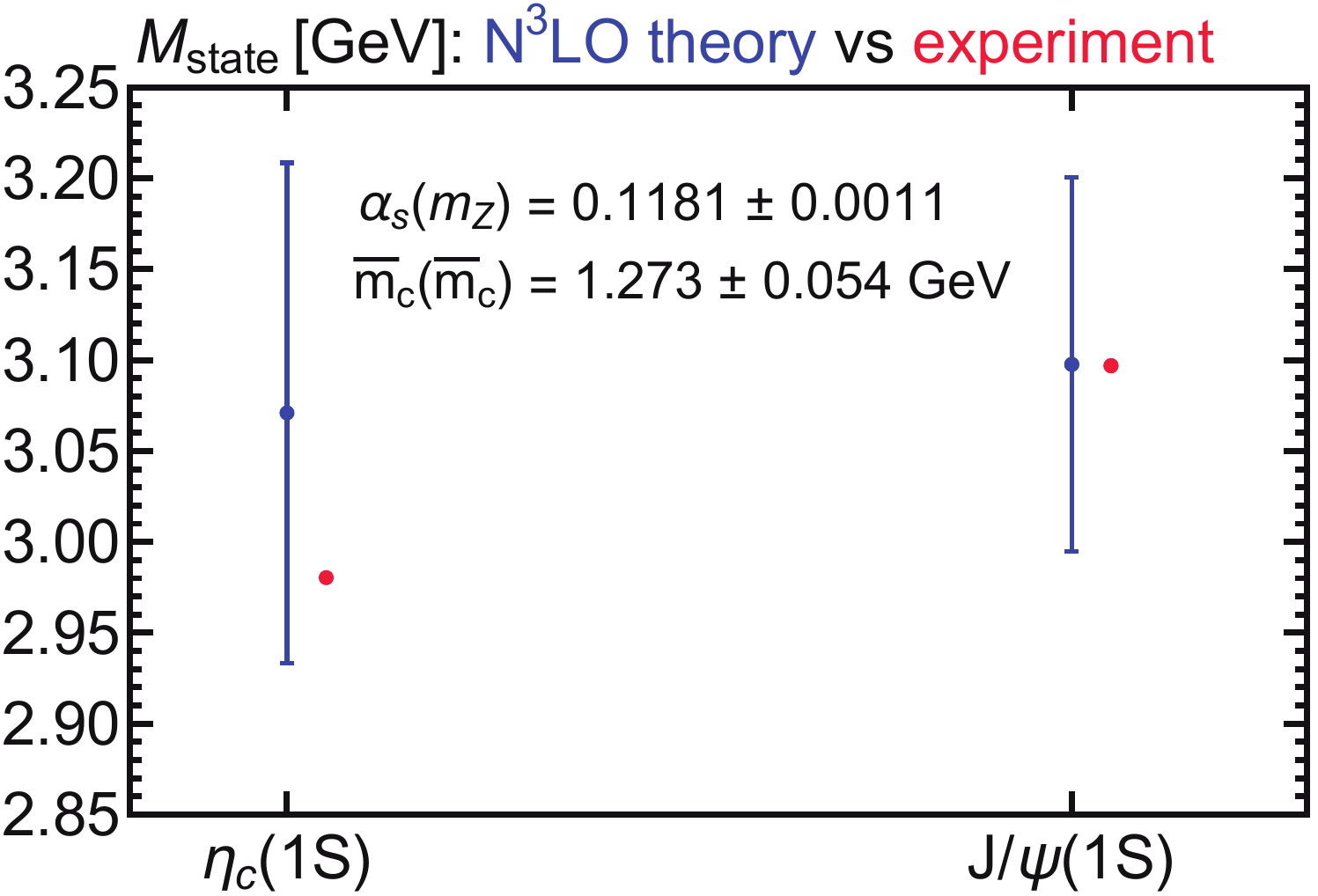}}
 \caption{\label{fig:charmRestuls} Left panel\,: Charm quark mass determination from the $\eta_c$ (blue),
$J/\psi$ (green) and both states together (red) as a function of the perturbative order N$^n$LO.
The PDG world average value is shown with a gray band.
Right panel\,: Prediction of the charmonium masses at N$^3$LO (in blue) vs the corresponding experimental values (in red).
Experimental error bars are too small to be visible.
Both panels\,: predictions in the MSRn scheme where the error bars correspond to the perturbative uncertainties only. }
\end{figure*}

One can use the master formula in Eq.~\eqref{eq:EXSD} in the MSR scheme to compute the masses of $c\bar{c}$ bound states.
One simplification with respect to bottomonium is that there are no lighter quarks with mass larger than $\LQCD$, therefore there
are no corrections from finite mases of lighter quarks. On the other hand the physical scales in the problem are smaller, hence one
expects much larger perturbative (as well as non-perturbative) uncertainties. As discussed in Sec.~\ref{sec:investigation}, we will
restrict our analysis to $n=1$ states and will vary $\mu$ and $R$ between $1.2\,$GeV and $4\,$GeV, which yields order-by-order
convergence and agreement with moderate perturbative uncertainties. We will perform individual (state by state) and global ($n=1$)
fits, which are summarized at N$^3$LO in Table~\ref{tab:mcmass}, whereas lower orders are shown graphically in
Fig.~\ref{fig:Charm-mass-orders}. We find that the result of the global fit is almost identical to the individual fit to the
$J/\psi$ state. This is nothing but expected given that the experimental uncertainty of $J/\psi$ is $80$ times smaller than that of
$\eta_c$. However, the combined fit yields very large $\chi^2_{\rm min}$ values, of the order of $45700$, indicating that the theory
accuracy is much worse than the experimental one. Therefore we apply the same rescaling procedure already applied in the bottom fits.

Our results are collected in Table~\ref{tab:mcmass} for the MSRn (left block) and MSRp (right) schemes. The error is divided into
various contributions\,: experimental, perturbative, and due to the uncertainties in $\alpha_s$ and $\mbar_b(\mbar_b)$. For the
latter we take the 2016 world average \mbox{$\mbar_b = 4.18_{-0.03}^{+0.04}\,$GeV}~\cite{Patrignani:2016xqp}.
Same as we did for the bottom analysis, we observe than theoretical uncertainties greatly dominate over $\alpha_s$ uncertainties,
and again experimental uncertainties (coming form the fit procedure) are negligibly small. The $\mbar_b$ uncertainty is, as
expected, tiny, as the dependence on the bottom  mass comes only from the $\alpha_s$ threshold matching from $5$ to $4$ flavors.
Within three-digit rounding, the total and perturbative errors are the same.

We observe that the difference between the MSRn and MSRp schemes is $1\,$MeV for the central values,
($0.4\%$) but the MSRn scheme yields perturbative uncertainties $9\,$MeV ($10\%$) smaller than the MSRp. Accordingly we again consider
the MSRn scheme as our default, and taking the global fit as the least biased we find our final result for the charm mass\,:
\begin{align}\label{eq:charmFinal}
\mbar_c(\mbar_c) & = 1.273 \pm 0.0005_{\rm exp} \pm 0.054_{\rm pert}\pm 0.006_{\alpha_s}\pm 0.0001_{\mbar_b}\,{\rm GeV}\\
& = 1.273 \pm 0.054\,{\rm GeV} \nonumber\,,
\end{align}
where we have symmetryzed the error due to the uncertainty of $\mbar_b$.
If one uses a correlated scale variation setting $\mu = R$ the central value does not change within three digits but the perturbative
error shrinks to $40\,$MeV. If we set $R=\mbar_c$ the result reads \mbox{$\mbar_c=1.310\pm 0.009_{\rm pert}\pm \ldots$}
If we change the $\Upsilon$-scheme counting to the non-relativistic counting (see discussion in Sec.~\ref{sec:comparison})
the central value becomes $20\,$MeV smaller. Similarly, if we employ the iterative method of Eq.~\eqref{eq:iterative}
the central value grows $20\,$MeV. These variations are much larger than what we find in the bottom analysis, reflecting that
perturbation theory is obviously less convergent for charmonium. Nevertheless they are well contained in our estimate for
higher-order corrections. Finally, if we add the difference between the results to individual fits to $J/\psi$ and $\eta_c$
to our error budget to estimate non-perturbative corrections, our combined uncertainty would increase to $68\,$MeV.

In Fig.~\ref{fig:CharmoniumPrediction} and Table~\ref{tab:Rescharm} we show the prediction for $n=1$ charmonium states
at N$^3$LO using Eq.~\eqref{eq:charmFinal} for charm quark mass. It appears very clear that the experimental precision is
far superior than the theoretical accuracy.
We also explore the dependence of the central value and perturbative error with $\alpha_s^{(\nf=5)}(m_Z)$. We again find a rather linear
dependence of the former while the latter is best approximated by a quadratic fit form\,:
\begin{align}\label{eq:charmFinalAlpha}
\mbar_c(\mbar_c) & = 1.273 - 5.793\, (\alpha_s - 0.1181)\pm 0.0005_{\rm exp}\pm 0.0001_{\mbar_b} \\
& \pm [\,0.054 + 12.66\, (\alpha_s - 0.1181) +  1219.11\, (\alpha_s - 0.1181)^2\, ]_{\rm pert}\,{\rm GeV}  .\nonumber
\end{align}
The situation is the opposite as for the bottom mass\,: the central value slowly decreases as $\alpha_s^{(\nf=5)}(m_Z)$ grows,
but the perturbative error rapidly increases [\,see Fig.~\ref{fig:Charm-mass-alpha}\,].

\subsection[Simultaneous Fits to $\mbar_b$ and $\alpha_s^{(\nf=5)}(m_Z)$]
{Simultaneous Fits to $\mathbf{\mbar_b}$ and $\mathbf{\alpha_s^{(\nf=5)}(m_Z)}$}\label{sec:alphaSFits}
The large value of $\chi^2$ per degree of freedom in fits to the bottomonium spectrum and the big amount
of precise data at our disposal suggests that one could fit another parameter. In this section we
briefly elaborate on the possibility of performing a simultaneous fit to both the bottom quark mass and
the strong coupling constant, as was first suggested in Ref.~\cite{Brambilla:2001qk}.
The first observation is that, for fixed values of the renormalization
scales, one only breaks the degeneracy between $\alpha_s$ and $\mbar_b$ if states with principal
quantum number $n=1$ and $n=2$ are included. For our default dataset $n \le 2$ and fixed values
of the renormalization scales we can determine both parameters with great precision\,:
$4.5\times 10^{-3}\,\%$ for the bottom mass and $9.7\times 10^{-3}\,\%$ for $\alpha_s$. The experimental
correlation between these is $\rho=-\,0.92$ in average.

This nice scenario deteriorates once we perform scale variation in the ranges explained in Sec.~\ref{sec:fits}, yielding sizable
perturbative uncertainties and correlating both parameters even more ($\rho = -\,0.99$). The average $\chi^2/{\rm d.o.f.}$
decreases by a factor of $2.6$ from $21300$ to $8045$. We find
\begin{align}
\mbar_b(\mbar_b) & = \,\,4.219 \,\pm \,\,0.0002_{\rm exp}\, \pm 0.062_{\rm pert}\,\,{\rm GeV}\,,\\
\alpha_s^{(\nf=5)}(m_Z) &= 0.1178 \pm 0.00001_{\rm exp} \pm 0.0050_{\rm pert}\,.\nonumber
\end{align}
Rescaling the experimental errors by $\sqrt{\chi^2/{\rm d.o.f.}}$ the uncertainties increase to $0.016\,$GeV and $0.0010$ for
$\mbar_b$ and $\alpha_s$, respectively.
Although the uncertainty on $\alpha_s$ is only $4\%$ (good in absolute terms), and our determination is certainly compatible
with the world average, it cannot compete with its current precision, which is at the percent level.


\setlength{\tabcolsep}{3.5pt}

\begin{table}[t!]\centering
\begin{tabular}{|l|ccccc|ccccc|}	\hline
	&\multicolumn{5}{c|}{MSRn} &\multicolumn{5}{c|}{MSRp}\\\hline
	States &  $\mbar_c$ & $\Delta^{\rm exp}$ & $\Delta^{\rm pert}$ & $\Delta^{\alpha_s}$ & $\Delta^{\mbar_b}$
	&  $\mbar_c$ & $\Delta^{\rm exp}$ & $\Delta^{\rm pert}$ & $\Delta^{\alpha_s}$ & $\Delta^{\mbar_b}$ \\\hline
	$\eta_c$ & $1.232$ &  $2\!\times\! 10^{-4}$ &  $0.070$  &  $0.005$ & ${}^{+\,\,7\,\times 10^{-5}}_{-\,\,9\,\times 10^{-5}}$ & $1.233$ &  $2\!\times\! 10^{-4}$  &  $0.078$ &  $0.005$ & ${}^{+\,\,6\,\times 10^{-5}}_{-\,\,8\,\times10^{-5}}$ \\
	$J/\psi$ & $1.273$ & $2.7\!\times\! 10^{-6}$ & $0.054$ &   $0.006$ & ${}^{+\,\,8\,\times 10^{-5}}_{-11\times 10^{-5}}$ & $1.274$ &  $2.7\!\times\! 10^{-6}$ & $0.063$  &  $0.006$ & ${}^{+\,\,8\,\times 10^{-5}}_{-11\times 10^{-5}}$ \\\hline
	$n=1$ & $1.273$ & $5\!\times\! 10^{-4}$ & $0.054$ & $0.006$ & ${}^{+\,\,8\,\times 10^{-5}}_{-11\times 10^{-5}}$ & $1.274$ & $5\!\times\! 10^{-4}$ & $0.063$ & $0.006$ & ${}^{+\,\,8\,\times 10^{-5}}_{-11\times 10^{-5}}$\\
	\hline
  \end{tabular}
\caption{\label{tab:mcmass} Results for $\mbar_c$ fits for different set of states at N$^3$LO.
The error is split by their different contributions. All masses and errors are expressed in GeV.
Columns 2 to 5 and 6 to 9 show the MSRn and MSRp schemes, respectively. Within each scheme block,
the uncertainty is split in columns second to fifth into experimental,
perturbative and due to uncertainty in $\alpha_s^{(\nf=5)}(m_Z)$ and $\mbar_b$, respectively.
For the $n=1$ fits the error associated to the experimental
uncertainty (that is, the error coming from the fit) is rescaled by the factor $\sqrt{\chi^2_{\rm min}}$.}
\end{table}

\setlength{\tabcolsep}{6pt}

\begin{table}
  \begin{center}
  \begin{tabular}{|l|ccll|}
    \hline 
    State & $n$ & $^{2s+1}\ell_j$ & $M_{\rm exp}$ [GeV] & $M_{\rm pert}$ [GeV]\\
    \hline
  $\eta_c(1S)$  & 1 & $^1S_0$ & $2.9834(5)$  & $3.07\pm 0.14$  \\
  $J/\psi(1S)$  & 1 & $^3S_1$ & $3.096900(6)$ & $3.10\pm 0.10$  \\
    \hline 
  \end{tabular}
  \end{center}

  \caption{\label{tab:Rescharm} Masses of the charmonium states with $n=1$, calculated with the perturbative
    formula of Eq.~\eqref{eq:EXpole}.
    The results are computed with the final charm quark fit value of $\mbar_c = 1.273$\,GeV.}
\end{table}

\section{Comparison with Previous Determinations}\label{sec:comparison}
We finish our analysis by comparing our results with recent previous determinations, either using quarkonium energy levels or
other observables. We split our comparison in bottom and charm analyses.

\begin{figure*}[t]
	\center
	\subfigure[]
    {\label{fig:BottomoniumComparison}\includegraphics[width=0.48\textwidth]{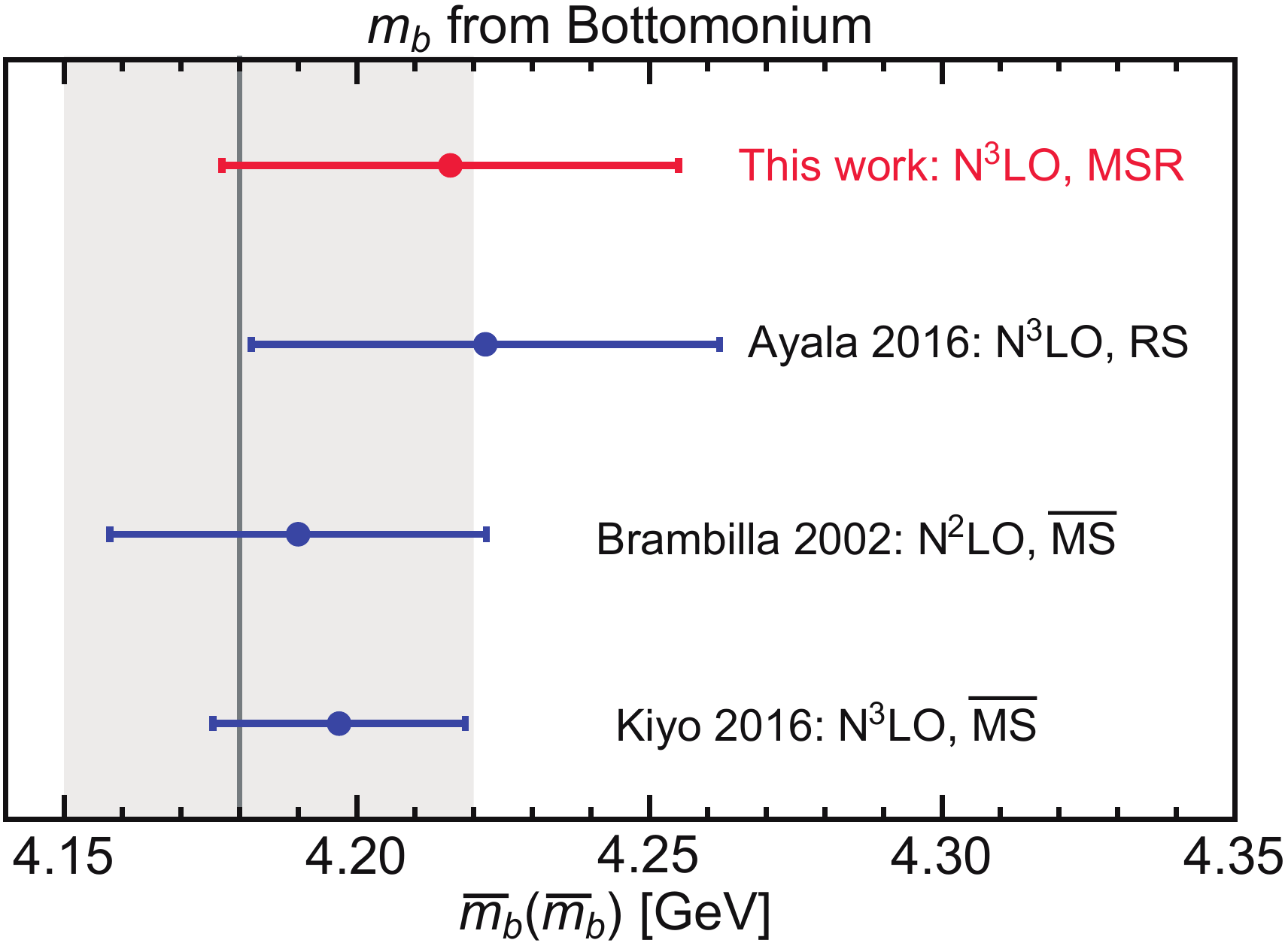}~~~}
	\subfigure[]
{\label{fig:BottomComparison}\includegraphics[width=0.467\textwidth]{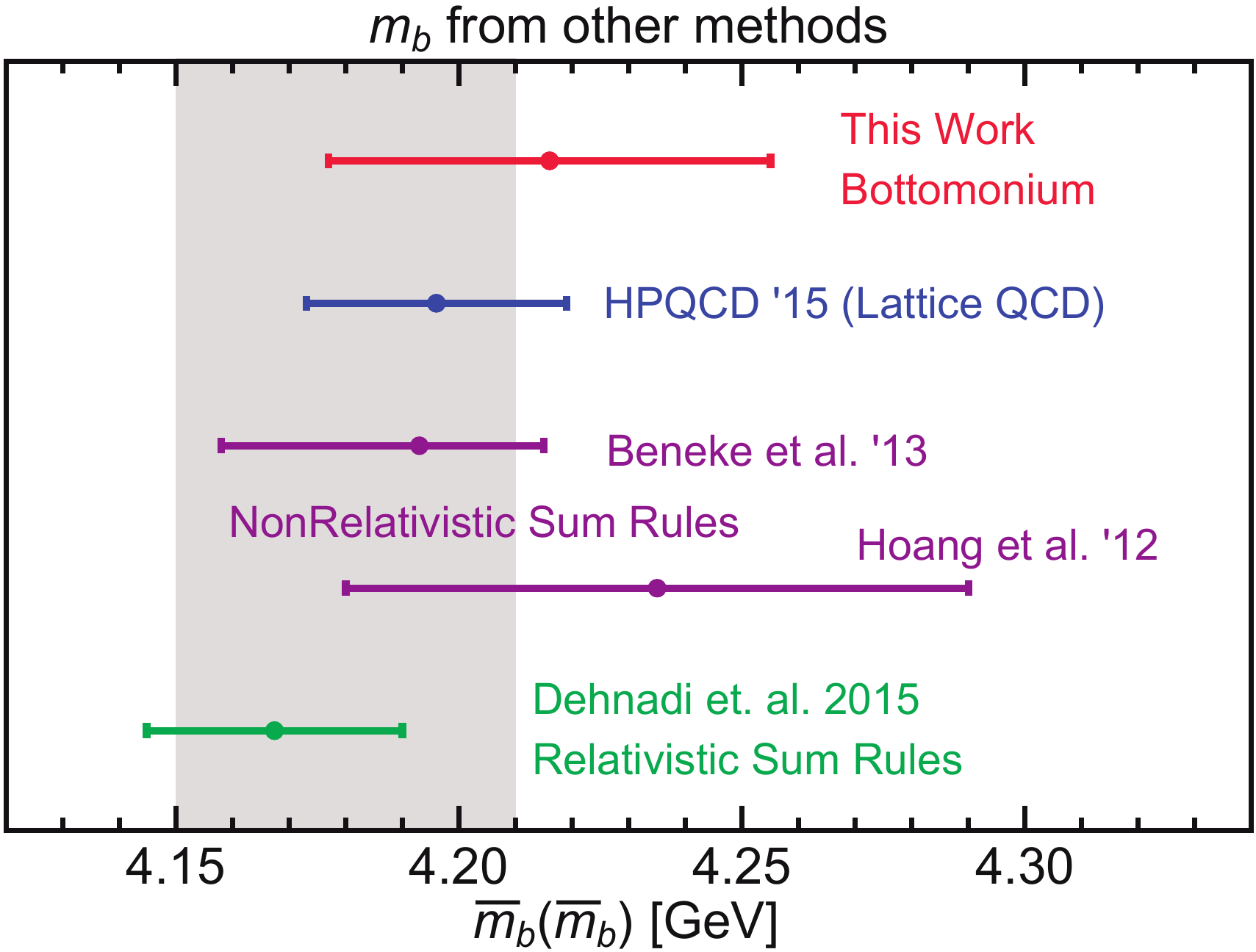}}
 \caption{\label{fig:comparebottom} Comparison of our determination of the bottom $\MSb$ mass (in red) with previous results from bottomonium
(left) and other methods (right). Panel (a) compares a result at N$^2$LO (second from the bottom in blue), with three results at
N$^3$LO. In these analyses various schemes for the bottom mass and different scale variations are used.
Panel (b) shows the outcome of our analysis, a lattice determination (blue), two results from
non-relativistic sum rules (purple) and one from relativistic sum rules (green). Both panels\,: The PDG world average value is shown as a vertical gray band. Note that the PDG world average has asymmetric uncertainties.}
\end{figure*}

\vskip 5mm
\noindent
{\bf Bottom Quark mass}\\[0.1cm]
In Fig.~\ref{fig:BottomoniumComparison} we compare our result for the bottom quark mass with other determinations that also use
bottomonium energy levels. We compare to the N$^2$LO analysis of Ref.~\cite{Brambilla:2001qk} and the N$^3$LO analyses of
Refs.~\cite{Ayala:2016sdn,Kiyo:2015ufa}. While our analysis and those of Refs.~\cite{Ayala:2016sdn,Brambilla:2001qk} use the
$\nl=3$ flavor scheme as the reference calculation, Ref.~\cite{Kiyo:2015ufa} uses $\nl=4$ for the main analyses and takes the
$\nl=3$ computation to estimate uncertainties due to the finite charm quark mass. While
Refs.~\cite{Brambilla:2001qk,Ayala:2016sdn} fit only to the $\Upsilon(1S)$ state and Ref.~\cite{Kiyo:2015ufa} considers both
states with $n=1$, we include all states with principal quantum number $n\le 2$. Ref.~\cite{Kiyo:2015ufa}
takes for their final determination the (unweighted) average of the individual determinations for $\Upsilon(1S)$ and $\eta_b(1S)$,
while we perform global fits which take into account theoretical correlations and weight each individual state by its
experimental uncertainty. Refs.~\cite{Brambilla:2001qk} and \cite{Kiyo:2015ufa} use the $\MSb$ scheme and vary scales according
to the principle of minimal sensitivity. Finally, Ref.~\cite{Brambilla:2001qk} uses $\alpha_s^{(\nf=5)}(m_Z)=0.1181\pm0.0020$,
\cite{Ayala:2016sdn} $\alpha_s^{(\nf=5)}(m_Z)=0.1184\pm0.0007$, and \cite{Kiyo:2015ufa} $\alpha_s^{(\nf=5)}(m_Z)=0.1185\pm0.0007$.

Our theoretical treatment is very similar to that of Ref.~\cite{Ayala:2016sdn}, since both
analyses use a low-scale short-distance mass (MSR vs RS). As for scale variation, for $n=1$ it is identical for $\mu$, while the lower
limit for $R$ is $1\,$GeV in Ref.~\cite{Ayala:2016sdn}, as opposed to our $1.5\,$GeV. On the other hand we vary both scales
independently on a square grid, while Ref.~\cite{Ayala:2016sdn} varies one scale at a time. Both our central values and theoretical
uncertainties are in excellent agreement.

The theoretical error found in Ref.~\cite{Kiyo:2015ufa} is only $20\,$MeV, half the size of ours, while their central value
is $19\,$MeV lower. This uncertainty estimate is based on the PMS, for which the canonical scales
are found to be $5.352\,$GeV and $6.157\,$GeV for the $\Upsilon(1S)$ and $\eta_b(1S)$ states, respectively. We already argued in
Sec.~\ref{sec:investigation} those values are unnaturally large for a non-relativistic system of bottom quarks. The uncertainty is
then estimated by comparing the results of the canonical scale with those that use twice that scale. Since the variation is only
towards larger scales, the variance found is rather small\,: $21\,$MeV and $18\,$MeV, respectively.
We also observe that this scale variation leads to a slight inconsistency between LO and N$^3$LO results.

The result of Ref.~\cite{Brambilla:2001qk} is $26\,$MeV lower than ours, but still in full agreement.
But being only an N$^2$LO analysis, it
claims an uncertainty due to missing higher order corrections of $25\,$MeV, which is $26\%$ smaller than
our estimate. This analysis also uses the PMS to fix the scale, but the perturbative uncertainty is obtained by estimating the
effect of different implementations of the $\Upsilon$-expansion scheme. In our code we can also use two counting schemes, depending
how one considers the scaling of the parameter $R$. The \mbox{$\Upsilon$-expansion} scheme considers \mbox{$R\sim m_Q$}, but in the
so called non-relativistic counting one takes \mbox{$R\sim m_Q\,\alpha_s$}. The difference between our results in the $\Upsilon$-scheme
and non-relativistic counting is $6\,$MeV. We can also compare our results with the expanded out and iterative methods of
Eqs.~\eqref{eq:expanded} and \eqref{eq:iterative}\footnote{These equations can be recast into an expression for the binding
energy of the state, and can therefore be implemented into our $\chi^2$ function.},
finding differences below $6\,$MeV. All these are much smaller than our perturbative uncertainty.

In Fig.~\ref{fig:BottomComparison} we compare the outcome of our analysis with determinations of the bottom mass from other observables.
We compare to a Lattice determination~\cite{Colquhoun:2014ica}, one relativistic~\cite{Dehnadi:2015fra} and two
non-relativistic~\cite{Beneke:2014pta,Hoang:2012us} sum rule analyses.\,\footnote{The result of
Ref.~\cite{Beneke:2014pta} has been updated in \cite{Beneke:2016oox} with the four-loop coefficient of the
$\MSb$-pole mass relation.} Ref.~\cite{Colquhoun:2014ica} uses a non-relativistic lattice
action to compute high moments of the vector correlator.\,\footnote{Another recent lattice determination can be found in
Ref.~\cite{Maezawa:2016vgv}, which computes the ratio $m_b/m_c$ on the lattice and determines $m_b$ from their $m_c$ lattice determination.
They obtain $\mbar_b = 4.184\pm 0.089\,$GeV. The same approach was used in Ref.~\cite{Chakraborty:2014aca} which found
$\mbar_b =  4.162\pm 0.048\,$GeV. These results are less precise than the direct determination of \cite{Colquhoun:2014ica}
$\mbar_b =  4.196\pm 0.023\,$GeV.} These are compared to relativistic continuum perturbation theory to extract
the bottom quark mass, a procedure that could be called into question. Refs.~\cite{Beneke:2014pta,Hoang:2012us} use non-relativistic
sum rules in a low-scale short-distance scheme, but the former uses fixed-order perturbation theory at N$^3$LO while the latter performs
large-log resummation at (partial) N$^2$LL in the framework of vNRQCD. Finally Ref.~\cite{Dehnadi:2015fra} uses the second moment of the vector
correlator in the $\MSb$ scheme and varies independently the scales associated to $\alpha_s$ and $\mbar_b$. The error from
relativistic sum rules are generically smaller since low-energy scales are not being probed.

\begin{figure}\centering
\includegraphics[width=.5\textwidth]{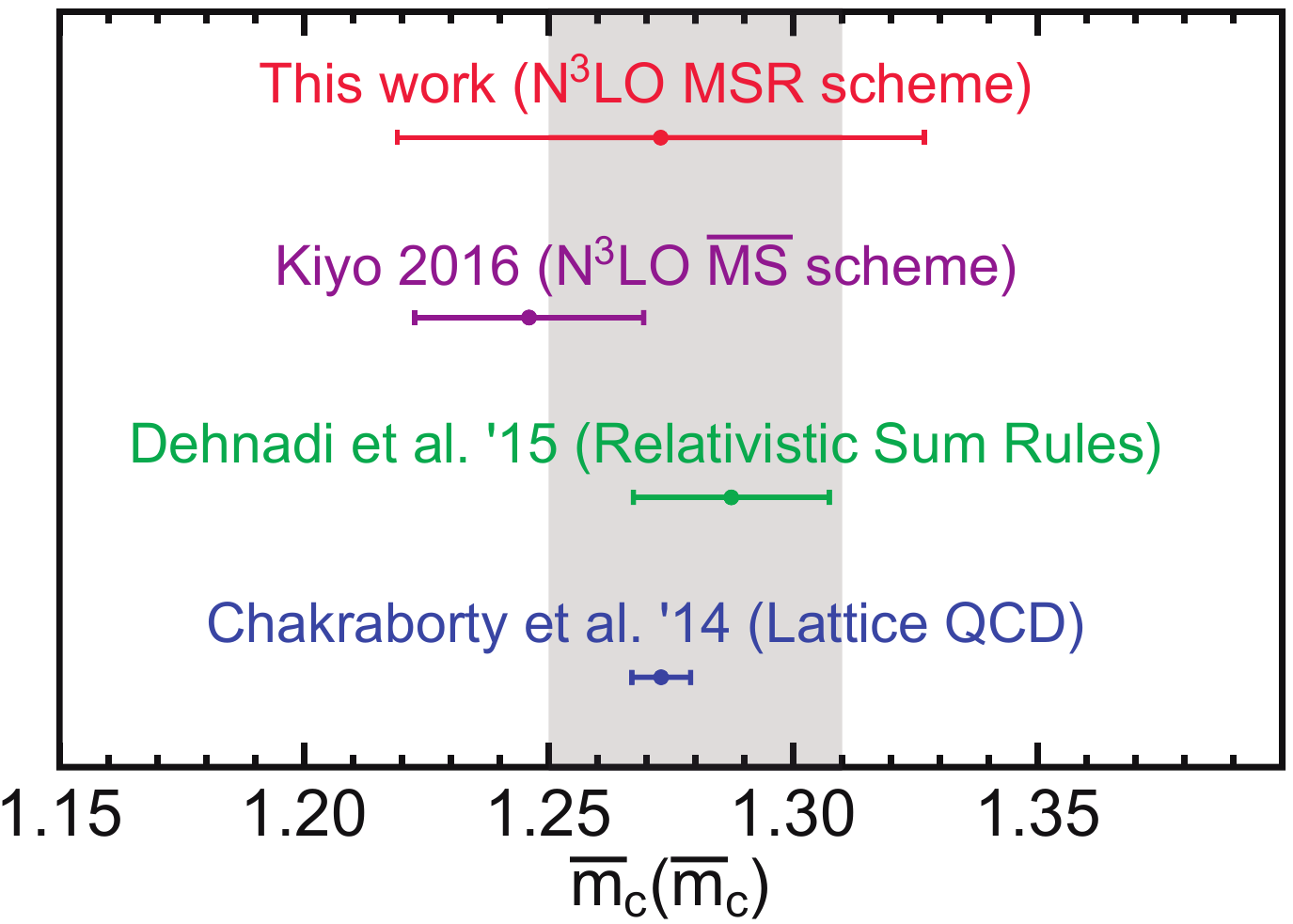}
\caption{Comparison of various charm quark mass determinations from charmonium fits [\,red (this work) and purple\,], relativistic
QCD sum rules (green) and Lattice QCD (blue). The PDG world average value is shown as a vertical gray band.\label{fig:compareCharm}}
\end{figure}

\vskip 5mm
\noindent
{\bf Charm Quark mass}\\[0.1cm]
In Fig.~\ref{fig:compareCharm} we compare our charm quark mass determination with a selection of previous results, which include
relativistic sum rules~\cite{Dehnadi:2015fra}, Lattice QCD~\cite{Chakraborty:2014aca} and fits to the lowest lying charmonium
states~\cite{Kiyo:2015ufa}. In the following we discuss the latter by itself, and two former together.

Let us start by comparing our result in Eq.~\eqref{eq:charmFinal} with the analysis of Ref.~\cite{Kiyo:2015ufa}, which also uses
non-relativistic QCD at N$^3$LO, but employing the $\MSb$ short-distance scheme. Furthermore, the principle of minimal sensitivity is
used to set the default renormalization scale $\mu$, finding $2.14\,$GeV and $2.42\,$GeV for $J/\psi$ and $\eta_c$, respectively. These
are much larger than what we find form requiring a small size for the perturbative logarithms ($\sim 1\,$GeV). Finally, the perturbative
uncertainty is estimated by setting the renormalization scale to twice the minimal sensitivity value. That yields for the average
of their individual determinations an uncertainty of $23\,$MeV, which is a factor of two smaller than our estimate. Their central value
is $27\,$MeV smaller than ours, but results are compatible within errors. Finally, while we perform a global fit two both $n=1$ states,
which takes into account theoretical correlations, Ref.~\cite{Kiyo:2015ufa} determines the charm mass from $J/\psi$ and $\eta_c$
individually and computes the (unweighted) average of those as their final determination.


The analyses of Refs.~\cite{Chakraborty:2014aca,Dehnadi:2015fra} use relativistic sum rules at $\mathcal{O}(\alpha_s^3)$ for the
vector and pseudo-scalar correlators, respectively.\footnote{A similar lattice analysis can be found in
Ref.~\cite{Maezawa:2016vgv} with the result $\mbar_c = 1.267\pm 0.012\,$GeV, which is less precise than the value obtained in
\cite{Chakraborty:2014aca}, $\mbar_c = 1.2715\pm 0.0095\,$GeV.}Although the analyses look similar at first glance, they differ in the way perturbative
uncertainties are estimated. While~\cite{Dehnadi:2015fra} varies renormalization scales associated to $\alpha_s$ and $\mbar_c$
independently in the range $\mbar_c$ to $4\,$GeV, Ref.~\cite{Chakraborty:2014aca} sets $\mu=3\,$GeV and makes an educated guess for
missing higher order corrections, resulting in a much smaller perturbative error. The investigations performed in
Ref.~\cite{Dehnadi:2015fra} suggest that the convergence of the pseudo-scalar perturbative series is significantly worse than of the
vector correlator, and that the theoretical uncertainty of Ref.~\cite{Chakraborty:2014aca} might be underestimated. The reader is
referred to Ref.~\cite{Dehnadi:2015fra} for more details on the analysis.


\section{Conclusions}\label{sec:conclusions}
In this work we have studied the bottomonium and charmonium spectra within Non-Relativistic QCD at the perturbative level up to
N$^3$LO. For the former we have considered corrections coming from the finite charm quark mass up to N$^2$LO, and studied setups
with $\nl=3$ and $\nl=4$, concluding that the former is very close to the decoupling limit and likely captures most of the finite charm
quark mass effects. To cancel the $\mathcal{O}(\LQCD)$ renormalon of the QCD static potential we employ the low-scale
short distance MSR mass, realized in its Practical and Natural versions. We make a dedicated study of scale setting by requiring that
the perturbative logarithms become minimal, and choose to vary around that canonical scale such that the size of the argument of
those logarithms departs $40\%$ from unity. We vary the renormalization scale $\mu$ and the infrared scale $R$ independently in the
same range. With this prescription we find that bottomonium states with principal quantum number
$n$ smaller than $3$ can be described within perturbation theory, but for $n=3$ states no convergence is found. For charmonium
we find that $n=1$ states can be predicted although the convergence of perturbation theory is significantly worse.
One could argue that a more appropriate short-distance mass for bottomonium physics is, for instance, the 1S
mass~\cite{Hoang:1998ng,Hoang:1998hm,Hoang:1999ye}, which does not explicitly depend on any additional low-energy
scale.\footnote{Other equally suitable masses include the PS mass~\cite{Beneke:1998rk}, which we have discarded due to its
dependence on a factorization scale at  $\mathcal{O}(\varepsilon^4)$,  the RS~\cite{Pineda:2001zq}, Kinetic~\cite{Czarnecki:1997sz}
and jet~\cite{Jain:2008gb,Fleming:2007tv} masses, which  we discard because either they are not know to the required accuracy or
do not include finite charm quark mass effects.} 
However $R$ variation accounts for the uncertainty associated in the conversion of the 1S mass into the $\MSb$ mass, in which the MSR mass
plays the role of a smooth interpolator between typical non-relativistic scales of the order of the inverse of the Bohr radius and
relativistic scales as the mass itself.

For bottomonium we include effects of the charm quark mass in the R-evolution of the MSR mass, what ensures there are no Heavy Quark
Symmetry breaking corrections for the Natural MSR mass. Furthermore, requiring that the massive R-anomalous dimension vanishes we
predict the uncalculated  $\mathcal{O}(\alpha_s^4)$ virtual massive quark corrections to the relation between the $\MSb$ and pole
masses, and obtain an (iterative) expression for these corrections in the asymptotic limit. We also define the MSR scheme with
$(\nl\, -\, 1)$ flavors, which is smoothly connected to the MSR$^{(\nl)}$ scheme at the scale $R=\mbar_q$. Of course the MSR$^{(\nl)}$ scheme
is also smoothly connected to the $\MSb$ scheme at $R=\mbar_Q$, what makes our connection between the MSR$^{(\nl-1)}$ and
$\MSb$ schemes smooth. In this context smooth means summing up large logarithms of ratios of scales.

To determine the bottom and charm quark masses we compare to individual states and perform global fits. Since theoretical uncertainties
are highly correlated among the various states we perform fits for given values of the renormalization scales, and scan over
those in a grid to determine
the uncertainties due to missing higher-order corrections. Since the lowest bound in which these scales are varied depends on the principal
quantum number, we correlated the values of $\mu$ and $R$ for each value of $n$ by rescaling the $n=1$ scales to the appropriate ranges.
For our final result we choose the most complete dataset that includes states with $n\le2$ for bottom and $n=1$ for charm. Taking the
MSRn scheme with $\nl=3$ active flavors in both cases we find\,:
\begin{align}\label{eq:bottomCharmFinal}
\mbar_b(\mbar_b) & = 4.216 \pm 0.039\,{\rm GeV} \,,\\
\mbar_c(\mbar_c) & = 1.273 \pm  0.054\,{\rm GeV} \nonumber\,,
\end{align}
where all errors have been added in quadrature, but are fairly dominated by the perturbative uncertainties. We have compared
our results with previous determinations which use either quarkonium spectroscopy or other methods, and find good agreement
with all of them and with the world average. Our estimate of the uncertainty due to missing higher-order corrections is larger
than those of Refs.~\cite{Brambilla:2001qk,Kiyo:2015ufa}, which use the principle of minimal sensitivity, but very similar to
that of Ref.~\cite{Ayala:2016sdn}, which varies both $\mu$ and $R$ scales independently (but one at a time) in a range similar
to ours.

We have also explored the possibility of performing simultaneous fits to $\mbar_b$ and $\alpha_s$, finding that if both $n=1$ and $2$
states are included a fit to both parameters is possible. We find\,:
\begin{align}
\mbar_b(\mbar_b) & = \,\,4.219\, \pm 0.064\,{\rm GeV}\,,\\
\alpha_s^{(\nf=5)}(m_Z) &= 0.1178 \pm 0.0051\,,\nonumber
\end{align}
where again all errors have been added in quadrature and are completely dominated by theoretical uncertainties. The bottom quark mass
uncertainty grows by $64\%$, and the uncertainty on $\alpha_s$, although good in absolute terms, cannot compete with recent
determinations that reach $\sim 1\%$.

Let us close this article discussing possible improvements on our analysis. Although the study performed in this article suggests that
non-perturbative effects are small for $n=1$ and $2$, a more systematic study of these could make our phenomenological observation
into a definite statement. Having higher-order corrections in the near future seems unrealistic,
but one could perform the analysis using a renormalization-group-improved theoretical expression, possibly in the context of pNRQCD.
This would sum up ultra-soft effects that already arise at $\mathcal{O}(\varepsilon^4)$, deteriorating the
convergence of the perturbative series. One would expect that such an RGE analysis would yield smaller perturbative uncertainties,
from which the charmonium analysis would particularly benefit.
On the other hand, one could incorporate the whole QCD static potential at the leading-order Hamiltonian and carry out perturbation
theory around it, as in the study of Ref.~\cite{Pineda:2013lta}. Finally one could use our estimate for the $\mathcal{O}(\alpha_s^4)$
charm mass correction to the $\MSb$-pole mass relation and the decoupling limit approximation to estimate the missing finite charm
quark mass corrections at $\mathcal{O}(\varepsilon^4)$, and check if this brings the $\nl=3$ and $\nl=4$ schemes closer to each another.

\section*{Acknowledgments}

This work has been partially funded by the Spanish MINECO ``Ram\'on y  Cajal'' program (RYC-2014-16022), the MECD grant FPA2016-78645-P, and 
the IFT ``Centro de Excelencia Severo Ochoa'' Program under Grant SEV-2012-0249.
P.G.O. acknowledges the financial support from Junta de Castilla y Le\'on and European Regional Development Funds (ERDF) under
Contract no. SA041U16.
We thank Antonio Pineda for comments on the draft and Pedro Ruiz Femen\'\i a for useful discussions.
\vspace*{0.3cm}

\begin{appendix}
 \section{Formulas}\label{app:formulas}

 In this appendix we provide some expressions related to the finite charm quark mass correction.
In the definition of the $\mathcal{O}(\varepsilon^2)$ correction to the binding energy $\delta E_X^{(1)}$ we need the expression
for the function $f_D(n,\ell,k,x)$, which can be written as
 \begin{equation}\label{eq:dE1fD}
  \begin{array}{l}
   f_D(n,\ell,k,x )=\frac{\df ^{2 n-1}}{\df x^{2 n-1}}\left[\left(2-x^4-x^2\right) x^{2 k+2 \ell+1} \left(
\begin{array}{cc}
 \frac{\cos ^{-1}\left(\frac{1}{x}\right)}{\sqrt{x^2-1}} & x>1 \\
 1 & x=1 \\
 \frac{\log \left(\frac{\sqrt{1-x^2}+1}{x}\right)}{\sqrt{1-x^2}} & x<1 \\
\end{array}
\right)\right]\,,
  \end{array}
 \end{equation}
where $\rho = (2\, n\, m_c^{\rm pole})/[R\, C_F\, \alpha^{(4)}_s(\mu)]$. Furthermore, the $\mathcal{O}(\varepsilon^3)$ charm-mass
correction to the binding energy $\delta E_X^{(2)}$ in the decoupling limit ($m_c\to\infty$) is
\begin{align}\label{eq:dEX2inf}
(\delta E_X^{(2)})_{m_c\to\infty} =& -\left(\frac{\alpha_s^{(4)}(\mu)}{4\pi}\right)^{\!\!2}
\frac{4C_F^2\,\alpha_s^{(4)}(\mu)^2m_b^{\rm pole}}{n^2}\bigg\{\log \left(\frac{\mu }{m_c^{\rm pole}}\right) \!\! \left(\frac{13}{18}-c_1^{(\nl-1)}-\frac{2\,\nl}{9}\right)\nonumber\\
&+L_n \left[\left(\frac{35}{2}-\nl\right) c_1^{(\nl-1)}+\left(\nl-\frac{33}{2}\right) c_1^{(\nl)}+\frac{2\, \nl-35}{3} \log \left(\frac{\mu }{m_c^{\rm pole}}\right)\right.\nonumber\\
&+\left.\frac{8\,\nl-79}{18}\right]+\frac{1}{3}\log^2 \left(\frac{\mu }{m_c^{\rm pole}}\right)-\frac{2}{3} (\nl-17)\, L_n^2-\frac{7}{12}  + c_2^{(\nl-1)}-c_2^{(\nl)}\bigg\},\nonumber\\
 L_n=& \log \left(\frac{n\, \mu }{C_F\, \alpha_s^{(4)}(\mu)\,m_b^{\rm pole}}\right)+H_{\ell+n}\,.
\end{align}

Finally, the three-loop correction from massive lighter quarks to the relation between the pole and the $\MSb$ masses,
$\Delta_3^{(g,\nl,n_h)}$ is parametrized with a Pad\`e approximants, as detailed in Sec.~\ref{sec:MSbarcharm}.
The terms of Eq.~\eqref{eq:Delt3split} have the following form\,:
 \begin{align}\label{eq:PadeEQ}
\Delta_3^{(g)}(\xi) \,=\,& \frac{\,\xi}{\,\xi +6.5849}  \bigg[121.713 \,\xi ^8+43.0875 \,\xi ^7+75.8986 \,\xi ^6-40.2407 \,\xi ^5\\
& +1143.85 \,\xi ^4+2861.3 \,\xi ^3+(26.0741 \,\xi +171.695) \,\xi ^3 \log ^3(\,\xi )+4621.34 \,\xi ^2 \nonumber\\
&+\xi ^3 \log ^2(\,\xi )\,(-19.545 \,\xi ^5-128.702 \,\xi ^4-54.5284 \,\xi ^3-359.064 \,\xi ^2-81.4937 \,\xi )\nonumber\\
&-536.628+(-6.2695 \,\xi ^8-76.1755 \,\xi ^7-194.437 \,\xi ^6+50.0976 \,\xi ^5-1217.64 \,\xi ^4\nonumber\\
&-1300.47 \,\xi^3-8551.79 \,\xi ^2-5659.48 \,\xi -7609.65) \log (\,\xi )-4387.21 \,\xi \nonumber\\
& +9379.48\bigg] +\frac{8}{3}\,\Delta_2^{\MSb}(\xi)\,,\nonumber\\
\Delta_3^{(\nl)}(\xi)\,=\,&\frac{\,\xi}{\,\xi +1.46064}  \bigg[0.430408 \,\xi ^8-0.657671 \,\xi ^7+5.88658 \,\xi ^6-
14.0718 \,\xi ^5-69.2551 \,\xi ^4\nonumber\\
&-18.3617 \,\xi ^3+(-4.74074 \,\xi -6.9245) \,\xi ^3 \log ^3(\,\xi )+(15.4074 \,\xi +22.5046) \,\xi ^3 \log ^2(\,\xi )\nonumber\\
&+36.737 \,\xi ^2+(-0.161088 \,\xi ^8-0.235292 \,\xi ^7-1.70667 \,\xi ^6-2.49282 \,\xi ^5+18.0613 \,\xi ^4\nonumber\\
&+\,96.5648 \,\xi ^3+102.513 \,\xi ^2+70.1839 \,\xi +102.513) \log (\,\xi )-45.6933 \,\xi -97.084\bigg],\nonumber\\
\Delta_3^{(n_h)}(\xi)\,=\,&\frac{\,\xi^2}{\,\xi -1.47336} \bigg[0.13153 \,\xi ^7-0.123376 \,\xi ^6+3.92059 \,\xi ^5-
5.77644 \,\xi ^4+7.36885 \,\xi ^3\nonumber\\
  &-29.5727 \,\xi ^2+\left(0.457144 \,\xi ^3-0.673537 \,\xi ^2+1.8963 \,\xi -2.79393\right) \,\xi ^4 \log ^2(\,\xi )\nonumber\\
  &+\left(-0.0772789 \,\xi ^7+0.11386 \,\xi ^6-3.60296 \,\xi ^5+5.30846 \,\xi ^4-11.1477 \,\xi ^3\right.\nonumber\\
  &\left. +\, 16.4246 \,\xi ^2 +0.0000225622 \,\xi -0.0000332423\right) \log (\,\xi )+41.7971 \,\xi -20.9544\bigg],\nonumber
\end{align}
where $\xi=\mbar_c/\mbar_b$.
 
\end{appendix}

\bibliography{NRQCD}
\bibliographystyle{JHEP}

\end{document}